\documentclass[eqsecnum,twocolumn,tightenlines,floats,floatfix,aps,preprintnumbers,nofootinbib]{revtex4}

\def\bea{\begin{eqnarray}}
\def\eea{\end{eqnarray}}

\def\st#1{{\kern-4pt} \not\!#1}

\def\sp{\kern +3pt}
\def\sm{\kern -3pt}

\def\be{\begin{equation}}
\def\ee{\end{equation}}
\def\ba{\begin{eqnarray}}
\def\ea{\end{eqnarray}}

\usepackage{graphics}
\usepackage{graphicx}
\usepackage{epsf} 
\usepackage{amsmath}
\usepackage{amssymb}
\usepackage{slashed}
\usepackage{bm}

\def\sfrac#1#2{{\textstyle \frac{#1}{#2}}}

\begin{document}

\phantom{0}
\vspace{-0.2in}
\hspace{5.5in}

\preprint{  }

\vspace{-1in}

\title
{\bf 
$\gamma^\ast N \to N^\ast(1520)$ form factors in the spacelike region}
\author{G.~Ramalho$^{1}$  and M.~T.~Pe\~na$^{2}$
\vspace{-0.1in}  }

\affiliation{
$^1$International Institute of Physics, Federal 
University of Rio Grande do Norte, Avenida Odilon Gomes de Lima 1722, 
Capim Macio, Natal-RN 59078-400, Brazil
\vspace{-0.15in}}
\affiliation{
$^2$Centro de F\'isica Te\'orica de Part\'iculas (CFTP), 
Instituto Superior T\'ecnico (IST), Universidade de Lisboa,
Avenida Rovisco Pais, 1049-001 Lisboa, Portugal}

\vspace{0.2in}
\date{\today}

\phantom{0}

\begin{abstract}
The covariant spectator quark model is applied 
to the $\gamma^\ast N \to N^\ast(1520)$ reaction in the spacelike region.
The spin quark core contributions to the 
electromagnetic form factors and helicity transition amplitudes
are estimated 
from the covariant structure of the $N^\ast(1520)$ 
wave function calibrated by the 
experimental data for large 
squared momentum transfer  $Q^2$.
The difference between the model results and the experimental data 
is then used to parametrize the low $Q^2$ behavior, 
where meson cloud effects are assumed to dominate.
This parametrization
can be very useful for future studies of the 
reaction, as well as for the extension of the 
transition form factors to the timelike region.
\end{abstract}

\vspace*{0.9in}  
\maketitle

\section{Introduction}

The electromagnetic structure 
of the hadrons and its connection to 
Quantum ChromoDynamics
is one of the most interesting topics of investigation 
in hadronic physics. 
Recently, accurate data 
involving nucleon resonances (baryons)
was extracted 
from experiments at low and high $Q^2$
($Q^2=-q^2$, where $q$ is the momentum transfer), 
in such facilities as Jefferson Lab (Jlab) and 
MAMI (Mainz) \cite{Aznauryan12a,NSTAR}, demanding 
theoretical interpretations.
These experiments access the electromagnetic structure of several baryons 
through the scattering  of
electrons off nucleons ($N$), inducing 
nucleon electro-excitation reactions  
($e\, N \to e'\, N^\ast$).
These electroproduction reactions proceed through the intermediate step
$\gamma^\ast N \to N^\ast$,
where $\gamma^\ast$ is a virtual photon, with a cross section 
that can be written 
in terms of electromagnetic form factors.

The pattern of the excitation of the nucleon resonances $N^\ast$ is observed
in the total cross section as a function of the $\gamma^\ast N$ 
invariant mass $W$, and
the first excitation is clearly characterized by 
the bump around $W \simeq 1.2$ GeV, identified as
the  state $\Delta(1232)$, which 
defines the first resonance region.
For a review about the $\Delta(1232)$ see 
Refs.~\cite{Aznauryan12a,Pascalutsa07,NDelta,NDeltaD}.
The second bump is a combination of several resonances,
dominated by the $N^\ast(1520)$ 
and $N^\ast(1535)$ states.
This last one was already studied in some detail
(see Ref.~\cite{S11} and references therein).
Here we will study the state $N^\ast(1520)$,
and the $\gamma^\ast N \to N^\ast(1520)$ transition.

The $N^\ast(1520)$ has spin 3/2 
and negative parity ($J^P= \sfrac{3}{2}^-$).
In the context of $\pi N$ 
scattering it contributes to the $D_{13}(1520)$ channel,
with isospin 1/2 and spin 3/2 and $\pi N$ relative orbital momentum $l=2$.
The $\gamma^\ast N \to N^\ast(1520)$ 
reaction is therefore characterized by 
three independent helicity amplitudes, usually defined 
in the final state rest frame: the two transverse
amplitudes $A_{1/2}, A_{3/2}$,  and the longitudinal amplitude
$S_{1/2}$.
Only recently was the longitudinal 
amplitude measured for the first time \cite{Aznauryan09,Mokeev12}.
In the timelike region ($Q^2 < 0$) the $N^\ast(1520)$ state 
also has a relevant contribution to
the dilepton decay reactions  ($\gamma^\ast N \to e^+ e^- N$)
\cite{Faessler03,Kaptari09,Weil12,Agakishiev14}.

The $\gamma^\ast N \to N^\ast(1520)$ was studied 
previously within the framework of nonrelativistic and 
relativistic
quark models~\cite{Close72,Koniuk80,Warns90,Aiello98,Merten02,Santopinto12a,Capstick95,Aznauryan12b,Ronniger13,Golli13},
the single quark transition model (SQTM) 
\cite{Close79,Burkert03,Burkert04}
and 
a collective model for baryons~\cite{Bijker96}. 
The electromagnetic structure of the $N^\ast(1520)$
was also estimated by the EBAC (Jlab) analysis, within a coupled-channel
dynamical model for the  meson-baryon systems~\cite{Diaz08}.
The study of the empirical charge density
distribution for the $N^\ast(1520)$ 
can be found in Refs.~\cite{Tiator09,Tiator11}.
From the experimental side, there are the MAID (Mainz) 
analysis~\cite{Drechse07,Tiator09,Tiator11}, the old data 
from DESY~\cite{DESY} and  NINA~\cite{NINA}, and the
recent data from CLAS (at Jlab) \cite{Aznauryan09,Mokeev12}.
For a review of results see Refs.~\cite{Aznauryan12a,Mokeev12,Aznauryan09}.

In this work we will study the  $\gamma^\ast N \to N^\ast(1520)$  transition
using the covariant spectator quark model 
\cite{Nucleon,Omega,ExclusiveR,Nucleon2,NucleonDIS}, which is
based on the so-called covariant spectator theory~\cite{Gross}. 
This model 
was already applied to the electromagnetic 
structure of the 
nucleon 
\cite{Nucleon,Nucleon2,NucleonDIS,Lattice}
and the $\Delta(1232)$
\cite{NDelta,NDeltaD,Lattice,LatticeD,Delta},
$N^\ast(1440)$ \cite{Roper},
$N^\ast(1535)$ \cite{S11,S11scaling},
$\Delta(1600)$ \cite{Delta1600}, the baryon
octet and decuplet \cite{Omega,Octet,OctetFF,Medium}
and other transitions 
\cite{Transitions,Octet2Decuplet}.
The model was also applied 
to the timelike regime for the $\Delta(1232)$ case, in particular 
to the calculation of the $\Delta$ dielectron Dalitz decay \cite{Timelike}.

In the calculations of the 
transition electromagnetic form factors of the $\gamma^\ast N \to N^\ast(1520)$ reaction
we use the relativistic  impulse approximation, as done in previous works \cite{NDelta,NDeltaD,Lattice,LatticeD,Delta,Omega,OctetFF,Medium} on different reactions.
In this approximation each quark interacts with the photon at a time, implying that the electromagnetic probe does not couple simultaneously with two or three quarks.
In our model the single quark electromagnetic form factor parametrizes 
the quark dressing from quark-antiquark pairs 
and gluons,
reproducing the quark charge and 
generating an anomalous magnetic moment.
This means that meson effects are effectively taken at the level 
of (dressing) one quark only, but processes where the 
meson is exchanged between different quarks,
and therefore is emitted and absorbed collectively by the three quarks, 
by the baryon as a whole, 
are not included~\cite{OctetFF,Octet2Decuplet}.  
Throughout this paper these are the effects 
we refer to when we use the term "meson cloud".
We discuss next the motivation to 
add these effects to the contributions from our covariant 
spectator quark model.

For the $\Delta(1232)$ excitation \cite{NDelta,NDeltaD,Lattice,LatticeD,Delta}
the comparison of our results to the data has shown that 
these meson cloud effects are important in the small $Q^2$ region. This
conclusion is shared with other constituent quark models
\cite{Aznauryan12a,Pascalutsa07}.
Namely, our results for the $\Delta$ electro-excitation are in line with the information on the  pion cloud extracted within a dynamical coupled-channel analysis of an extensive collection of data \cite{Diaz08,Diaz07}. 
Besides, the same conclusion was obtained by 
a less phenomenological calculation, 
the dynamical quark calculation based on the 
Dyson-Schwinger framework~\cite{Gernot} 
which used an underlying dynamics
to generate the diquark propagation, 
and included the photon coupling to the diquark. 
Even with these features, that calculation could not describe 
the experimental data 
for the $\gamma^\ast N \to \Delta(1232)$ magnetic form factor 
in the small $Q^2$ region,
pointing to the importance of meson cloud effects
at the baryon level, as our calculation did.

Turning to the $\gamma^\ast N \to N^\ast(1520)$ reaction, 
in this work we start by writing the $N^\ast(1520)$ wave function 
in spin-flavor and momentum space by imposing 
the  correct symmetries as described 
in Refs.~\cite{Aznauryan12a,Capstick00}.
This wave function is
the superposition of two configurations: one configuration
where the quark core is a $S=1/2$ spin state,
and another  where the quark core is a
$S=3/2$ spin state. 
The mixture coefficient for  the $S=3/2$ configuration
has been estimated to be $\sin \theta_D \approx 0.1$,
suggesting a dominance of the $S=1/2$ configuration 
\cite{Aznauryan12a,Koniuk80}. In our work we confirmed 
the importance of the $S=1/2$ configuration.

Our results for the helicity amplitudes 
at low $Q^2$ are too small when compared to the data.
This seems 
to indicate that the meson cloud effects not included 
in our quark core model play 
a relevant role for the $N^*(1520)$, 
as they do for the $\Delta(1232)$.
The $N^*(1520)$  decay 
to $\pi N$ (60\%) and $\pi \pi N$ (40\%) \cite{Aznauryan12a}, 
gives us already an indication
that diagrams where the photon couples 
to a meson in flight in an intermediate  
baryon-meson state
may very well be important for the $\gamma^\ast N \to N^\ast(1520)$ 
reaction, and point to 
the importance of the meson cloud contributions at the baryon level.

By comparing our results to the data, we 
extract a simple parametrization of their difference, that we interpret then as
meson cloud contributions. 
This parametrization will be very useful in the extension 
of our calculation to the timelike region, which will enable 
us to interpret dilepton production data from $NN$ collisions
\cite{Weil12,Agakishiev14}.

This article is organized as follows:
In Sec.~\ref{secFormalism} we present the formalism 
required to parametrize the electromagnetic structure 
of the $\gamma^\ast N \to N^\ast(1520)$ reaction.
The  covariant spectator quark model 
for the baryon quark cores 
and the baryon effective wave functions 
necessary to calculate the transition
are presented in Secs.~\ref{secCSQM} and~\ref{secBaryonWF}.
In Sec.~\ref{secFF} we derive the results for the 
form factors and helicity amplitudes, 
and discuss how to parametrize the meson cloud contributions.
Details are presented in Appendices~\ref{apStateP1} 
to \ref{apFF-P3}.
The numerical results for form factors and 
helicity amplitudes are presented in Sec.~\ref{secResults}.
Final conclusions are presented in Sec.~\ref{secConclusions}.

\section{Formalism}
\label{secFormalism}

We start by introducing the formalism required for the 
study of the $\gamma^\ast N \to N^\ast(1520)$ transition.
In what follows we will often represent $N^\ast(1520)$ 
as $R$ (from resonance), and will use $M$ for the nucleon mass and $M_R$ 
for the resonance $R$ mass.
We will present the definitions of the 
helicity amplitudes
$A_{3/2}, A_{1/2}$ and $S_{1/2}$ 
(which are experimentally determined)
together with their relation 
to the electromagnetic transition form factors, $G_M$, $G_E$ and $G_C$.

\subsection{Helicity amplitudes}

The electromagnetic transition 
from a $J^P= \frac{1}{2}^+$ state to a 
$J^P= \sfrac{3}{2}^-$ state is 
described by three
amplitudes. They are functions of $Q^2$, and
in the $R$  rest frame they are defined as
\cite{Aznauryan12a}:
\ba
A_{3/2}&=&
\sqrt{\frac{2\pi \alpha}{K}}
\left< R, S_z'=+\sfrac{3}{2} \right|
\epsilon_+ \cdot J \left| N,S_z=+\sfrac{1}{2} \right> 
\nonumber  \\
& &
\label{eqA32} \\
A_{1/2}
&= &\sqrt{\frac{2\pi \alpha}{K}}
\left< R, S_z'=+\sfrac{1}{2} \right|
\epsilon_+ \cdot J \left| N,S_z=-\sfrac{1}{2} \right> 
\nonumber \\
\label{eqA12} \\
S_{1/2}&=&
\sqrt{\frac{2\pi \alpha}{K}}
\left< R, S_z'=+\sfrac{1}{2} \right|
\epsilon_0 \cdot J \left| N,S_z=+\sfrac{1}{2} \right>\frac{|{\bf q}|}{Q},
\nonumber \\
& &
\label{eqS12}
\ea
where $S_z'$ ($S_z$) is the final (initial) 
spin projection,
${\bf q}$ is the photon three-momentum in the rest frame of $R$,
$Q=\sqrt{Q^2}$,  $\epsilon_\lambda^\mu$ ($\lambda=0,\pm 1$) are
the photon polarization vectors 
and $J^\mu$  is the 
electromagnetic transition current in 
proton charge $e$ units.
In the previous equations
 $\alpha \simeq 1/137$ 
is the fine-structure constant  and 
$K= \frac{M_R^2-M^2}{2M_R}$
is the magnitude of the photon momentum (and nucleon)
when $Q^2=0$.
 In the rest frame of $R$ the magnitude of the nucleon three-momentum is
 $|{\bf q}|$, and reads
\be
|{\bf q}|= \frac{\sqrt{Q_+^2Q_-^2}}{2M_R},
\label{eqq2}
\ee
where $Q_\pm^2= (M_R \pm M)^2 + Q^2$.

The transition current $J^\mu$ for the reaction 
$\gamma^\ast N \to N^\ast(1520)$ between an initial nucleon state
with momentum $P_-$, 
and a  final $R$ state 
with momentum $P_+$ 
can be represented in terms of the matrix elements
$J^\mu_{NR}$ defined by the asymptotic states, and given by
\ba
J^\mu_{NR} &\equiv  &
\left< R | J^\mu | N \right> \nonumber \\
&=& 
\bar u_\beta(P_+) \Gamma^{\beta \mu} u (P_-),
\label{eqJ11}
\ea
where $u_\beta$, $u$ are respectively
the Rarita-Schwinger and Dirac spinors. 
The operator $\Gamma^{\beta \mu}$ 
has the general Lorentz structure
\ba
\Gamma^{\beta \mu}=
G_1 q^\beta \gamma^\mu + 
G_2 q^\beta P^\mu + 
G_3 q^\beta q^\mu -
G_4 g^{\beta \mu},
\label{eqJ12}
\ea
where $q=P_+-P_-$ is the transferred momentum and  
$P= \sfrac{1}{2}(P_+ + P_-)$.
In the previous equation $G_i$ ($i=1,..,4$) 
are form factor functions that depend on $Q^2$.
Because of current conservation only three of the four  
$G_i$ form factors are independent, and
one is free to choose which three are to be taken as independent.
For instance, from the knowledge of $G_i$ ($i=1,..,3$), $G_4$ is determined 
by the current conservation condition
$q_\mu J^\mu=0$ as
\ba
G_4=(M_R-M) G_1 + \frac{1}{2}(M_R^2-M^2)G_2 -Q^2 G_3.
\label{eqG4}
\ea
Note that there are alternative
representations of the operator $\Gamma^{\beta \mu}$.
They are all however equivalent  to the ones
given by Eq.~(\ref{eqJ12}) \cite{Aznauryan12a,Jones73}.
Using  (\ref{eqJ11})-(\ref{eqJ12}) we 
can write the amplitudes (\ref{eqA32})-(\ref{eqS12}) as 
\cite{Aznauryan12a,Devenish76}
\ba
\hspace{-.7cm}
& &
A_{1/2}=2  {\cal A}
\left\{
G_4 -\left[
(M_R-M)^2 +Q^2\right]
 \frac{G_1}{M_R} \right\}, \label{eqA12b} \\
\hspace{-.7cm}
& &
A_{3/2}=2 \sqrt{3} {\cal A}
G_4,  
\label{eqA32b} \\
\hspace{-.7cm}
& &
S_{1/2}= - \frac{1}{\sqrt{2}}
\frac{|{\bf q}|}{M_R} {\cal A} \,
g_C,  
\label{eqS12b}
\ea
where
${\cal A}= \frac{e}{4} \sqrt{\frac{(M_R+M)^2+Q^2}{6 M M_R K}}$,
and
\ba
g_C&=& 4M_R G_1 + (3 M_R^2+M^2+Q^2)G_2 \nonumber \\
& & + 
2 (M_R^2-M^2-Q^2) G_3.
\label{eqGCsmall}
\ea

The obtained formulas for the helicity amplitudes suggest 
that  for the  $\gamma^\ast N \to N^\ast(1520)$ reaction
it is convenient to choose  as independent functions the three form factors
$G_1,G_4$ and $g_C$. Equations (\ref{eqG4}) and (\ref{eqGCsmall}) 
can be used to express
$G_2$ and $G_3$ in terms of those three quantities. 
In addition, one concludes that if $G_4=0$ 
then $A_{3/2}$  vanishes identically. 
Experimentally one has that $A_{3/2}\ne 0$, particularly at low $Q^2$, and
therefore the data demand $G_4 \ne 0$.

\subsection{Electromagnetic form factors}

Instead of the helicity 
amplitudes defined in the  rest frame of $R$, 
or instead of the form factors $G_i$,
one may also use the three so-called 
multipole electromagnetic form factors.
Those are, in the present case the 
magnetic dipole $G_M$,
the electric quadrupole $G_E$ and the Coulomb quadrupole $G_C$. 
They can be written as combinations 
of the helicity amplitudes or the form factors 
$G_i$ defined above, as
\cite{Aznauryan12a} 
\ba
\hspace{-.7cm}
G_M &=& - F\left( 
\frac{1}{\sqrt{3}} A_{3/2} - A_{1/2}
\right) \nonumber \\
&= &-
{\cal R} \left[ (M_R-M)^2 +Q^2 \right] \frac{G_1}{M_R}, 
\label{eqGM}\\
\hspace{-.7cm}
G_E &=& - F\left( 
\sqrt{3} A_{3/2} + A_{1/2}
\right) \nonumber \\
&= &-
{\cal R} \left\{ 2 G_4 - 
\left[ (M_R-M)^2 +Q^2 \right] \frac{G_1}{M_R} \right\}, 
\label{eqGE} \\
\hspace{-.7cm}
G_C &=& 2 \sqrt{2} \frac{M_R}{|{\bf q}|} 
F \,S_{1/2} =  - {\cal R} g_C,
\label{eqGC}
\ea
where 
$F= \frac{1}{e} \frac{M}{|{\bf q}|} \sqrt{\frac{M K}{M_R} 
\frac{(M_R-M)^2+Q^2}{(M_R-M)^2}}$
and ${\cal R} = 2{\cal A} F$.
We can also write 
$ {\cal R}=  \frac{1}{\sqrt{6}}\frac{M}{M_R-M}$.

Combining (\ref{eqGM}) and (\ref{eqGE}), one obtains
\ba
G_M + G_E= - 2 {\cal R} G_4.
\label{eqGEGM}
\ea
  
When the form factors $G_M$ and $G_E$ are known the helicity amplitudes become
\ba
& &
A_{1/2} = + \frac{1}{4F} \left( 3 G_M - G_E \right) 
\label{eqA12d}\\
& &
A_{3/2} = - \frac{\sqrt{3}}{4F} \left( G_M + G_E \right).
\label{eqA32d}
\ea

Defining 
\ba
G_4^\prime=  - 2 {\cal R} G_4, 
\label{eqG4c}
\ea
one has
\ba
& &
A_{1/2} =  \frac{1}{F} G_M + \frac{1}{4F} G_4^\prime 
\label{eqA12c}
\\
& &
A_{3/2} =  \frac{\sqrt{3}}{4F} G_4^\prime.
\label{eqA32c}
\ea

From Eqs.~(\ref{eqGM}), ~(\ref{eqGE}), and (\ref{eqGEGM}),  
we conclude that  $G_E$ 
and $G_M$ are determined 
by  $G_1$ and $G_4$ only;
$G_1$ fixes $G_M$; $G_4$ fixes the sum 
 $G_M+G_E$. We conclude, as it happened for the helicity amplitudes, 
that it is also convenient for the description
 in terms of $G_E$, $G_M$ and $G_C$, 
to  choose as independent functions $G_1$  (or $G_M$, 
since they are proportional), $G_4$ and $g_C$.
Additionally, a result to be retained from these formulas 
is that when $G_4=G_4^\prime=0$, 
one has $G_M= - G_E$ (which is equivalent to $A_{3/2} \equiv 0$) 
for any value of $Q^2$.
Note that the relation $G_M= - G_E$ 
is not confirmed experimentally (because $A_{3/2} \ne 0$).

\section{Electromagnetic current}
\label{secCSQM}

In the calculation of the
baryon transition electromagnetic form factors
we use the relativistic  impulse approximation.
In this approximation only one quark interacts with 
the photon while the other two quarks are spectators, but the electromagnetic
interaction is distorted by the initial and final state baryon vertices, 
defining a Feynman diagram with one loop integration. First,
one notes that within impulse approximation 
the relative momentum of the two quarks not interacting 
with the photon can be integrated over, since it does not depend
on the electromagnetic interaction. 
(This is why for the calculation of the impulse diagram one may
start with an effective wave function with a quark-diquark structure.)
Second, when performing that integration we apply
the covariant spectator theory to reduce in a covariant way the 
four dimensional integration to a three dimensional one.

This reduction amounts to select for the energy integration only
the positive energy poles of the two quarks in the diquark. 
The consequence is that after the internal diquark three-momentum 
is integrated out, one ends up with on-mass-shell 
diquark with an averaged invariant mass $m_D$~\cite{Nucleon,Omega,Nucleon2}.
The selected quark poles dominate in the energy integration.
The residues of the other poles (from the two propagators 
of the interacting quark, before and after 
its interaction with the photon) give the 
off-mass-shell diquark contribution. 
Their contribution is indeed small due to the large value 
of the baryon mass that decisively determines the relative 
location of all the poles in the complex plane \cite{poles}.

If one represents the initial 
and final baryon wave functions by $\Psi_N(P_-,k)$ 
and $\Psi_R(P_+,k)$, where
$P_-$ is the $N$ momentum,
$P_+$  is the $R$ momentum, and $k$ the diquark momentum, 
the  electromagnetic current 
in relativistic impulse approximation is given by
\cite{Nucleon,Nucleon2,Omega} 
\ba
J_{NR}^\mu=
3 \sum_{\Gamma} \int_k \overline \Psi_R(P_+,k) j_q^\mu \Psi_N(P_-,k),
\label{eqJ2}
\ea 
where $j_q^\mu$ is the quark current 
associated with one quark only (the factor 3 takes into account 
the contributions of the other quarks demanded by symmetry),
and the sum is over the diquark spin states $\Gamma$, including
a diquark scalar component 
and a diquark vector component with polarization $\Lambda =0,\pm$.
The integral symbol $\int_k  \equiv \int \frac{d^3 {\bf k}}{2 E_D (2\pi)^3}$
stands for  the covariant integration 
in the diquark three-momentum ${\bf k}$,
with $E_D= \sqrt{m_D^2 + {\bf k}^2}$.
The single constituent quark current $j_q^\mu$ is decomposed 
into two terms
\ba
j_q^\mu = j_1 \hat \gamma^\mu + j_2 \frac{i \sigma^{\mu \nu} q_\nu}{2M},
\label{eqJsub}
\ea
where $M$ is again the nucleon mass, 
$j_1$ and $j_2$ are the Dirac and Pauli quark operators 
and
\ba
\hat \gamma^\mu = \gamma^\mu - \frac{{\not \! q} q^\mu}{q^2}.
\label{eqJQ}
\ea
The inclusion of the last term is equivalent
to using the Landau prescription for the
electromagnetic current 
and ensures the conservation of $J_{NR}^\mu$~\cite{Gilman02,Kelly98,Batiz98}.
Equation (\ref{eqJsub}) 
is a simple prescription that builds in current conservation
in calculations within impulse approximation for inelastic processes with a pure
phenomenological  description of the final and initial states.
It overcomes the difficulty that these states
and the consistent interaction current are not calculated
from an underlying dynamics~\cite{Gilman02}.
Reference~\cite{Kelly98} also shows that the inclusion
of the term $-{\not \! q}q^\mu/q^2$
does not affect the results for the observables because
it is orthogonal to the lepton current.

The quark form factors $j_i$ ($i=1,2$)
have an isoscalar and an isovector 
component, given respectively by the functions 
$f_{i+}$ and $f_{i-}$  (of $Q^2$),
\ba
j_i = \frac{1}{6} f_{i+} + \frac{1}{2} f_{i-} \tau_3.
\ea

The explicit forms of the Dirac and Pauli quark form factors,   
$f_{1\pm}$ and $f_{2\pm}$ respectively, 
are chosen to be consistent with the vector 
meson dominance (VMD) mechanism, being
parametrized as 
\cite{Nucleon,NDelta,Omega}
\ba
& &
f_{1\pm}(Q^2)=\lambda_q
+ (1-\lambda_q) \frac{m_v^2}{m_v^2 +Q^2} +
c_\pm \frac{M_h^2 Q^2}{(M_h^2+Q^2)^2} \nonumber \\
& &
f_{2\pm}(Q^2)=
\kappa_\pm \left\{ d_\pm
\frac{m_v^2}{m_v^2+ Q^2} 
+(1-d_\pm) \frac{M_h^2}{M_h^2 +Q^2}
\right\}, \nonumber \\
& &
\label{eqQcurrent}
\ea
where $m_v$ is a light vector meson mass, $M_h$ 
is a mass of an effective heavy vector meson,
$\kappa_\pm$ are quark anomalous magnetic moments,
$c_\pm,d_\pm$ are mixture coefficients and  $\lambda_q$ is 
a parameter related with the quark density number in 
deep inelastic scattering. 

The quark form factors are normalized according to
$f_{1\pm}(0)=1$ and $f_{2\pm}(0)= \kappa_\pm$, with
the quark isoscalar ($\kappa_+$)
and isovector ($\kappa_-$) magnetic moments
given in terms of the $u$ and $d$ quark 
anomalous moments as 
$\kappa_+= 2 \kappa_u - \kappa_d$ 
and $\kappa_-= \sfrac{1}{3}(2 \kappa_u + \kappa_d)$.
In the applications we took $m_v= m_\rho$ ($\simeq m_\omega$)
to include the physics associated with the $\rho$-pole
and $M_h= 2 M$ (twice the nucleon mass) to 
take into account effects of meson resonances with a larger mass.
We consider here the parametrization that is consistent 
with the model for the nucleon
labeled model II in Ref.~\cite{Nucleon}.
The current parameters are $c_+=4.16$, $c_-=1.16$,
$d_+=d_-=-0.686$, $\lambda_q=1.21$,
$\kappa_+= 1.639$ and $\kappa_-= 1.833$.

\section{Baryon wave functions}
\label{secBaryonWF}

In the covariant spectator quark-diquark model 
the diquark  states are described in terms of diquark polarization vector states
 $\varepsilon_{\Lambda P}^\alpha$,
where $\Lambda=0,\pm$ are the polarization indices, which
are expressed in the basis of fixed-axis states
\cite{Nucleon,FixedAxis,NDelta}.
For a resonance $R$
with momentum $P=(E_R,0,0,P_z)$ the diquark polarization states read
\ba
& &
\varepsilon_{\pm P}^\alpha = \mp \frac{1}{\sqrt{2}} (0,1,\pm i,0) \nonumber \\ 
& &
\varepsilon_{0 P}^\alpha = \frac{1}{M_R}(P_z,0,0,E_R),
\label{eqEPS}
\ea
where $E_R=\sqrt{M_R^2 + P_z^2}$  is 
the resonance energy. 
The same form applies to the nucleon if one
replaces $M_R \to M$. 
Note that the polarization vectors
depend on both  the baryon mass and the baryon momentum, 
and satisfy the condition 
$\varepsilon_{\Lambda P} \cdot P=0$.

The core spin  3/2 state 
are represented by the Rarita-Schwinger vector state $u_\alpha$,
and the core spin 1/2 are represented by the combination 
of a spin-1 (diquark) and a Dirac spin 1/2 states that reads
\ba
U^\alpha_R(P,s) = \frac{1}{\sqrt{3}} \gamma_5
\left(\gamma^\alpha - \frac{P^\alpha}{M_R} \right) u_R(P,s),
\label{eqUR}
\ea
where $u_R$ is the Dirac spinor for the particle $R$~\cite{NDelta,Nucleon}.
Within this formalism, 
the wave functions for several 
baryon systems 
can be written in terms of the states
$U_R^\alpha$, $u_R$ and $u_\beta$ 
\cite{NDelta,Nucleon,Nucleon2,Omega,ExclusiveR}.
These building blocks make possible 
the construction of baryon wave functions that are
explicitly covariant and 
have the correct nonrelativistic limit~\cite{Nucleon,NDelta}. 

Next we will review the formulas
for the nucleon wave function,
and we will obtain the $N^\ast(1520)$ wave function.

\subsection{Nucleon wave function}

In the simplest covariant spectator model for the nucleon wave function, 
one takes an $S$-state for the quark-diquark configuration.
In that configuration, the nucleon wave function 
has a form imposed by demanding that the full wave function 
is symmetric under the exchange of any two quarks
in momentum-spin and flavor space.
One has then~\cite{Nucleon,FixedAxis,Nucleon2} 
\ba
\Psi_N(P,k) =
\frac{1}{\sqrt{2}}
\left[
\phi_I^0 u(P) - \phi_I^1 (\varepsilon_{\Lambda P}^\ast)_\alpha 
U^\alpha (P)
\right] \psi_N(P,k), \nonumber \\
\label{eqPsiN}
\ea
The first and the second terms 
are, respectively, the contributions 
from the  scalar (spin-0, isospin-0) and 
from the  axial vector (spin-1, isospin-1) diquark  states.
In addition, $\phi_I^{0,1}$ are the nucleon isospin states \cite{Nucleon}, 
$u$ is the Dirac spinor, 
and $U^\alpha$ is the state defined by Eq.~(\ref{eqUR})
in the special case of the  nucleon ($M_R \to M$).
It corresponds to the coupling of the 
spectator quark with a spin-1 vector diquark state 
to a three-constituent quark core state of spin
1/2.
The vector $\varepsilon_{\Lambda P}$, where 
$\Lambda =0,\pm 1$, is the diquark polarization state, introduced before.
Finally, $\psi_N$ is a radial wave function which encodes
the information on the quark-diquark relative momentum distribution. 
It was determined phenomenologically~\cite{Nucleon}.

The nucleon spin and isospin projections
and  the diquark polarization index $\Lambda$ were not included explicitly 
in Eq.~(\ref{eqPsiN}), to keep a short-hand notation. 
The spin projections were also omitted 
in the spin states $u$ and $U^\alpha$.

\subsection{$N^\ast(1520)$ wave function}

Since the $N^\ast(1520)$ has intrinsic negative parity 
and total spin $J=3/2$,
its wave function is a mixture of two contributions, 
$\Psi_{R1}$ and $\Psi_{R3}$, 
with total orbital angular momentum $L=1$, 
coupled respectively to
states with core spin 1/2 and states with core spin 3/2.
One writes then for the $N^\ast(1520)$
wave function~\cite{Aznauryan12a,Koniuk80,Capstick95}
\ba
\Psi_R(P,k) =
\cos \theta_D \Psi_{R1}(P,k) - \sin \theta_D \Psi_{R3}(P,k),
\label{eqPsiNX}
\ea
where the two components are normalized.
The admixture parameter, given by the angle $\theta_D$ 
depends on the model for the quark-quark interaction,
and can be determined from the radiative decay of 
the resonances $N^\ast(1520)$ and $N^\ast(1700)$
(both $\sfrac{3}{2}^-$ states)~\cite{Burkert03,Isgur77,Hey75}.
The most common estimate is  $\sin \theta_D \simeq 0.11$ 
\cite{Capstick00,Burkert03}.

To write the components $\Psi_{R1}$ and $\Psi_{R3}$ 
in the covariant spectator quark model we will 
start with the nonrelativistic form, 
and discuss afterward how the nonrelativistic structure
(in the rest frame) is obtained, 
and written in a covariant form valid in an arbitrary frame.

\subsubsection{Nonrelativistic wave functions}
\label{wavereduced}

In what follows, and as usual in the literature, 
we denote each wave function component
by its symmetry labels $(\rho, \lambda)$. These labels coincide with 
the symmetry labels of the two Jacobi momentum states $(k_\rho, k_\lambda)$,
respectively antisymmetric and symmetric 
in the change of quarks (12), and defined as
\ba
k_\rho &=&  \frac{1}{\sqrt{2}}(k_1 -k_2), \nonumber \\
k_\lambda&= & \frac{1}{\sqrt{6}} 
( k_1 + k_2 - 2 k_3) \nonumber \\
 &=& \sqrt{\frac{2}{3}} (k_1 + k_2) - \frac{1}{\sqrt{6}} P,
\label{eqKl1}
\ea
where $k_i$ is the individual momenta
($i=1,2,3$), and $P= k_1 + k_2 + k_3$, is the center of mass momentum.
In the nonrelativistic framework all the
momenta introduced are three-vectors, although
we will use later the same notation 
to represent their four-vector counterparts.

In the center of mass  frame, ${\bf P}={\bf 0}$,
$k_\lambda$ becomes proportional to  $k_1+ k_2$, and
one can describe the system by
the variables $r$ and $k$, as
\ba
& &
k_\rho \to r= \frac{1}{2} (k_1-k_2) \nonumber \\
& &
k_\lambda \to k= k_1 + k_2.
\ea

In the construction of the wave function components
$\Psi_{R1}$ and $\Psi_{R3}$, instead of 
using the basis of states corresponding 
to the Jacobi-momentum states of the three-quark system, 
we follow the usual practice in calculations of the baryon spectra, 
and take a representation of those baryon states in a basis 
that combines the Jacobi-momentum states into four 
(orthogonal) states with mixed-symmetry in the Jacobi momentum-spin
and isospin variables~\cite{Capstick00}. These four combinations are 
built to be either symmetric or antisymmetric in the 
interchange of quarks 1 and 2.
In such a basis the  
$N^\ast(1520)$ corresponds to the 
following isospin-spin-momentum combination~\cite{Capstick00},
\ba
\Psi_{Ri} =
N_{Ri}
\left[
\phi_I^0 X_\rho + 
\phi_I^1 X_\lambda 
 \right]\tilde \psi_{Ri},
\label{eqPsiNR1}
\ea
where $Ri$ stands for $R1$ and $R3$ and
$N_{Ri}$ is a normalization factor,
$\phi_I^{0,1}$ are isospin states 
(the same isospin states as for the nucleon, 
because both particles have isospin $1/2$), and
$\tilde \psi_{Ri}$ is a radial phenomenological wave function 
depending on $k$ and $r$.
The functions $X_\rho$ and $X_\lambda$ are states which couple
spin and orbital motion, and are respectively asymmetric and symmetric 
in the interchange of quarks (12).
Since $l=1$ is the angular momentum that is to be coupled with 
the core spin $1/2$ or $3/2$,
the coupled orbital-spin states $X_\rho$ and $X_\lambda$ 
contain the spherical harmonics 
$Y_{1m}(r)$ and $Y_{1m}(k)$ 
($m=0,\pm$) coupled to the spin states of the three-quarks. 
It is convenient to write the 
spherical harmonics in terms of the spherical 
components of the three-momentum $k$ and $r$.
For instance for $Y_{1m} (k)$, one has 
$k_0=k_z$, $k_\pm = \mp \sfrac{1}{\sqrt{2}} (k_x \pm i k_y)$, 
and we can write \cite{S11}
\ba
Y_{1m} (k) = \sqrt{\frac{3}{4\pi}} N_k  k_m,
\label{eqY1}
\ea
where $N_k=1/|{\bf k}|$,
and ${\bf k}$ is the diquark three-momentum. 
The form of the functions $X_\rho$ and $X_\lambda$
depends on the spin core state ($R1$ or $R3$), and their derivation 
is detailed in Appendices~\ref{apStateP1} and \ref{apStateP3}.

Here we just briefly describe how to accommodate the needed 
internal $l=1$ 
orbital angular momentum state of the diquark sub-structure 
(given by the $Y_{1m}(r)$  spherical harmonics 
that depends on the diquark internal relative momentum $r$).
This is an important point
because the inclusion of a diquark state with $l \ne 0$ implies 
that the diquark is not considered as pointlike particle
(this was already encountered in the nucleon case \cite{Nucleon2}
and we solve it here in the same fashion).
To realize it,
remember that because 
we are using impulse approximation, 
the internal variable $r$ is integrated out, i.e., 
the full three-body wave function
$\tilde \psi_R(r,k)$ is integrated in
$r$. This integration is equivalent to averaging the 
full wave function in $r$ and to generating 
an effective radial wave function corresponding to a quark-diquark structure,
$\psi_R(P,k)$, that depends on  the diquark momentum $k$ only.
Now, we may write $Y_{1m}(r)$ in terms 
of the spherical components of $r$, as in 
Eq.~(\ref{eqY1}) with $k \to r$.
That form exhibits the vector character of $Y_{1m}(r)$, and makes clear
that the average over the
diquark internal states associated with $l=1$ is not simply a scalar.
Instead, 
after the full three-body wave function 
is averaged in $r$, the vector structure of $Y_{1m}(r)$ originates
a polarization vector $\zeta_m^\nu$~\cite{Nucleon2}.
This new polarization vector is orthogonal to 
the diquark polarization vector $\varepsilon_{\Lambda P}^\alpha$, and
satisfies
\ba
\sum_\nu \zeta_m^\nu \zeta_{m'}^{\nu \ast} = \delta_{m m'}.
\label{vectorzeta}
\ea
The integration in the variable $r$ amounts then to the replacement\footnote{
In Ref.~\cite{Nucleon2} 
a factor $|k|$ was included 
in the replacement (\ref{eqIntR}),
but that factor was canceled by a factor $1/|k|$
included in the radial wave function.}~\cite{Nucleon2}
\ba
 Y_{1 m}(r) \to c\, \zeta_m^\nu ,
\label{eqIntR}
\ea
where we should set $c=1$ in order to
recover the result obtained when the explicit 
integration in $r$ is performed, in the nonrelativistic limit.
In addition,
we replace also $\tilde \psi_R $ by
$\psi_R$, where  $\psi_R$ is now a function of $k$ only.

\subsubsection{Relativistic generalization}

The relativistic form of the wave function is obtained 
by extending the nonrelativistic quantities  
to their relativistic description.  
For example, since the nonrelativistic wave function was written 
in terms of the quark-diquark relative momentum
momentum $k= k_1+k_2$, we have to construct the 
corresponding four-momentum. The general procedure
involves the baryon momentum $P$,
through
the substitution $k \rightarrow \tilde k$ with
\ba
\tilde k^\alpha= k^\alpha -\frac{P \cdot k}{M_R^2} P^\alpha,
\label{eqTildeK}
\ea
where $P$ is the resonance momentum. 
In the rest frame of the resonance, $P=(M_R,0,0,0)$, and  
$\tilde k = (0, {\bf k})$, and the formula above reduces
to its three-dimensional components. 
The radial wave function $\tilde \psi_R$
will then be replaced by its relativistic form $\psi_R(P,k)$. 

This procedure also helps us to establish the 
relativistic form for  $Y_{1m}(k)$, 
by using Eq.~(\ref{eqY1}). 
The replacement
$k \to \tilde k$ defined
by Eq.~(\ref{eqTildeK}), extends
the orbital angular momentum states from the rest frame to any frame, 
according to
 \cite{S11}
\ba
Y_{1 m} (k) \to - N_{\tilde k} 
(\varepsilon_{m} \cdot \tilde k),
\label{eqYrel1}
\ea 
with $\varepsilon_m^\alpha \equiv \varepsilon_P^\alpha (m)$ and 
$
N_{\tilde k} = \frac{1}{\sqrt{- \tilde k^2}}.
$

The polarization states $\zeta_m^\nu$ of 
Eq.~(\ref{eqIntR}) which will enter into the coupled 
spin-orbit states are also replaced by their
relativistic generalization,
normalized according to
$\zeta_\Lambda \cdot \zeta_{\Lambda'}^\ast= - \delta_{\Lambda \Lambda'}$.

To finish the relativistic generalization we need to replace 
the two coupled orbital-spin coupled states, $X_\rho$ and $X_\lambda$,
by their full corresponding relativistic form. 
This is done in Appendices~\ref{apStateP1} and \ref{apStateP3}, 
respectively for the $R1$ and $R3$ components of the wave function.

In Appendix~\ref{apStateP1} 
we obtain that the 
final expression for $\Psi_{R1}$ is
\ba
\Psi_{R1}(P,k)&=&  
\frac{1}{2} 
\Big\{
(- {\cal T}_R \phi_I^0 + \phi_I^1) u_\zeta^\nu (P)  
 \nonumber \\
& &
-
N_{\tilde k}
\left(
\phi_I^0 + {\cal T}_R \phi_I^1\right) 
\tilde k^\beta
u_\beta(P)
\Big\} \psi_{R1}(P,k), \nonumber \\
\label{eqPsiP1}
\ea
where 
$\psi_{R1}$ is a radial wave function,
\ba
\hspace{-.5cm}
u_\zeta^\nu(P,s) =
 \sum_{s'} 
\left< 1 \sfrac{1}{2};  (s-s') \, s' | \sfrac{3}{2} s \right> 
\zeta_{s-s'}^\nu u_R(P,s'),
\label{eqZeta1}
\ea
with $\zeta_m^\nu$ 
a spin-1 state introduced in Eq.~(\ref{eqIntR}), and
\ba
{\cal T}_R
= \frac{1}{\sqrt{3}}
(\varepsilon_{\Lambda P}^\ast)_\alpha
\gamma_5 
\left( \gamma^\alpha - \frac{P^\alpha }{ M_R} \right).
\ea

It is relevant to interpret the meaning of each of  
the terms of 
Eq.~(\ref{eqPsiP1}).
The terms in $u_\zeta^\nu$ contain the states where 
the diquark is in an internal $P$-state (note the presence of  
$\zeta_m^\nu$ in $u_\zeta^\nu$),  
while the terms in $u_\beta$ contain the orbital quark-diquark 
$P$-state for the wave function 
(note  $\tilde k^\beta$ in the combination $\tilde k^\beta u_\beta$).

The terms in $u_\zeta^\nu(P)$
will not interfere with the nucleon wave function 
and have therefore no contribution to the 
transition current.

The radial wave function $\psi_{R1}$ will be 
constrained in order to assure the orthogonality 
with the nucleon wave function as discussed below.

In Appendix~\ref{apStateP3} we obtain that the component $\Psi_{R3}(P,k)$  
of the $N^\ast(1520)$ wave function is
\ba
\Psi_{R3}(P,k)= 
- \psi_{R3}(P,k)(\varepsilon_{\Lambda P}^{\ast})^\beta (W_{R3})_\beta(P,s),
\label{eqPsiP3}
\ea
where $\psi_{R3}$ is the radial wave function, 
\ba
& & 
\hspace{-.8cm}
(W_{R3})_\beta(P,s)= \nonumber \\
& &
\hspace{-.8cm}
\frac{1}{\sqrt{2}} \gamma_5 
\left[
\phi_I^0 (V_\zeta^\nu)_\beta (P,s) - 
\phi_I^1 N_{\tilde k} \tilde k_\alpha
(V_{\varepsilon}^\alpha)_\beta (P,s) 
\right], 
\label{eqPsiP3B}
\ea
and
\ba
& &
(V_\zeta^\nu)_\beta (P,s)=
\sum_{s'} 
\left< 1 \sfrac{3}{2}; (s-s')\, s' | \sfrac{3}{2} s \right>
\zeta_{s-s'}^\nu u_\beta(P,s') \nonumber  \\
& & 
\label{eqVzeta}\\
& &
(V_{\varepsilon}^\alpha)_\beta (P,s)=
\sum_{s'} 
\left< 1 \sfrac{3}{2}; (s-s')\, s' | \sfrac{3}{2} s \right>
\varepsilon_{s-s'}^{\alpha} u_\beta(P,s'). \nonumber \\
& &
\label{eqVeps}
\ea
In the equation for $(W_{R3})_\beta(P,s)$ the factor $\gamma_5$
gives the needed relativistic form 
for a state with negative parity.

Note in Eq.~(\ref{eqPsiP3B})  that without the 
terms associated with the diquark internal $P$-states
(the ones that contain $\zeta^\nu$) 
only isospin-1 contributions remain,
and therefore the charge of the state 
would differ from $\frac{1}{2}(1 + \tau_3)$.
This shows that the diquark cannot be pointlike. 
Its internal structure, and particularly its $l=1$ relative angular momentum
has to be taken into account, 
since it plays an important role in the baryon properties.

\subsection{Orthogonality conditions
 and the phenomenological radial functions}
\label{secScalarWF}

Within the quark-diquark  picture 
of a baryon with total momentum $P$ 
and diquark momentum $k$, 
the covariant spectator quark model 
wave function for a baryon 
includes, 
not only the spin-flavor structure, but also a radial function 
for the momentum distribution of the quark-diquark system. 
The forms for these functions
are described in Appendix~\ref{radial}.

In this work the radial  baryon wave functions 
are not determined through a dynamical calculation, 
and are purely phenomenological.
For the nucleon,
the parameters were  determined 
by the study of the nucleon form factors 
on Ref.~\cite{Nucleon} (model II). 
That parametrization was successfully applied 
to predict the transition form factors 
of the photo-excitation of the nucleon to other $N^{*}$'s.

The radial wave functions $\psi_X$ ($X=N,R1,R3$)  for the nucleon 
and the components of the wave function  $N^*(1520)$ 
are normalized according to  
\ba
\int_k |\psi_X(\bar P,k)|^2=1,
\label{eqNorma}
\ea
where $\bar P$ is the baryon momentum at the rest frame.
This condition correctly fixes 
the baryon charge.
For instance, for the nucleon we obtain 
\ba
3 \sum_\Gamma \int_k 
{\overline \Psi_N} (\bar P,k) j_1 \gamma^0 \Psi_N (\bar P,k)
=e_N
\left[ \int_k |\psi_N(\bar P,k)|^2 \right],
\nonumber \\
\ea
where $j_1 \gamma^0$ is the quark charge operator,
and $e_N= \sfrac{1}{2}(1+ \tau_3)$ is the nucleon charge.
One realizes that
the normalization condition (\ref{eqNorma}) 
is required to obtain the  right charge 
of the nucleon.

Another remark has to be done at this point.
The wave function components in  Eqs.~(\ref{eqPsiP1}) and (\ref{eqPsiP3})
depend on the mass $M_R$ of the system
(for instance $u_\beta$, $u_R$ and $U_R^\alpha$ depend on $M_R$).
This implies that when the particles in the final and initial state 
have different masses their states are necessarily defined in different frames.
Therefore, if no additional
condition is imposed on the phenomenological radial function 
to enforce orthogonality,
it becomes possible that states 
orthogonal in the nonrelativistic limit 
become not orthogonal in their relativistic generalization. 
An example found in previous studies was
the $N^\ast(1535)$ state of negative parity~\cite{S11}.
The same happens here for the $N^*(1520)$ state.

We impose then the condition that 
the $R1$ and $R3$ components of the resonance wave function are orthogonal to 
the nucleon wave function, i.e., 
\ba
3 \sum_\Gamma \int_k 
{\overline \Psi_{Ri}} (\bar P_+,k) j_1 \gamma^0 \Psi_N (\bar P_-,k)
=0,
\label{eqOrthCondition0}
\ea
when  $Q^2=0$
($\bar P_+$ and $\bar P_-$  are the baryon momenta when $Q^2=0$).
In particular in the resonance rest frame
one has $\bar P_+ = (M_R,0,0,0)$ 
and $\bar P_-=(E_N,0,0,-|{\bf q}|)$, 
where $E_N= \frac{M_R^2+ M^2}{2M_R}$ 
and $ |{\bf q}|= \frac{M_R^2-M^2}{2M_R}$.
For the wave functions 
defined in this section
Eq.~(\ref{eqOrthCondition0})  
leads to 
\ba
\int_k N_{\tilde k} (\varepsilon_{0\bar P_+} 
\cdot \tilde k) \psi_{Ri} (\bar P_+,k) 
\psi_{N} (\bar P_-,k) =0.
\label{eqOrthCondition}
\ea
This equation is used to fix 
the free parameters of the radial wave functions 
$\psi_{R1}$ and $\psi_{R3}$ respectively (see Appendix~\ref{radial}).
All the numerical values of the wave function parameters are given 
in Sec.~\ref{secResults}
where the numerical results are presented. 

It is important to realize that the need to impose the orthogonality
conditions (\ref{eqOrthCondition0}) is a consequence of relativity. 
For $Q^2 =0$,
in the nonrelativistic limit there are no recoil effects, and therefore
both particles are considered in their rest  frames. 
In this limit then
the overlap integral $\int_{\Omega_{\bf k}} Y_{10}(\hat {\bf k}) \psi_R \psi_N$, 
between the two baryon wave functions at $Q^2=0$,
vanishes. In the relativistic case, however,
because the nucleon and $R$ have different masses, 
they cannot be simultaneously in their rest frame when $Q^2= 0$.
Then at least one of the wave functions 
is distorted  by a boost, which induces 
a dependence on the direction of ${\bf k}$, and
leads to $\int_{\Omega_{\bf k}} Y_{10}(\hat {\bf k}) \psi_R \psi_N \ne 0$.

\section{Transition form factors}
\label{secFF}

In this section we will present the algebraic results obtained
for the transition form factors and helicity amplitudes from
our quark core model.
First, we will derive the separate contributions to 
the $\gamma N \rightarrow N^*(1520)$ transition form factors from the 
$R1$ and $R3$ components
of the wave function.

Second, we will see that for small $Q^2$
we obtain amplitudes that 
are small when compared with the data.
The reasons for this are identified. 
This result will give us an indication that meson cloud 
effects have to be considered, 
and therefore we finish this section by also giving a
parametrization to describe them.

\subsection{Quark core contributions}

\subsubsection{Contribution from the $R1$-component}

To calculate the the $R1$-state contribution to the transition form factors 
we use the definition of the current (\ref{eqJ2}),
with $\Psi_R$ given by $\Psi_{R1}$ [see Eq.~(\ref{eqPsiP1})].
Strictly speaking in Eq.~(\ref{eqJ2}) 
we should add the sum in the index $\nu$ 
to take into account the dependence 
on $\zeta^\nu$ in the $R$ wave function.
However, since those terms do not interfere 
with the nucleon wave function 
that is not necessary.

The details of the calculation are included in 
the Appendix~\ref{apFF-P1}. 
The final results are
\ba
G_1^{R1}&=& -\frac{3}{2\sqrt{2} |{\bf q}|} \cos \theta_D \nonumber \\
& & \times
\left[
 \left( j_1^A + \frac{1}{3} j_1^S\right)+ 
\frac{M_R+M}{2M}\left( j_2^A + \frac{1}{3} j_2^S\right) 
\right] I_z^{R1} \nonumber \\
& & \label{eqG1_P1}\\
G_2^{R1} &=& 
 \frac{3}{2\sqrt{2} M |{\bf q}|}   \cos \theta_D
\nonumber \\
& & \times
\left[
j_2^A 
+ \frac{1}{3}  \frac{1- 3 \tau}{1 + \tau}   j_2^S
+ 
\frac{4}{3} \frac{2 M}{M_R +M} \frac{1}{1 + \tau} j_1^S 
\right] I_z^{R1} \nonumber \\
& &  \\
G_3^{R1}&=& -\frac{3}{2\sqrt{2} |{\bf q}|} 
\frac{M_R-M}{Q^2}  \cos \theta_D
\nonumber \\
& &
\times \left[
j_1^A + \frac{1}{3}\frac{\tau-3}{1+ \tau} j_1^S 
+ \frac{4}{3} \frac{M_R +M}{2M} \frac{\tau}{1+ \tau} j_2^S
\right] I_z^{R1}, \nonumber \\
\label{eqG3_P1}
\ea 
where $\tau=\frac{Q^2}{(M_R+M)^2}$,
\ba
I_z^{R1}= - \int_k  
N_{\tilde k} (\varepsilon_{0 P_+} \cdot \tilde k) \psi_{R1}(P_+,k) 
\psi_N(P_-,k). 
\ea
$j_i^S =\frac{1}{6}f_{i+} + \frac{1}{6}f_{i-} \tau_3 $ 
and $j_i^A= \frac{1}{6}f_{i+} - \frac{1}{2}f_{i-} \tau_3$ 
with $i=1,2$.

In addition, one has
\ba
G_4^{R1}=0.
\ea

From these results
one can calculate $G_M^{R1}$ and $G_E^{R1}$ using 
Eqs.~(\ref{eqGM}) and (\ref{eqGE}), as well as 
$A_{1/2}^{R1}$ and $A_{3/2}^{R1}$, 
using Eqs.~(\ref{eqA12b}) and (\ref{eqA32b}).
As $G_4^{R1}=0$, we have
\ba
G_E^{R1}= - G_M^{R1},
\ea
and consequently 
\ba
A_{3/2}^{R1}=0.
\ea
Therefore there is no contribution of  the $R1$ component to 
$A_{3/2}$.

Equations (\ref{eqG1_P1})-(\ref{eqG3_P1}) show
the proportionality between the form factors and the 
overlap integral $I_z^{R1}$.
For $Q^2=0$ one has $I_z^{R1}(0)=0$ according to
the orthogonality condition (\ref{eqOrthCondition}).
The consequence  is that $G_M^{R1}(0)=G_E^{R1}(0)=0$
and $A_{1/2}^{R1}(0)=A_{3/2}^{R1}(0)=0$.
However, $G_C^{R1}(0)$ is not zero and is finite,
according to Eqs.~(\ref{eqGCsmall}), (\ref{eqGC})
and (\ref{eqG3_P1}),  because
$G_C^{R1} \propto I_z^{R1}/Q^2$ 
and by construction (through the orthogonality
condition) 
$I_z^{R1} \propto Q^2$ when $Q^2 \to 0$.

\subsubsection{Contribution from the $R3$-component}

The calculations of the contributions of the $R3$ component 
in the wave function to the transition form factors 
and the helicity amplitudes are 
detailed in Appendix~\ref{apFF-P3}.

The results for the form factors are
\ba
G_M^{R3} &=& 
  \frac{3}{\sqrt{5}} {\cal R}  f_v I_z^{R3} 
\sin \theta_D\\
G_E^{R3}&=& 3 G_M^{R3} =
         \frac{9}{\sqrt{5}} {\cal R}  f_v I_z^{R3} 
\sin \theta_D\\
G_C^{R3} &=&  
  2   \sqrt{\frac{2}{15}}  I_z^{R3} 
\sin \theta_D \nonumber \\
& &
\times \left( 
\frac{M M_R}{Q^2} j_1^S + \frac{M_R}{2(M_R-M)} j_2^S 
\right),
\ea
where we recall that ${\cal R}= \frac{1}{\sqrt{6}} \frac{M}{M_R-M}$,
\ba
f_v &=& 
j_1^S - \frac{M_R-M}{2M} j_2^S,
\ea
and
\ba
I_z^{R3}= - \int_k N_{\tilde k} (\varepsilon_{0P_+} \cdot \tilde k)
\psi_{R3}(P_+,k) \psi_{N}(P_-,k).
\ea

The $R3$-component contributions to $G_E(0)$ and $G_M(0)$
vanish because of the
orthogonality condition between the initial and final states, 
i.e.~$I_z^{R3}(0)=0$. Only $G_C$ is nonzero 
for $Q^2=0$, as it happened for the $R1$-component.

The contribution from the $R3$-state to $G_4^\prime$
(which is proportional to $A_{3/2}$)  is
\ba
(G_4^\prime)^{R3}= - \frac{6}{\sqrt{5}} {\cal R}
f_v I_z^{R3}.
\ea

For the helicity amplitudes 
one obtains
\ba
& &
A_{3/2}^{R3}= 
- \frac{3}{\sqrt{5}}
\sqrt{\frac{2 \pi \alpha}{K}}
 N_q
f_v I_z^{R3} \sin \theta_D,   \\
& &
A_{1/2}^{R3} = 0, \\
& &
S_{1/2}^{R3} =    
\sqrt{ \frac{2}{15}  }  
\sqrt{\frac{2\pi \alpha}{K}} 
\bar f_v^\prime
N_q |{\bf q}|
 I_z^{R3} \sin \theta_D,
\ea
where 
\ba
N_q &=& \sqrt{\frac{(M_R+M)^2+ Q^2}{4MM_R }}, \\
\bar f_v^\prime &=& 
 \frac{M_R-M}{Q^2} j_1^S + \frac{j_2^S}{2M}.
\ea

The helicity amplitudes and 
the form factors obtained from the $R3$-component alone are proportional to 
$\sin \theta_D$, estimated to be $\approx 0.1$
in some models \cite{Burkert03}.
For this reason we do not 
expect a significant effect  
from the $R3$-component.
It is nevertheless interesting to note 
that only this component gives a finite contribution 
to $A_{3/2}$. 
Although small, 
it could in principle be important
to understand the falloff 
of the amplitude $A_{3/2}$ for large $Q^2$. 
Also in contrast to the $R1$-component, $R3$ 
does not contribute at all to $A_{1/2}$.

\subsubsection{Summary and discussion of the quark core contributions}
\label{DiscussionA3/2}

We can summarize the obtained results 
for the  spin quark core contributions to the form factors 
in the following formulas
 (the index $b$ stands for bare):
\ba
& &
G_M^b= G_M^{R1} + G_M^{R3}  \\
& &
G_E^b= -G_M^{R1} + 3 G_M^{R3}  \\
& &
G_C^b= G_C^{R1} + G_C^{R3}.
\ea
For $G_4$ only the $R3$ component of the wave function contributes:
\ba
(G_4^\prime)^b= (G_4^\prime)^{R3}.
\ea

Alternatively, 
for the helicity amplitudes,
we have,
following Eqs.~(\ref{eqA12c})-(\ref{eqA32c}) and (\ref{eqGC}):
\ba
& &A_{1/2}^b= \frac{1}{F} G_M^b + \frac{1}{4F} (G_4^\prime)^{b} \\
& &A_{3/2}^b = \frac{\sqrt{3}}{4F} (G_4^\prime)^{b}  \\
& &S_{1/2}^b = \frac{{\cal K}}{F}G_C^b,
\ea
where 
${\cal K}= \frac{1}{2\sqrt{2}} \frac{|{\bf q}|}{M_R} $.

Because the contributions from $R1$ are 
proportional to $\cos \theta_D \approx 0.99$,
and the contributions from $R3$ are 
proportional to $\sin \theta_D \approx 0.11$,
we can anticipate a small
quark core contribution to $A_{3/2}$.

Using Eqs.~(\ref{eqA12d}) and (\ref{eqA32d}) we conclude that, 
from the $R1$ contribution,  $G_E= -G_M$ 
($A_{3/2}=0$), while from the $R3$ contribution, $G_E= 3G_M$ ($A_{1/2}=0$).
As the $R3$ admixture is small, we obtain
$A_{3/2} \propto G_4^\prime$ also small, and
we can expect an almost correlation between $G_E$ 
and $G_M$ with $G_E  \simeq  - G_M$.
In addition, only $G_C$ has a non zero contribution for 
$Q^2=0$, from both $R1$ and $R3$ core spin states.

We may discuss a bit further these following general features of our results:
\begin{enumerate}
\item
The reason why in our model the $R1$  state does 
not contribute to $A_{3/2}$, and consequently  
$A_{3/2} \approx 0$ for any $Q^2$ range,
lies in the form of the $R1$-state, 
and in particular in the specific structure of 
the diquark polarization vectors.
In the fixed-axis representation~\cite{FixedAxis}  the diquark momentum 
is averaged along the direction 
of the reaction, defined by the three-momentum part of
$\sfrac{1}{2}(P_+ + P_-)$. 
This is very successful 
to construct (orbital) angular momentum state components
of the wave function, as shown for the nucleon 
and the $\Delta$ \cite{Nucleon,Nucleon2,NDelta,NDeltaD}.
However, it may happen that it is
incomplete for the structure of the
$P$-state excitations, in spite of
the model being successful in the description of 
the state $N^\ast(1535)$  characterized also by
$P$-state excitations \cite{S11}.
\item 
Our result that the contribution from the  $R1$-component
to the form factors $G_M$ and $G_E$ 
vanishes at $Q^2=0$ is consistent with
a nonrelativistic framework:
since the initial and final states
have the same core spin ($S=1/2$),
they are necessarily orthogonal
because of the orthogonality 
between spherical harmonics.
In the relativistic case, since the 
initial and the final state have different masses,
and the wave functions  are defined 
in different frames, the boosts cannot be neglected and
the orthogonality condition is not automatically satisfied.
We are then forced to impose the orthogonality condition 
(\ref{eqOrthCondition})
between the initial and the final state ---
which gives at $Q^2=0$ a zero contribution to $G_E$ and $G_M$, 
while making $G_C(0)$ finite.
This form of imposing the orthogonality 
was already considered in the $\gamma^\ast N \to \Delta (1232)$ 
reaction and was there also responsible for the 
generation of nonzero and finite contributions to $G_C$~\cite{NDeltaD}. 
In the case
of $R3$-component, 
the core spin state is different from the nucleon state
($S=1/2$ for the nucleon and $S=3/2$ for the $R3$-component),
and therefore $R3$ is also orthogonal to the nucleon wave function
in the nonrelativistic limit.
However as that  does not happen in 
our relativistic generalization of the states 
for $N^\ast(1520)$,
we are again forced 
to impose the orthogonality between the 
nucleon and the $R3$-component.
\item
In this work the nucleon wave function 
is reduced to an
$S$-state configuration.
It is possible that the states $R1$ and $R3$ 
interfere with nucleon $P$-states, as
already proposed in Ref.~\cite{Nucleon2}.
Nevertheless, we expect those contributions to be small
due to the small $P$-state admixture in the nucleon wave function.
\end{enumerate}

\subsection{Parametrization of the meson cloud}
\label{secPC}

As explained already in the introduction, 
although our quarks are dressed, there are still meson cloud effects 
that can give extra contributions to the 
transition form factors. 
In general meson cloud contributions are expected to be significant 
at low $Q^2$~\cite{Aznauryan12a,NSTAR}.
It is then natural to assume that meson cloud effects can 
give important contributions for the helicity amplitudes 
at low $Q^2$, in particular 
to the amplitude $A_{3/2}$, where our quark core model 
predicts only small contributions,
 because, as we have seen, 
the $R1$-state contribution vanishes and 
the $R3$-state is itself strongly 
suppressed by the weight factor $\sin \theta_D$.
The small results of our model for $A_{3/2}$ 
are consistent with the conclusions
of several authors 
\cite{Aznauryan12a,NSTAR} 
and EBAC estimations \cite{Diaz08} that
the meson cloud effects for $A_{3/2}$
must be sizeable at least at low $Q^2$.

Because the pion is the lightest meson, 
one may assume that pion cloud contributions dominate
over  heavier meson contributions, and also that
heavy meson effects fall off faster with $Q^2$ than pion effects.
Another natural assumption on meson cloud effects 
is that diagrams where the
photon couples with the meson in flight 
--- the leading order contribution 
according to 
chiral perturbation theory~\cite{Kaiser93,Cloet03,Arndt04}, 
give larger contributions than diagrams where 
the photon couples with the whole baryon, 
while the meson in flight is dressing the baryon.

Assuming then that the meson cloud 
can be added to the core quark effects, 
and that the pion is the dominant contribution,
we take the following 
structure for the form factors
\ba
& &G_M= G_M^b + G_M^\pi \\
& &G_4^\prime = (G_4^\prime)^b + G_4^\pi 
\simeq G_4^\pi
\label{eqG4v2}
 \\
& &G_C= G_C^b + G_C^\pi,
\ea
where $G_M^\pi$, $G_4^\pi$ and $G_C^\pi$ 
are the pion cloud contributions 
or the form factors, to be parametrized and extracted from 
the difference between the experimental data and 
the quark model results.
The approximation in Eq.~(\ref{eqG4v2})
corresponds to neglecting the $R3$ contributions. 
The impact of this approximation
is small since the weight of $R3$ in the wave function is small.

From the previous relations we may write for $G_E$
\ba
G_E &\equiv &- G_M - G_4^\prime \nonumber \\
& \simeq & - G_M^b  - (G_M^\pi + G_4^\pi),
\ea
and the helicity amplitudes become
\ba
& &
A_{1/2} \simeq \frac{1}{F} G_M^b +  
\frac{1}{F} G_M^\pi +  \frac{1}{4F} G_4^\pi  \label{eqA12pi}  \\
& & 
A_{3/2} \simeq 
\frac{\sqrt{3}}{4F} G_4^\pi  \\
& &
S_{1/2} =
\frac{{\cal K}}{F}
 (G_C^b + G_C^\pi).
\label{eqS12pi}
\ea 
In the approximation of neglecting the $R3$-state contributions,
$A_{3/2}$ is reduced to the dominant $G_4^\pi$ term. 
The fact that $A_{3/2}$ is only connected to $G_4$, and does 
not mix other form factors is most fortunate, since 
then the meson cloud effect on $G_4$  
can be directly read off from $A_{3/2}$ only. 
Similarly,
the meson cloud effect on $G_C$  can be separately read 
off from the $S_{1/2}$ data. 
Finally,  the $A_{1/2}$ amplitude will mix that contribution 
with the contribution to $G_4$.
These results motivate the use of the helicity amplitudes data 
to breakdown the meson cloud effects into three independent terms, 
contributing respectively to $G_4$, $G_C$, and finally $G_M$.
The numerical results will be shown in the next section. 
The information on the three pion cloud terms was based 
on functional forms  and  inspired
in previous studies
of the pion cloud contribution
in the timelike regime \cite{Timelike,Frohlich10}.

The parametrization of the pion cloud contributions that we use here is then
\ba
G_4^\pi &=& \lambda_\pi^{(4)} \left(\frac{\Lambda_4^2}{\Lambda_4^2+ Q^2}\right)^3
F_\rho \; \tau_3 \label{eqG4pi}\\
G_M^\pi &=& (1 + a_M Q^2) \times \nonumber \\
& & 
\lambda_\pi^{M} \left(\frac{\Lambda_M^2}{\Lambda_M^2+ Q^2}\right)^3
F_\rho \; \tau_3 
\label{eqGMpi} \\
G_C^\pi &=& \lambda_\pi^{C} \left(\frac{\Lambda_C^2}{\Lambda_C^2 + Q^2}\right)^3
F_\rho \; \tau_3,
\label{eqGCpi}
\ea
with
\ba
F_\rho = \frac{m_\rho^2}{m_\rho^2 + Q^2  + 
\frac{1}{\pi} \frac{\Gamma_\rho^0}{m_\pi} Q^2 \log \frac{Q^2}{m_\pi^2}},
\label{eqFrho}
\ea
where $m_\rho$ and $m_\pi$ are the $\rho$ and pion mass,
and 
$\Gamma_\rho^0= 0.149$ GeV~\cite{Timelike}. 
The isospin operator $\tau_3$ gives the isospin dependence 
of  the diagram for the 
direct coupling between the photon 
and the pion~\cite{Octet,Kaiser93}. 
For the reaction starting with the proton the isospin dependence
gives  a $+$ sign, while the neutron case brings in a $-$ sign.

The adjustable parameters in the parametrization 
of the pion cloud are the strength coefficients
 $\lambda_\pi^{(4)}, \lambda_\pi^{M},\lambda_\pi^{C}$ and 
the cutoff parameters $\Lambda_4,\Lambda_M,\Lambda_C$, as well as
the coefficient $a_M$. 

One important motivation for the forms 
from Eqs.~(\ref{eqG4pi})-(\ref{eqGCpi})
is the expected leading order behavior from pQCD: 
$G_M \propto 1/Q^4$, $G_4^\prime \propto 1/Q^6$ 
and $G_C \propto 1/Q^6$ based in similar reactions 
\cite{Carlson,Carlson2},
corrected by an factor $1/Q^4$ 
due to the additional $q\bar q$ contribution, 
The extra factor $1/Q^4$ is a consequence 
of the estimation of the behavior of the 
leading order form factor $G$ given by $G \propto 1/(Q^2)^{(N-1)}$,
where $N$ is the  number of constituents, and comes from
replacing  $N=3$ (3 quarks) by $N=5$ 
(3 quarks + 1 quark-antiquark pair from the meson)~\cite{Carlson}. 
In this work, the extra factor  $1/Q^4$, 
due to  the  $q \bar q$  contribution,  
is slightly smoothened, and replaced by $F_\rho \propto 1/(Q^2 \log Q^2)$.
See Ref.~\cite{Timelike} for more details.

In the parametrization of $G_M^\pi$ 
in Eq.~(\ref{eqGMpi}), 
apart from the 
falloff with $Q^2$ of the simple multipole function,
we included an extra term
with a higher power dependence in $Q^2$, when compared to the terms
used for $G^{\pi}_4$ and $G^{\pi}_C$.
Different from the other form factor parametrizations, 
this term has no fundamental justification.
It simply allows more flexibility 
in the phenomenological description of the pion cloud effects.
While the function $G_4^\pi$ that
fixes $A_{3/2}$ 
and the function $G_C^\pi$ that fixes $S_{1/2}$  
have a simple falloff behavior,
as we will see next 
in the graph for $A_{3/2}$ in Fig.~\ref{figAmp1},
the pion cloud contributions to $G_M$
require a more complex analytic form. 

Note that although we adopt for simplicity 
a meson cloud parametrization with a structure corresponding to the pion cloud,
we cannot exclude that 
it effectively contains contributions from heavier mesons, 
given the phenomenological fitting procedure.

\section{Results}
\label{secResults}

Here we present the numerical results
of the covariant spectator quark model 
for the $\gamma^\ast N \to N^\ast(1520)$ reaction.
We calculated the  
quark core model contributions to the helicity amplitudes
and form factors, by applying the equations in Sec.~\ref{secFF}.
First, we will compare our results to the experimental data, and
in the sequel we will describe the difference 
between our quark core model contributions 
and the data by the meson cloud 
parametrization presented in  the Sec.~\ref{secPC}.

In the comparison to the data 
we use the PDG results for $Q^2=0$ 
\cite{PDG} for the amplitudes $A_{1/2}$ and $A_{3/2}$, 
and the CLAS data for $Q^2=0.3-4.2$ GeV$^2$ 
\cite{Aznauryan09} (pion production data) 
and for $Q^2=0.3-0.6$ GeV$^2$ \cite{Mokeev12} 
(double pion production data).
At the end we also 
discuss and make predictions
for the very large $Q^2$ region
which may be measured after the Jlab 12-GeV upgrade.

\subsection{Quark core effects}

Our calculation of the quark core contributions 
includes the contributions from both  $R1$ and $R3$, 
the quark core spin $1/2$ and $3/2$, respectively.
As mentioned already, the first ones are proportional 
to $\cos \theta_D$ and the second ones 
proportional to $\sin \theta_D$.
Because we are not using a dynamical model
starting with a well defined quark-quark interaction, 
we cannot calculate $\theta_D$, and
we use then the most common estimation 
in the literature $\theta_D \simeq 6.3^\circ$
(with $\cos \theta_D \simeq 0.994$ and
$\sin \theta_D=0.110$) \cite{Koniuk80,Capstick95,Capstick00,Burkert03}.

In the following calculations we 
use the range parameters of the nucleon 
radial wave function (model II in Ref.~\cite{Nucleon})
defined by Eq.~(\ref{eqPsiNrad}).
In particular we use 
$\beta_1= 0.049$, $\beta_2= 0.717$, 
corresponding to a normalization constant $N_0=3.35$.

\begin{figure}[t]
\vspace{.3cm}
\centerline{
\mbox{
\includegraphics[width=2.8in]{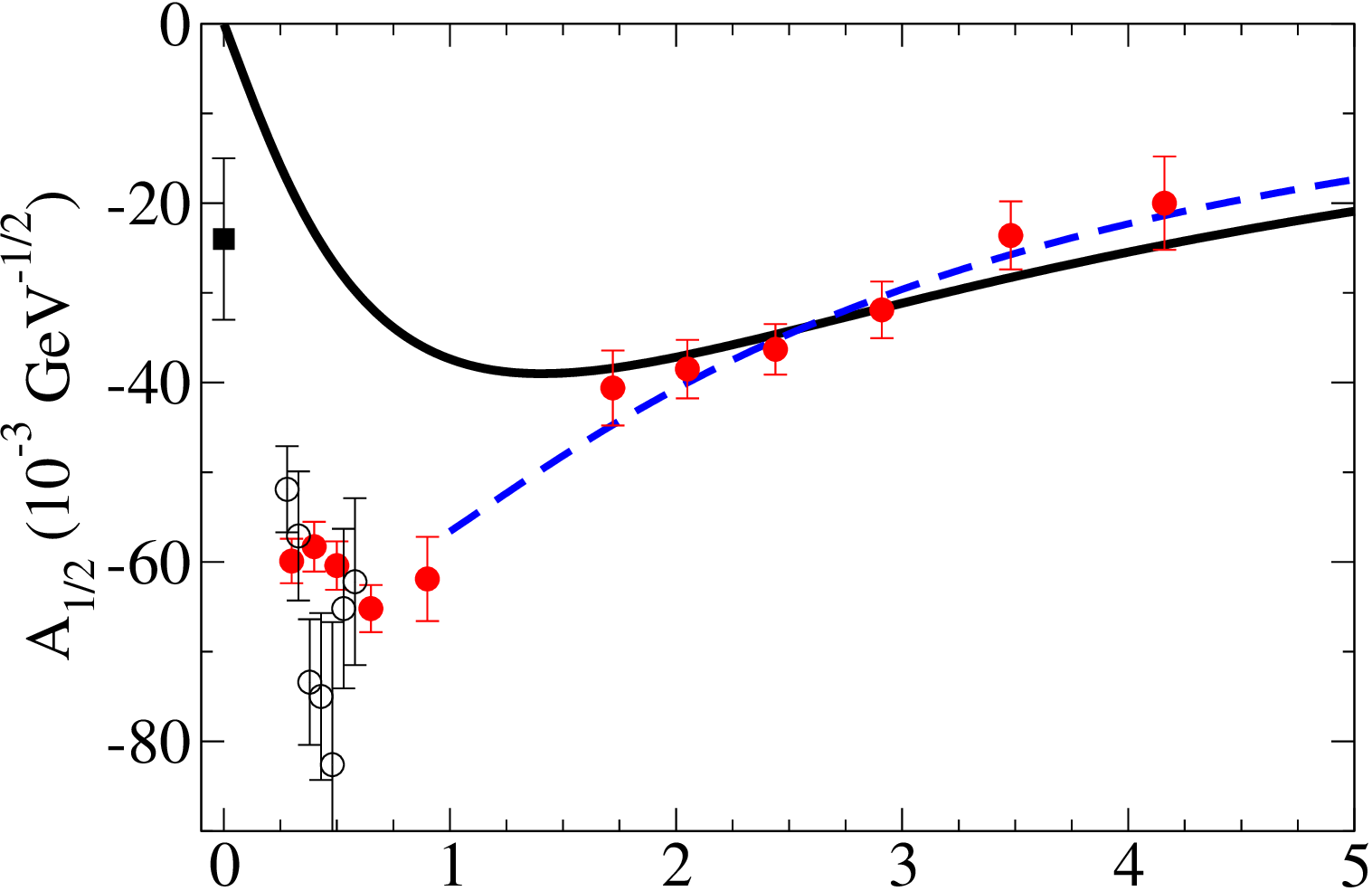}
}}
\centerline{
\vspace{.5cm} }
\centerline{
\mbox{
\includegraphics[width=2.8in]{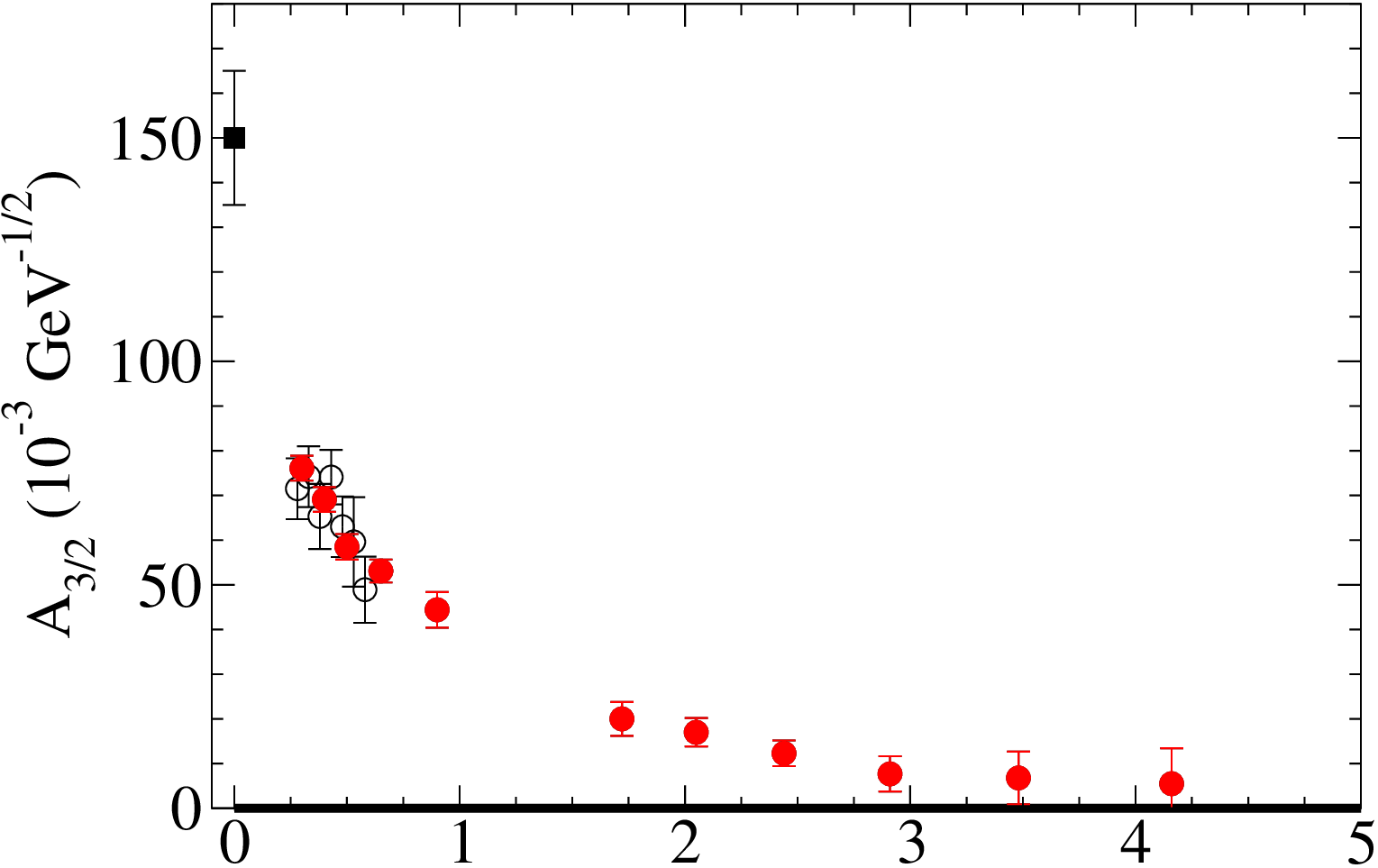}
}}
\centerline{
\vspace{.5cm} }
\centerline{
\mbox{
\includegraphics[width=2.8in]{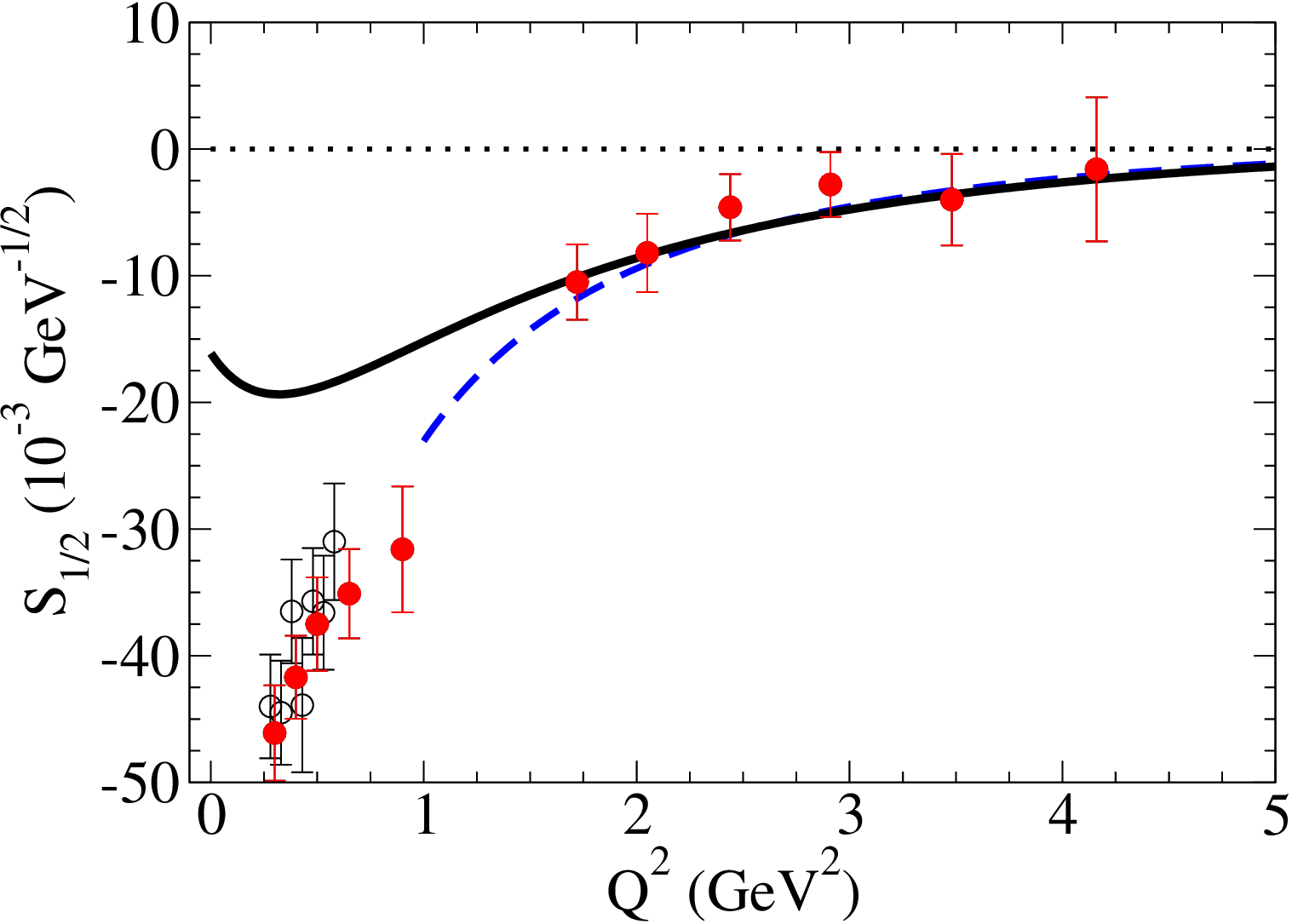}
}}
\caption{\footnotesize{
Quark core contributions to the helicity amplitudes.
The dashed line line is the model 0.
The solid line is the result 
from model 1
[fit of the parameter $\beta_3$ for $Q^2> 1.5$ GeV$^2$].
Data from Ref.~\cite{Aznauryan09} (full circles),
Ref.~\cite{Mokeev12} (empty circles) and 
PDG \cite{PDG} (square).
}}
\label{figAmp1}
\end{figure}

\begin{figure}[t]
\vspace{.3cm}
\centerline{
\mbox{
\includegraphics[width=2.8in]{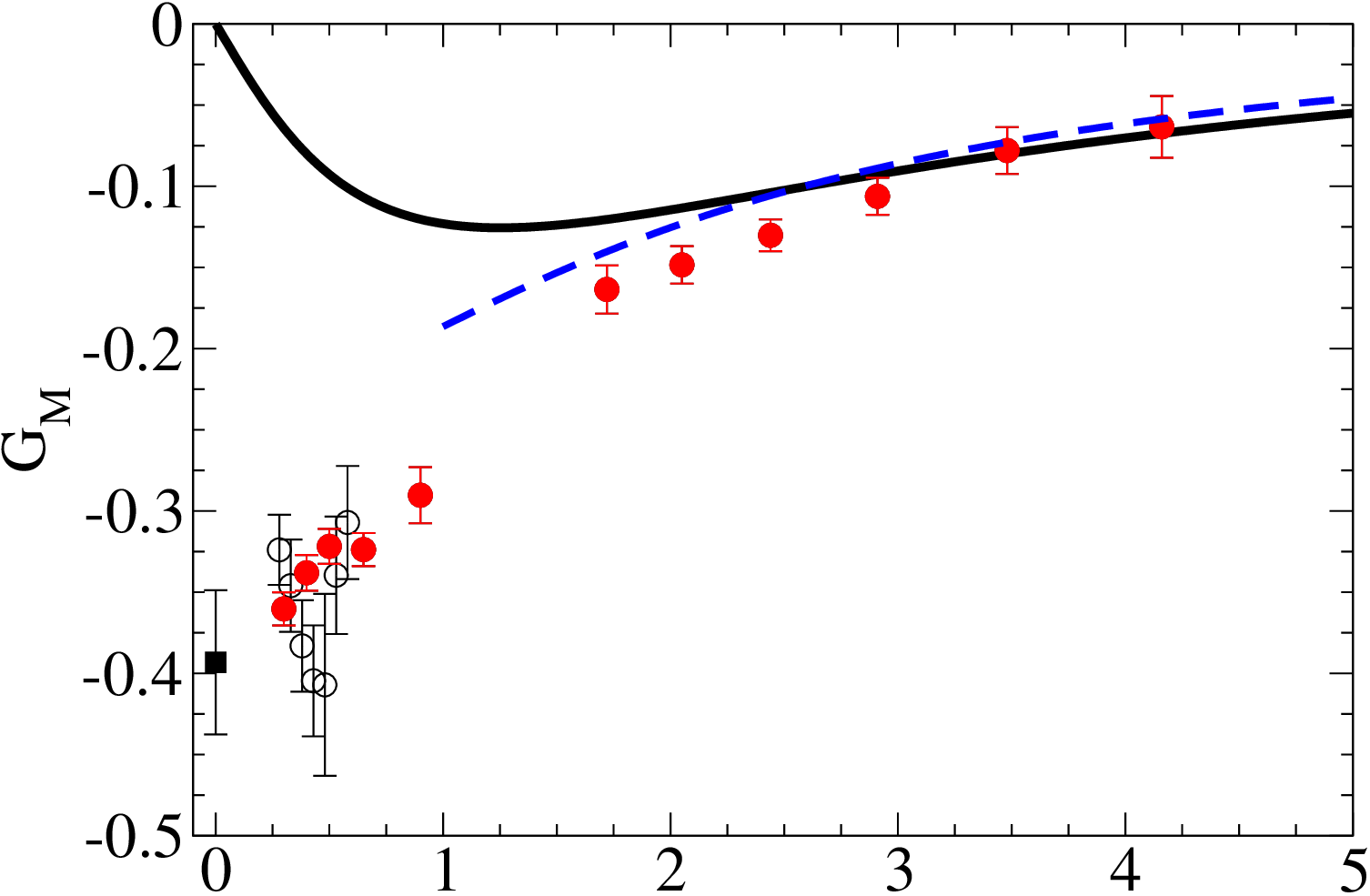}
}}
\centerline{
\vspace{.5cm} }
\centerline{
\mbox{
\includegraphics[width=2.8in]{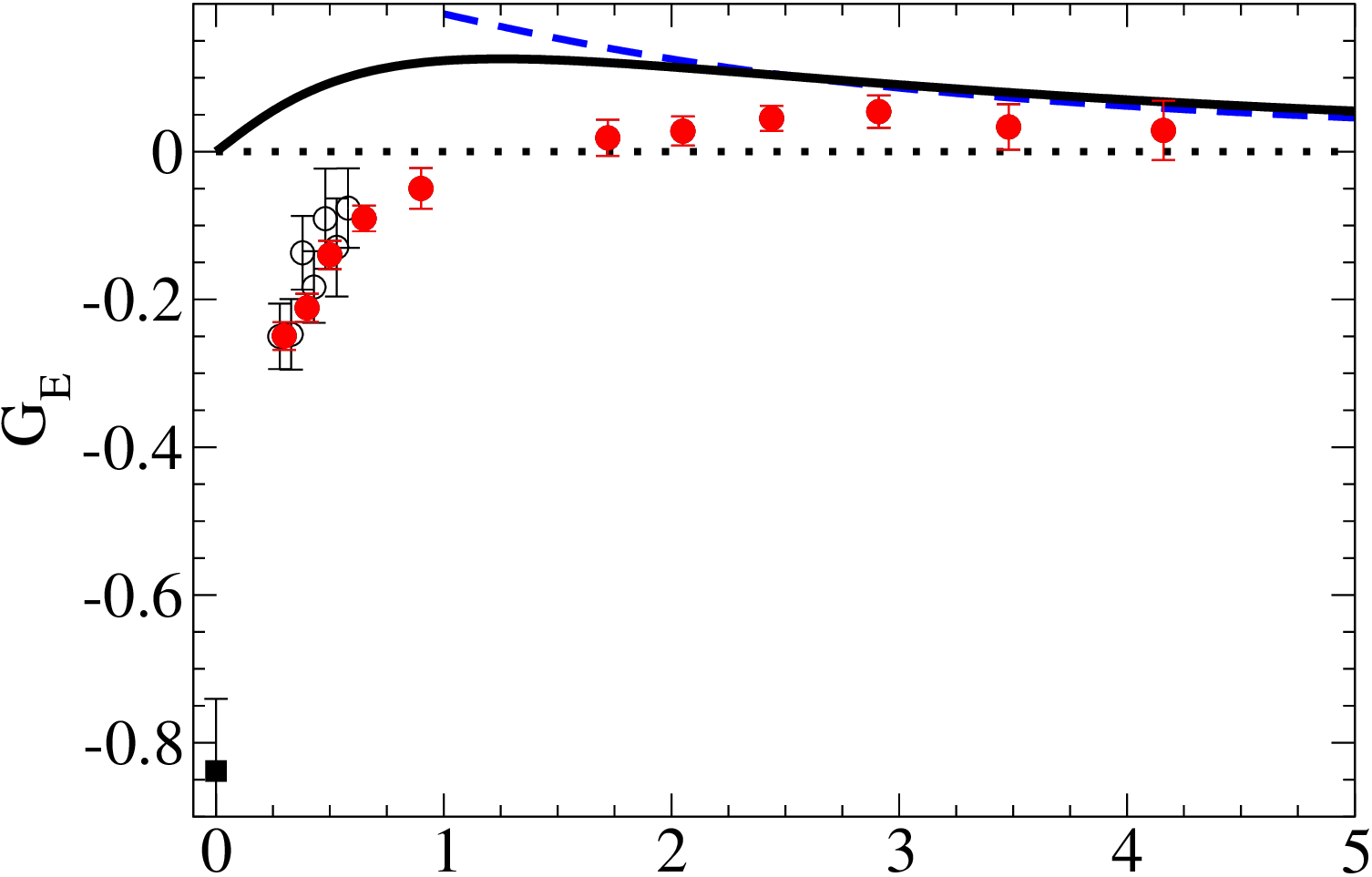}
}}
\centerline{
\vspace{.5cm} }
\centerline{
\mbox{
\includegraphics[width=2.8in]{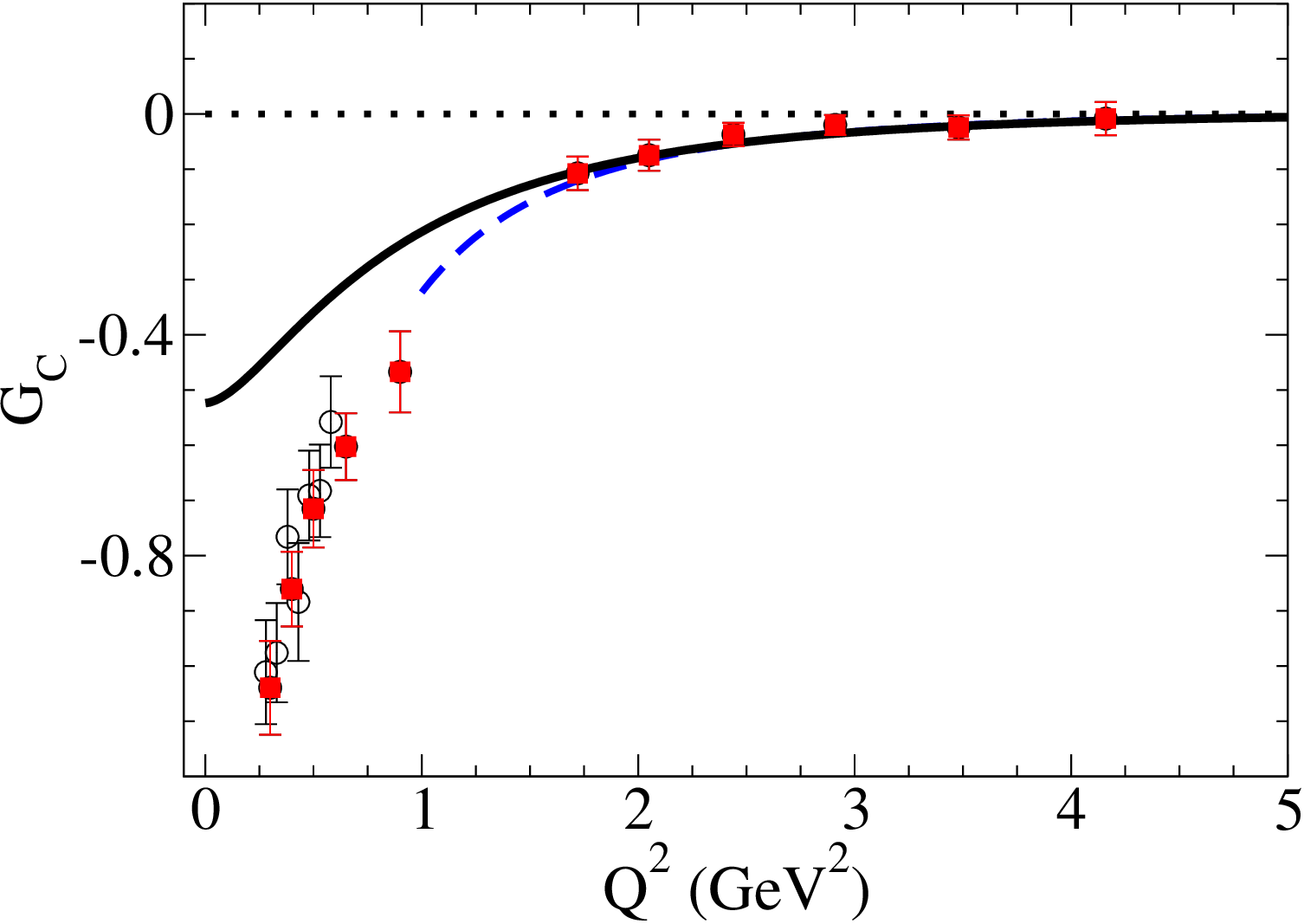}
}}
\caption{\footnotesize{
Quark core contributions to the form factors.
The dashed line line is the model 0.
The solid line is the result 
from model 1
[fit of the parameter $\beta_3$ for $Q^2> 1.5$ GeV$^2$].
Data from Ref.~\cite{Aznauryan09} (full circles),
Ref.~\cite{Mokeev12} (empty circles) and 
PDG \cite{PDG} (square).
}}
\label{figFF1}
\end{figure}

We start with the results obtained with the $R1$-state component 
only in the wave function.
To test the sensitivity of the results 
to the radial wave function parametrization 
we consider first a model where 
the $R$ radial wave function ($\psi_{R1}$) 
is taken to be identical to
the nucleon radial wave function ($\psi_N$),
written in terms of the $R$ variables.
Note that this model has no adjustable parameters.
We label this model as model 0.
The results are presented by the dashed lines, 
in Figs.~\ref{figAmp1} and \ref{figFF1},
respectively
for the helicity amplitudes and form factors.
Because in this toy model 
the nucleon and $R$ wave functions are not orthogonal 
in the relativistic formulation,
it fails necessarily for low $Q^2$, and therefore
we plot the results only for $Q^2> 1$ GeV$^2$.

It is interesting
that the results from model 0
are very close to the data for $A_{1/2}$ and $S_{1/2}$.
This suggests that the naive model gives a good 
first approximation to the $R$ wave function, 
at least for high $Q^2$.
It means that in their inner core, probed 
in the high momentum transfer region,
the baryons have a very similar structure.  
It also means that in principle the toy model can be improved by
re-adjusting the radial wave function.

We took therefore the radial wave function 
given by Eq.~(\ref{eqPsiRP1}),
where a new range parameter $\beta_3$ can be chosen 
to obtain an improved description 
of the high $Q^2$ data,
a region where the meson cloud effects are 
expected to be very small.
The orthogonality with the nucleon state 
is then imposed using the  Eq.~(\ref{eqOrthCondition}).
See the discussion in Sec.~\ref{secScalarWF}.
The parameter $\beta_3$ is determined by the fit to the 
data for  $Q^2> 1.5$ GeV$^2$.
The minimization of $\chi^2$,
gives us $\beta_3= 0.257$.
The corresponding values for $\lambda_{R1}$ 
and the normalization constant are 
$\lambda_{R1}= 0.519$ and $N_1= 12.68$.
We label this new model as model 1.

The results from model 1
for the helicity amplitudes
in the resonance rest frame, and obtained with 
the $R1$-state component only in the wave function, are presented in 
the Fig.~\ref{figAmp1}.
We conclude that the 
quark core contributions
give a good description of the 
$A_{1/2}$ and $S_{1/2}$ data 
for $Q^2> 1.5$ GeV$^2$, but fail in 
the low $Q^2$ region for all the helicity amplitudes. 
These results justify our motivation 
to describe the low $Q^2$ region 
and the amplitude $A_{3/2}$ using 
an effective parametrization of the meson cloud 
effects. 

The results for the 
electromagnetic form factors are presented 
in Fig.~\ref{figFF1}. 
The same trend of the amplitudes is observed for the
form factors, except that the discrepancy 
between the model and the data is larger for $G_E$. This happens 
because in this quark core model the amplitude $A_{3/2}$ is too small,
and, according to Eq.~(\ref{eqGE}) this amplitude
has a relevant weight for $G_E$ (3 times larger than the weight for $G_M$).

\begin{figure}[t]
\vspace{.3cm}
\centerline{
\mbox{
\includegraphics[width=2.8in]{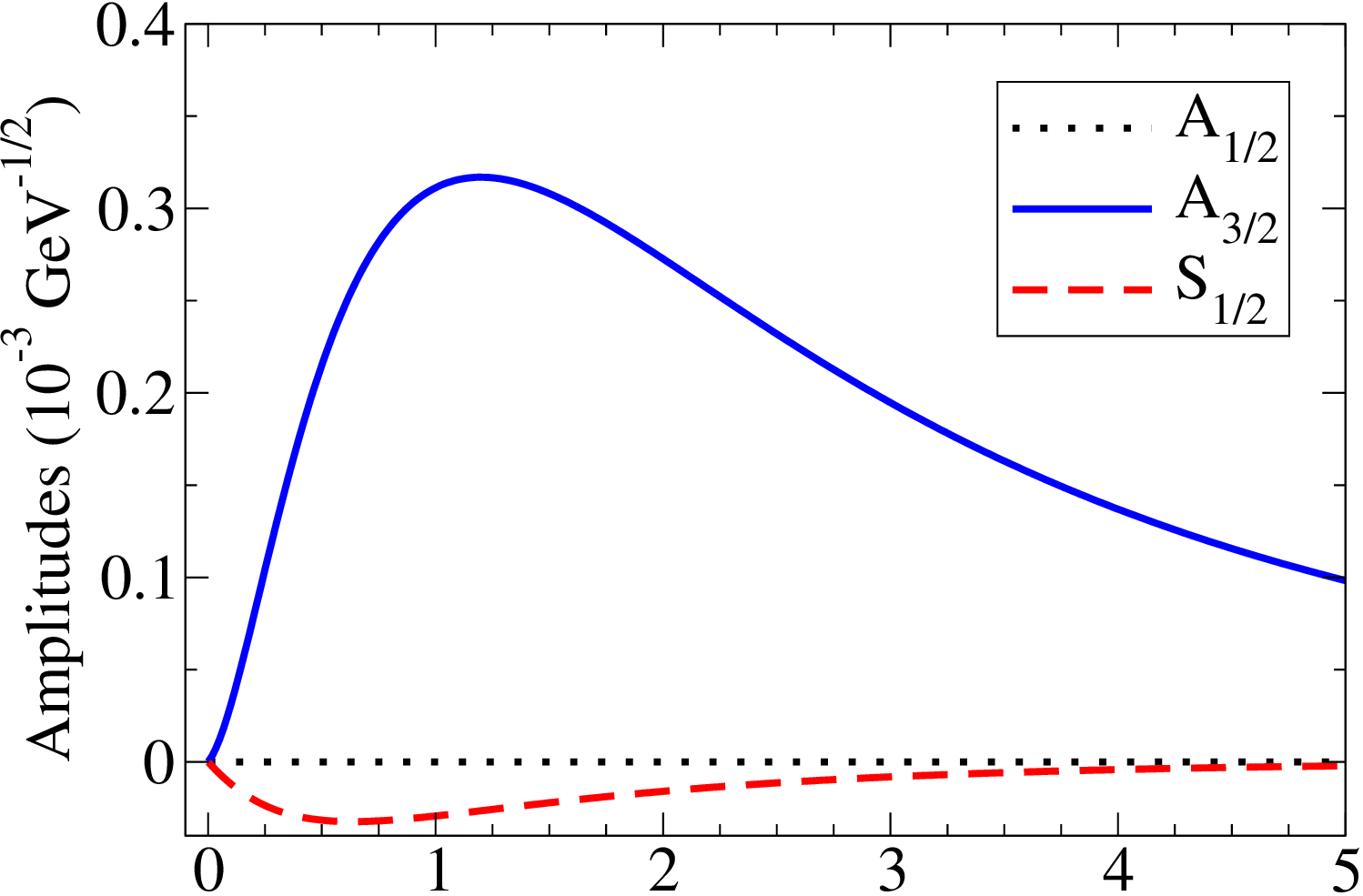}
}}
\vspace{.8cm}
\centerline{
\mbox{
\includegraphics[width=2.8in]{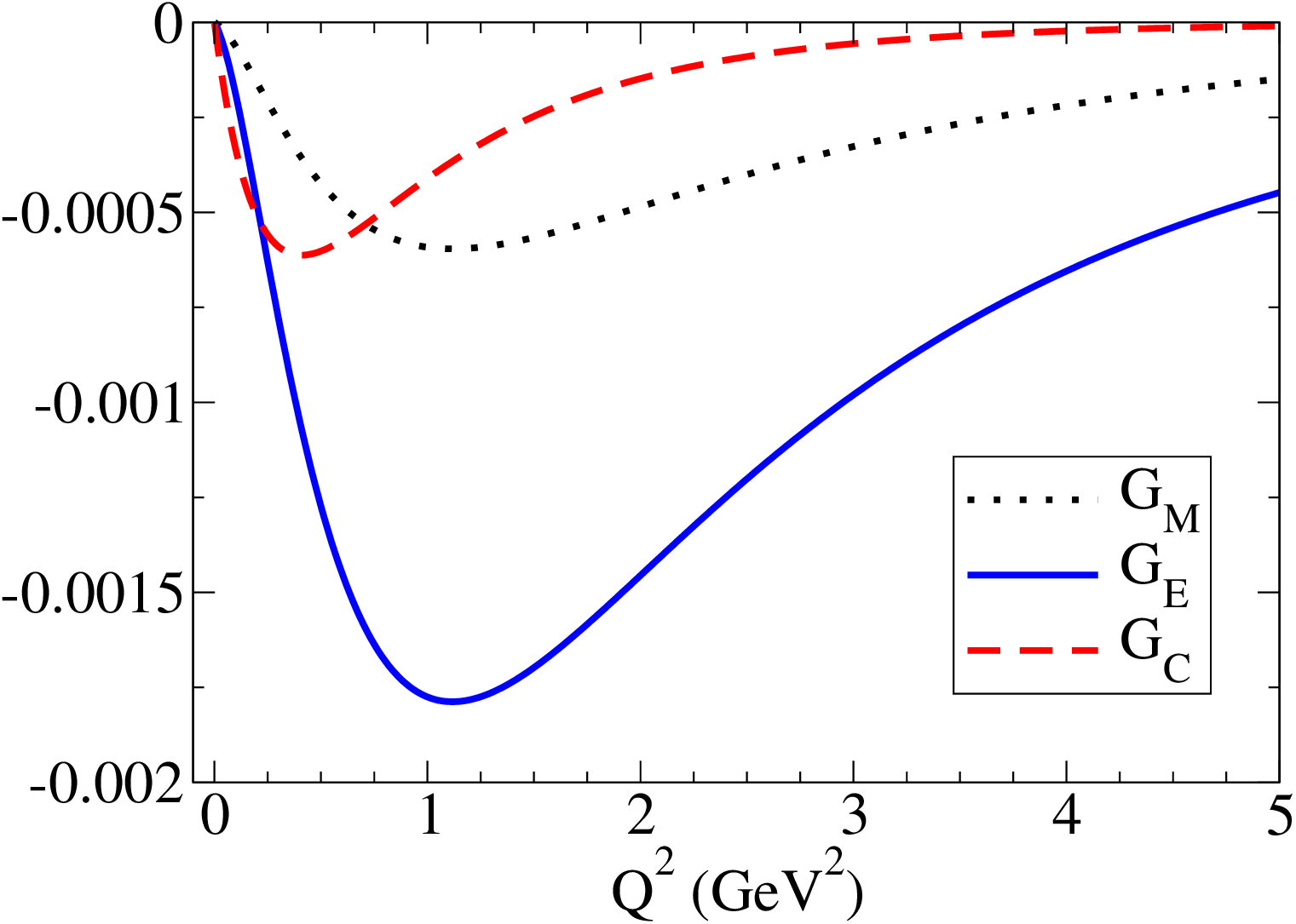}
}}
\caption{\footnotesize{
Contributions from the state $R3$ to the 
amplitudes (at the top) and form factors 
(at the bottom).
Note that those contributions are about 
2 or 3 orders of magnitude smaller that 
the contributions from $R1$ or 
the experimental data (see Figs.~\ref{figAmp1} and \ref{figFF1}).
}}
\label{figP3}
\end{figure}

Next, we included  the $R3$-state in the resonance wave function. 
The contribution 
from this component was calculated using the 
radial wave function
(\ref{eqPsiP3R}) with the value $\alpha_1= 0.337$ 
determined in Refs.~\cite{NDeltaD,LatticeD} 
for $\Delta(1232)$ (a resonance with core $S=3/2$).
Imposing that the $R3$-state is orthogonal 
to the nucleon initial state, one obtains
$\lambda_{R3}=0.557$ and $N_3=7.16$.
The results for the helicity amplitudes
and form factors are presented in Fig.~\ref{figP3}.
The results are about 2 or 3 orders 
of magnitude smaller 
than the contributions from the $R1$-state component.
The main reason is the magnitude of the 
admixture coefficient, $\sin \theta_D \simeq 0.11$,
but the smallness of the isospin coefficients 
$j_i^S= \sfrac{1}{6}(f_{i+} -f_{1-})$ 
also helps to suppress the $R3$ contributions.
Note that $j_1^S(0)=0$ and 
$j_2^S(0)= \sfrac{1}{6}(\kappa_+ - \kappa_-) \simeq 0.03$.

Figures 1 and 3 show that the $R1$-state component 
dominates in the quark core effects.
The smallness of the $R3$-state contributions
is the reason why we did not adjust
a new range parameter to the corresponding radial function.
The results from the full model
are indeed not very sensitive 
to the $R3$-state radial wave function.

\subsection{Quark core and meson cloud effects combined}

We have assumed that the decomposition 
given by Eqs.~(\ref{eqA12pi})-(\ref{eqS12pi}) is valid.
Then we were able to use
the parametrization 
from Eqs.~(\ref{eqG4pi})-(\ref{eqGCpi}) to describe 
the difference between the data and our quark core results --
which we interpret as due to contributions from the meson cloud.
We show now the results from combining the quark core effects 
with these meson cloud effects introduced in Sec.~\ref{secPC}.

\begin{table}[t]
\begin{tabular}{c c c c}
\hline
\hline
$a_M$  &  $\lambda_\pi^{(4)}$ & $\lambda_\pi^{M}$ &  $\lambda_\pi^{C}$ \\
4.934  &    1.354            &   $-0.404$              &  $-1.851$ \\
\hline
       &  $\Lambda_4^2$      &  $\Lambda_M^2$   &   $\Lambda_C^2$  \\
       &    20.0             &   1.663               &  1.850 \\         
\hline
\hline
\end{tabular}
\caption{
Model parameters for the pion cloud parametrization.
$a_M$ has units GeV$^{-2}$.
The coefficients $\lambda_\pi$ have no dimensions.
The cutoffs are in units GeV$^2$. }
\label{tableFIT}
\end{table}

The parameters of the best fit are in Table~\ref{tableFIT}.
We fit only the parameters related with 
the pion cloud dressing, 
since if we perform a combined fit of the  
valence plus pion cloud contributions we would 
lose control of the quark core content,
and artificially very large pion cloud contributions 
emerge as numerical solutions.
In this work we do not have an indirect way of 
calibrating  
the quark core contributions, 
as we did in our studies of the nucleon, Roper and the 
$\Delta(1232)$, where lattice QCD data 
are available~\cite{Lattice,LatticeD,Roper}. 
The only way here
to check that the quark core content 
is under control, is to extrapolate
our model to  the large $Q^2$ regime.

Concerning the fit, 
we note that the $A_{3/2}$ amplitude 
is determined only by the function $G_4^\pi$,
but $A_{1/2}$ depends on both $G_M^\pi$ and  $G_4^\pi$.
Nevertheless, an  overall and simultaneous fit
of the two amplitudes $A_{1/2}$ and $A_{3/2}$ 
(or $G_4^\pi$ and $G_M^\pi$) is better constrained 
than the two-step procedure of fitting
$A_{3/2}$ to fix separately $G_4^\pi$ first, followed by an independent fit of
$A_{1/2}$ to fix $G_M^\pi$.
Finally, $G_C^\pi$ is only constrained by the data from 
the amplitude $S_{1/2}$ (or form factor $G_C$). 

In the fitting procedure we noticed that 
the best fit for $A_{3/2}$ is achieved when we fix  
the cutoff parameter $\Lambda_4^2$ in Eq.~(\ref{eqG4pi})
at an extremely large value,
such that the multipole factor  in that formula
behaves as a constant.
To preserve a multipole falloff 
for very high $Q^2$ we took  $\Lambda_4^2=20$ GeV$^2$, 
allowing  the multipole factor 
{\it to behave  like} a constant in the $Q^2$ regime 
under study, although behaving as $1/Q^6$ for much larger values of $Q^2$.

\begin{figure}[t]
\vspace{.3cm}
\centerline{
\mbox{
\includegraphics[width=2.8in]{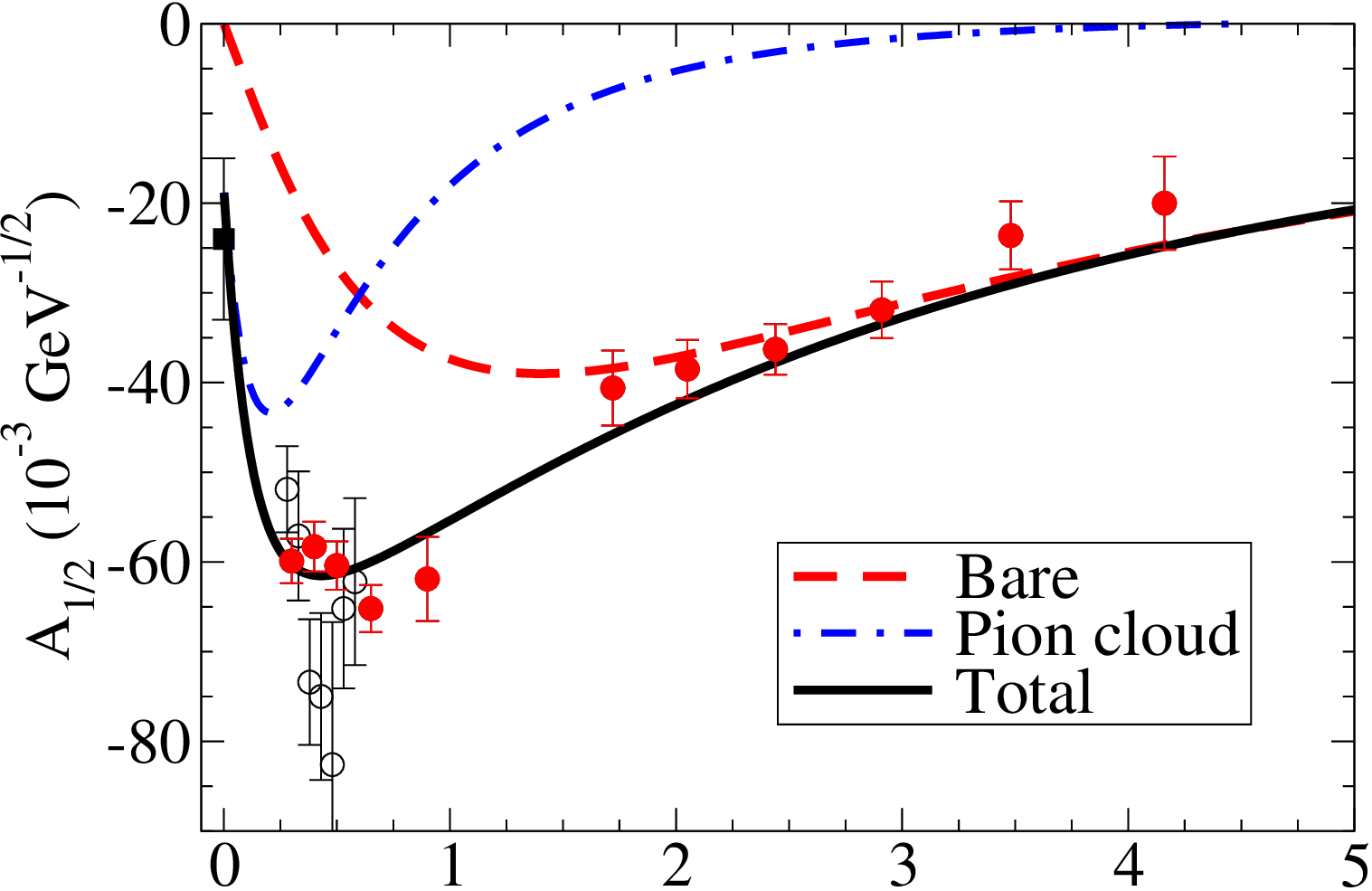}
}}
\centerline{
\vspace{.5cm} }
\centerline{
\mbox{
\includegraphics[width=2.8in]{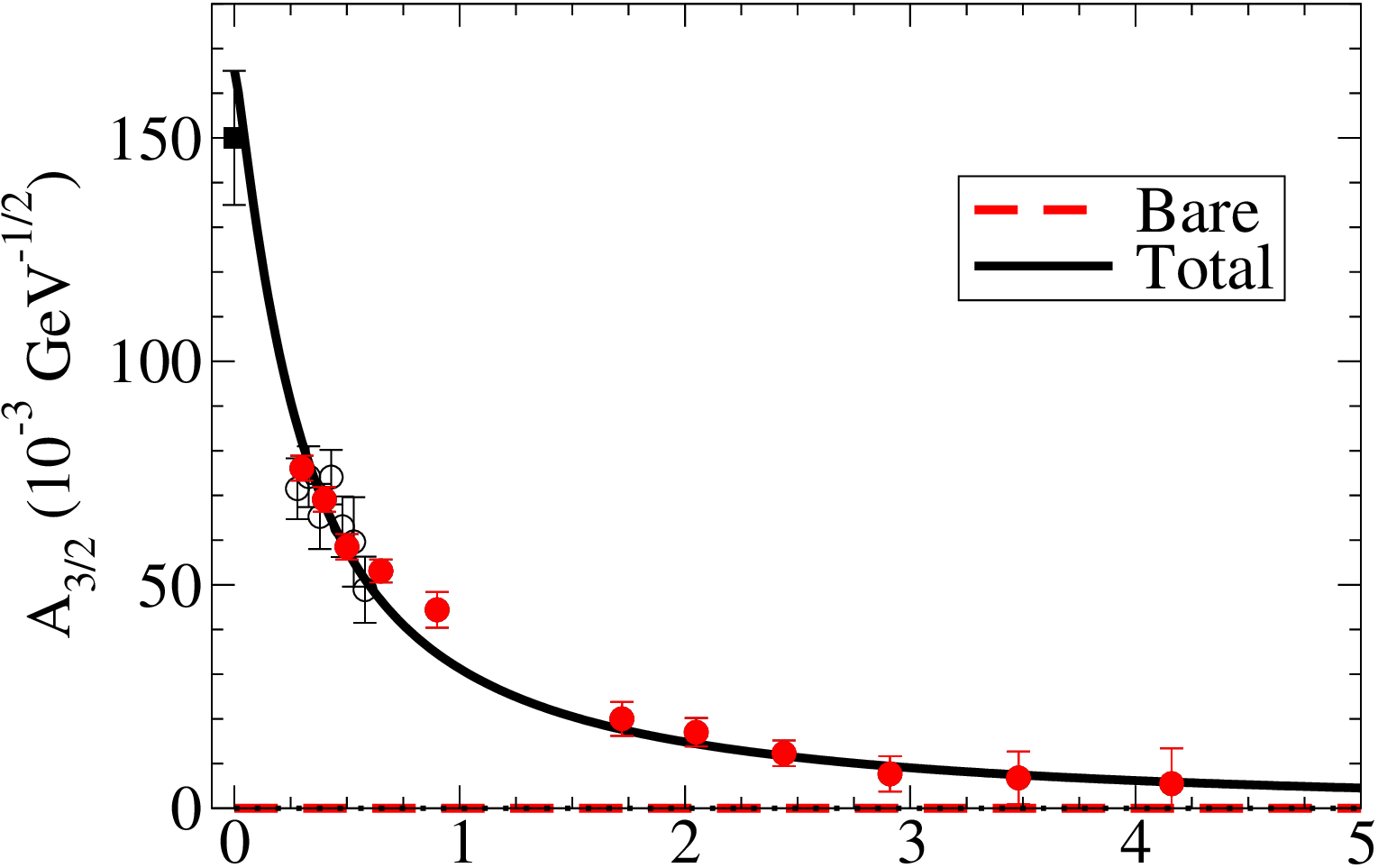}
}}
\centerline{
\vspace{.4cm} }
\centerline{
\mbox{
\includegraphics[width=2.8in]{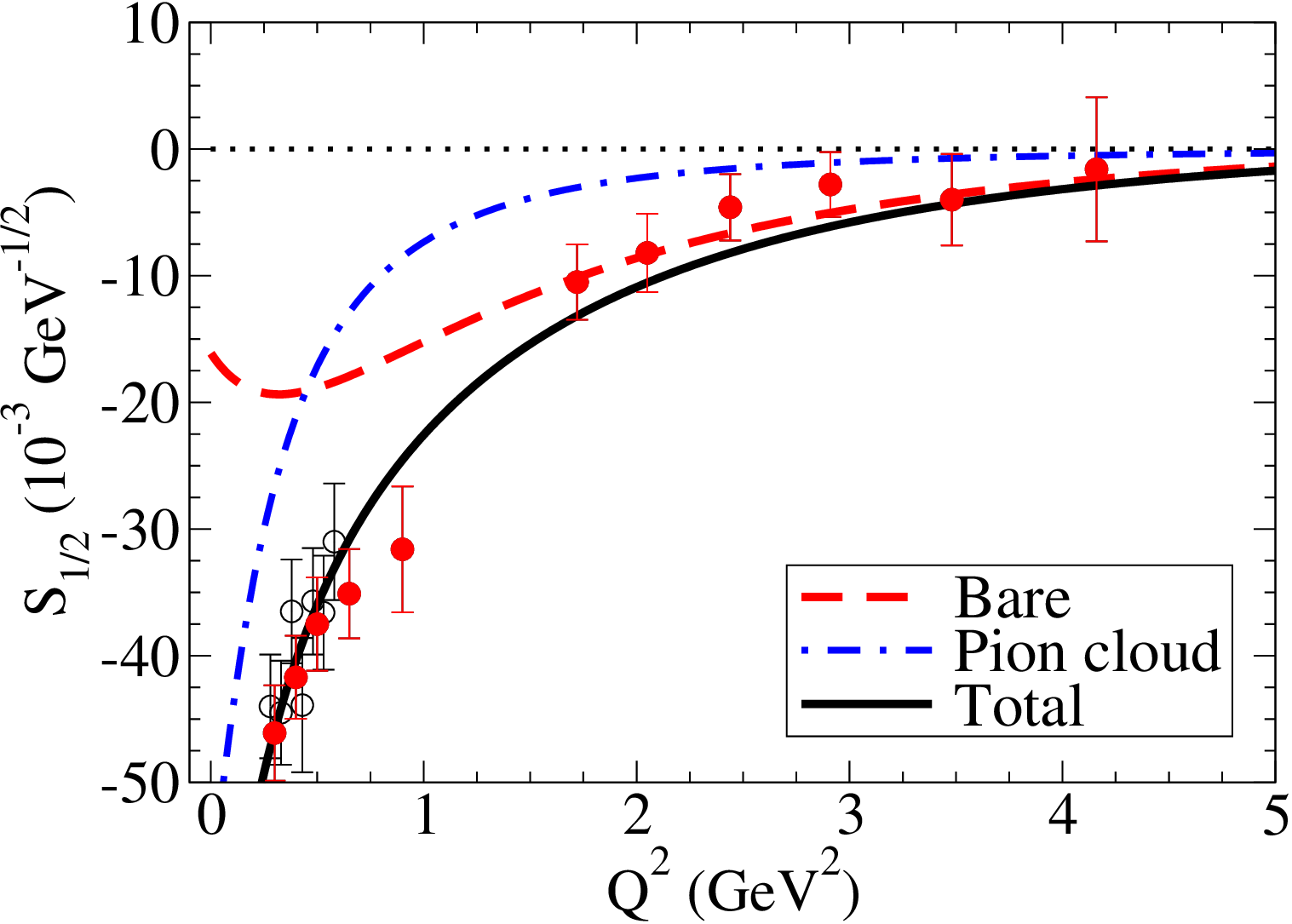}
}}
\caption{\footnotesize{
Quark core plus pion cloud contributions to 
the helicity amplitudes. 
For the amplitude $A_{3/2}$ the pion cloud contribution 
coincides with the total.
Data from Ref.~\cite{Aznauryan09} (full circles),
Ref.~\cite{Mokeev12} (empty circles) and PDG~\cite{PDG} (square).
}}
\label{figAmp2}
\end{figure}

\begin{figure}[t]
\vspace{.3cm}
\centerline{
\mbox{
\includegraphics[width=2.8in]{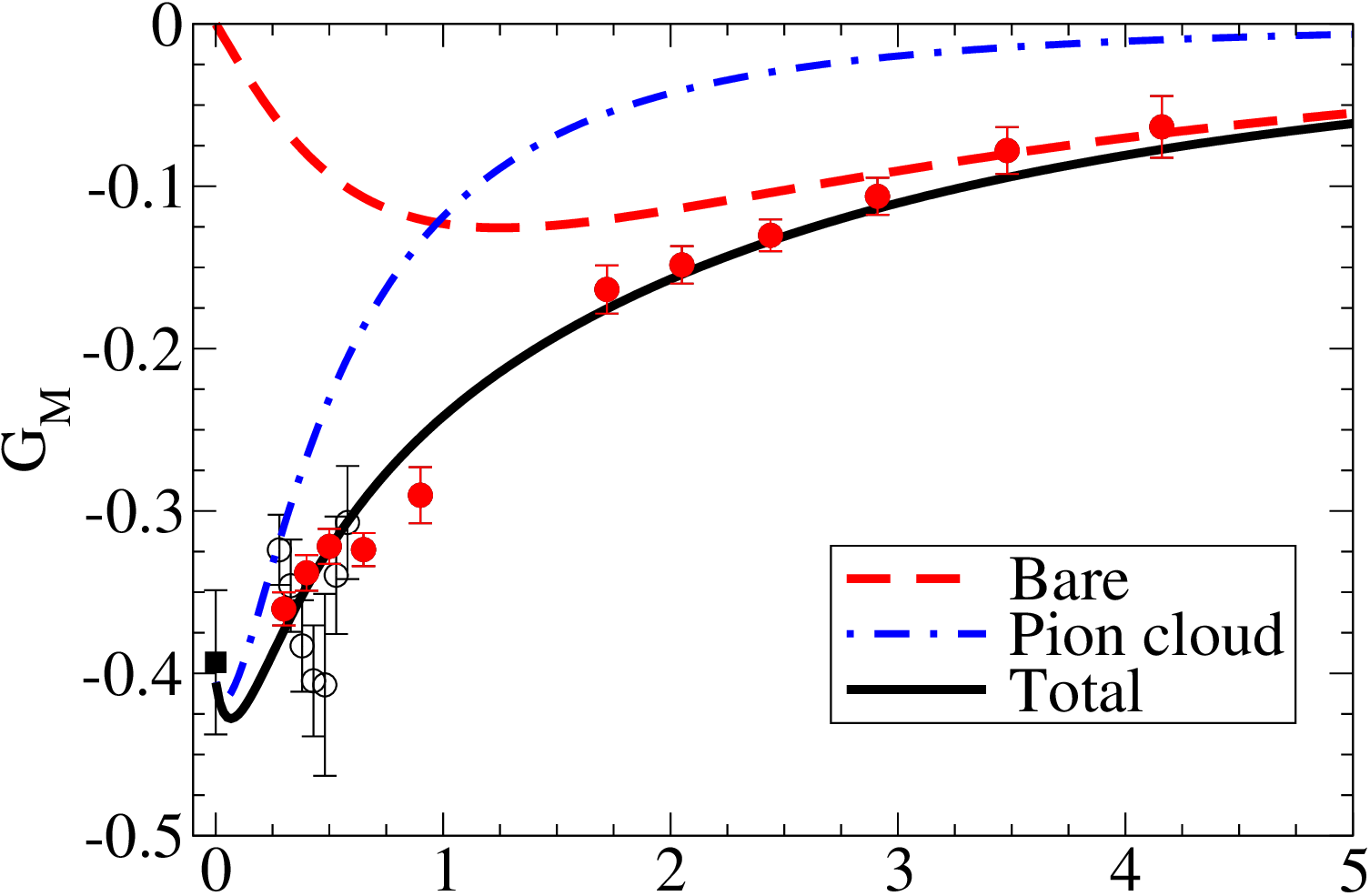}
}}
\centerline{
\vspace{.5cm} }
\centerline{
\mbox{
\includegraphics[width=2.8in]{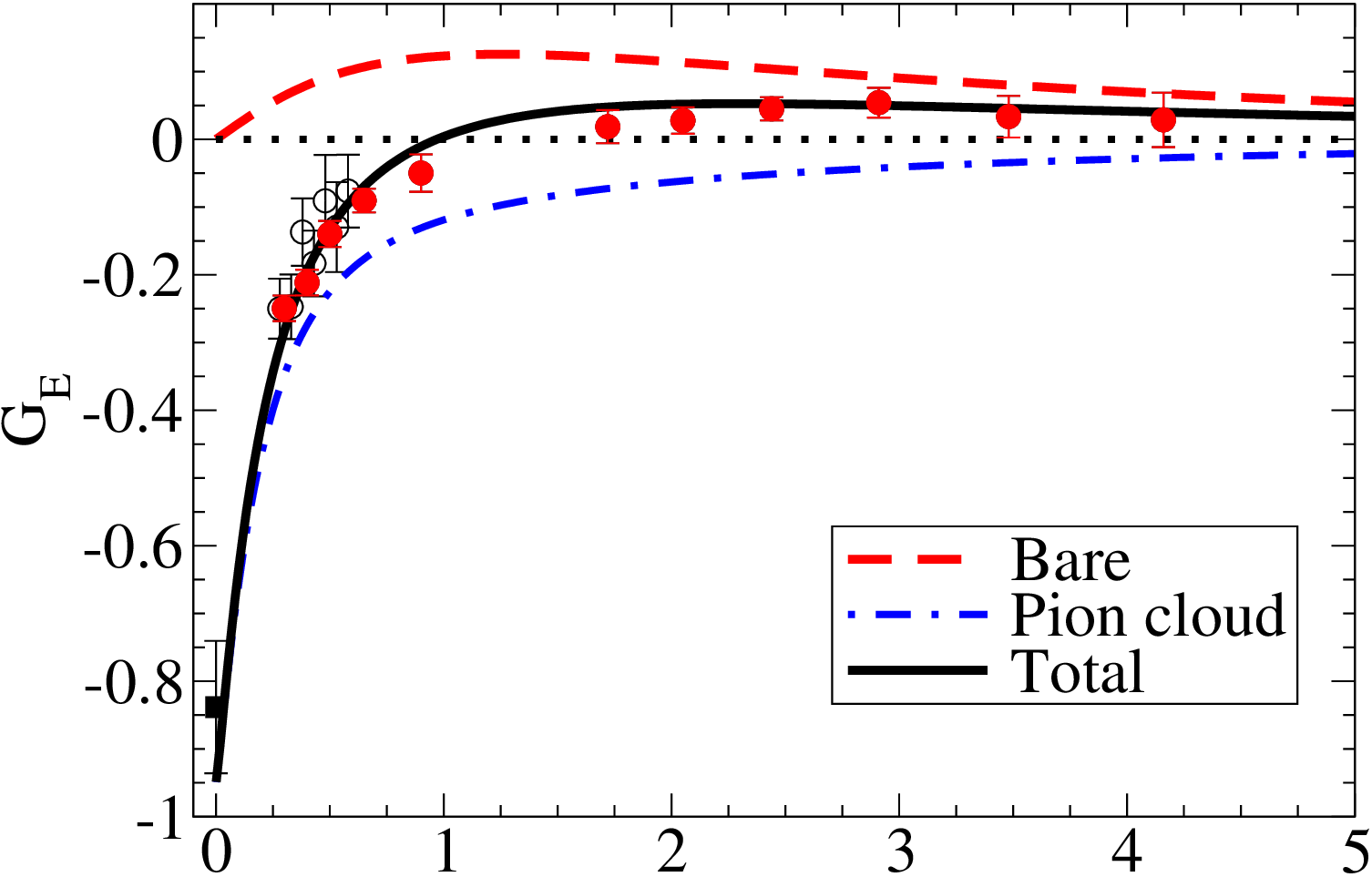}
}}
\centerline{
\vspace{.5cm} }
\centerline{
\mbox{
\includegraphics[width=2.8in]{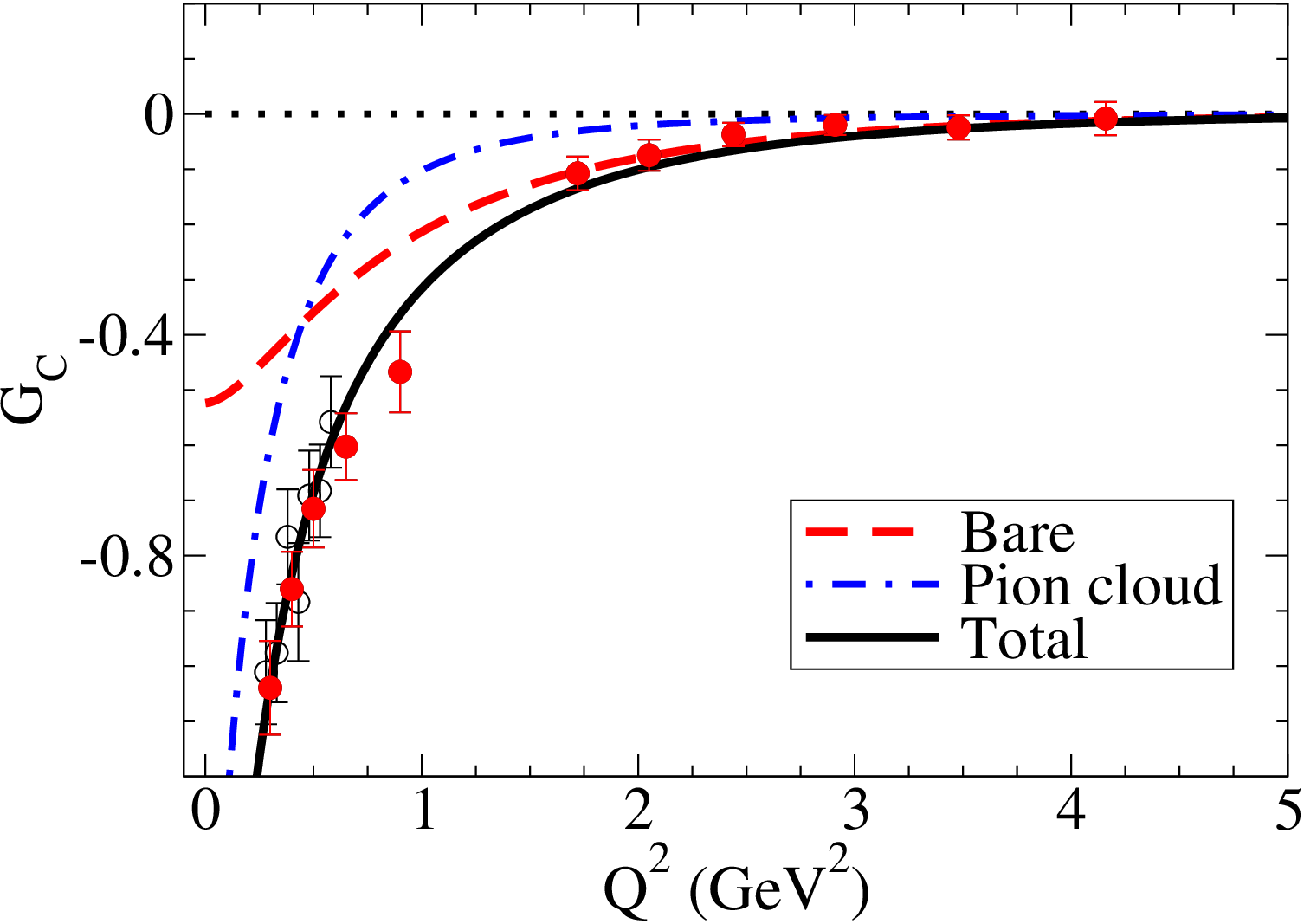}
}}
\caption{\footnotesize{
Quark core
 plus pion cloud contributions to
the form factors. 
Data from Ref.~\cite{Aznauryan09} (full circles),
Ref.~\cite{Mokeev12} (empty circles) and PDG~\cite{PDG} (square).
}}
\label{figFF2}
\end{figure}

Using the parameters of the best fit 
(Table~\ref{tableFIT}) one can write 
the final form for the pion cloud contributions 
to the helicity amplitudes as 
\ba
A_{3/2}^\pi &=& 
165 \, \sqrt{1 + \tau} 
\left( \frac{\Lambda_4^2}{\Lambda_4^2+ Q^2}\right)^3 F_\rho \tau_3 \\
A_{1/2}^\pi &=&  - 114 \, (1 + a_M Q^2) 
\sqrt{1 + \tau} 
\left( \frac{\Lambda_M^2}{\Lambda_M^2+ Q^2}\right)^3 F_\rho \tau_3 
\nonumber \\
& & + 
\frac{1}{\sqrt{3}} A_{3/2}^\pi  
\label{eqA12piN} \\
S_{1/2}^\pi &=&
-1094 \, \sqrt{1 + \frac{Q^2}{(M_R-M)^2}} 
\left( \frac{\Lambda_C^2}{\Lambda_C^2+ Q^2}\right)^3 F_\rho \tau_3.
\nonumber \\
& & 
\label{eqS12piN}
\ea
Where the numerical coefficients are in units of 10$^{-3}$ GeV$^{-1/2}$.

The results for the 
combination of the quark core and pion cloud contributions
are presented in  Fig.~\ref{figAmp2},
for the helicity amplitudes, 
and in  Fig.~\ref{figFF2}, for the form factors.

In   Fig.~\ref{figAmp2}, there is an excellent description of $A_{3/2}$, 
obtained with
the parametrization of $G_4^\pi$.
From the results for $A_{1/2}$ we conclude that the parametrization 
of the pion cloud for $G_M^\pi$ is important to obtain 
a good description of the data at low $Q^2$, and in particular, it
is responsible for the minimum near $Q^2 \approx 0.2$ GeV$^2$.
It is this minimum that demands the
inclusion of the factor $(1 + a_M Q^2)$ 
in the $G_M^\pi$ parametrization.
The shape of $G_M^\pi$ is shown in the panel for $G_M$ in Fig.~\ref{figFF2}.
The results for the $S_{1/2}$ amplitude show that
the pion cloud effects are large at low $Q^2$ [see  Eq.~(\ref{eqS12piN})],
but fall off very fast with $Q^2$.
In general, meson cloud effects explain well
the low $Q^2$ behavior of all helicity amplitudes.

In Fig.~\ref{figFF2} the form factor results encode the 
same information as the helicity amplitudes
but in a different perspective.
We can make two remarks on Fig.~\ref{figFF2}.
Our first remark is 
the fast falloff of the pion cloud contributions 
to $G_M$ and $G_C$. The second remark is that for $G_E$, the difference 
between the quark core contributions
and the experimental data is still meaningful for $Q^2> 2$ GeV$^2$,
and that
the effect of the pion cloud comprises a 
significant fraction of the full result.  This is a consequence of
our results for $A_{3/2}$ 
being in great part determined by the pion cloud.

In the comparison of our work to the literature,
our results agree with the general conclusion 
that at very high $Q^2$ the $A_{1/2}$ amplitude
is the dominant helicity amplitude.
This dominance of $A_{1/2}$ is equivalent  
to have in that regime $G_E \simeq - G_M$~\cite{Aznauryan09,Koniuk80,Close72}.
[See Eqs.~(\ref{eqGM}) and  (\ref{eqGE}).]
Our calculations are also consistent with 
the findings that in the low $Q^2$ regime 
the meson cloud contributions are decisive for the description of the data,
as suggested by Ref.~\cite{Diaz08} 
within a coupled-channel formalism.

Compared to some estimations from 
constituent quark models~\cite{Koniuk80,Close72,Warns90,Aiello98,Merten02,Santopinto12a,Capstick95}, 
our model  gives results for $A_{3/2}$  in the low $Q^2$ 
region that are too small.
Although in 
Refs.~\cite{Warns90,Aiello98,Merten02,Santopinto12a,Ronniger13,Golli13}, 
the result for $A_{3/2}$ near $Q^2=0$
is typically $1/3$ of the experimental result,
Ref.~\cite{Capstick95}  predicts 
a result for $A_{3/2}(0)$ that is very 
close to the PDG value for small $Q^2$.
The possible explanations for our $A_{3/2} \approx 0$ result  
were already given in the discussion made in Sec.~\ref{DiscussionA3/2}.

We may also compare our result
for the form factor $G_1$  (or $G_M$ [see Eq.~(\ref{eqGM})]) with the 
estimations of the light-front quark model from Ref.~\cite{Aznauryan12b}.
This model,
contrary to our model, gives a good 
description of the $Q^2<1.5$  GeV$^2$ data,
even without meson cloud effects.
But in that work, for the estimation of the effects 
of the meson cloud
using high $Q^2$ data, the strength of the 
quark contributions 
was reduced about 20\%.
If we consider a similar reduction in our quark model
we also improve our description of the data.
Focusing on the graph for $A_{1/2}$ from Fig.~\ref{figAmp2}
or the graph for $G_M$ from Fig.~\ref{figFF2},
the suppression of 20$\%$ in the quark core contributions
would shift our model results 
to be almost on top of the high $Q^2$ data,
improving the result from the quark core contribution.
Unfortunately, as our meson cloud estimation 
is phenomenological, and not determined together with 
the quark core wave function, 
we do not have a simple method to 
estimate the effect of the meson cloud 
in the normalization of our wave function.
It is nevertheless encouraging to notice the 
convergence of the two different works.

\subsection{Jlab and MAID parametrizations 
and extrapolation of our results to larger $Q^2$ values}
\label{secLargeQ2}

\begin{figure}[t]
\vspace{.3cm}
\centerline{
\mbox{
\includegraphics[width=2.8in]{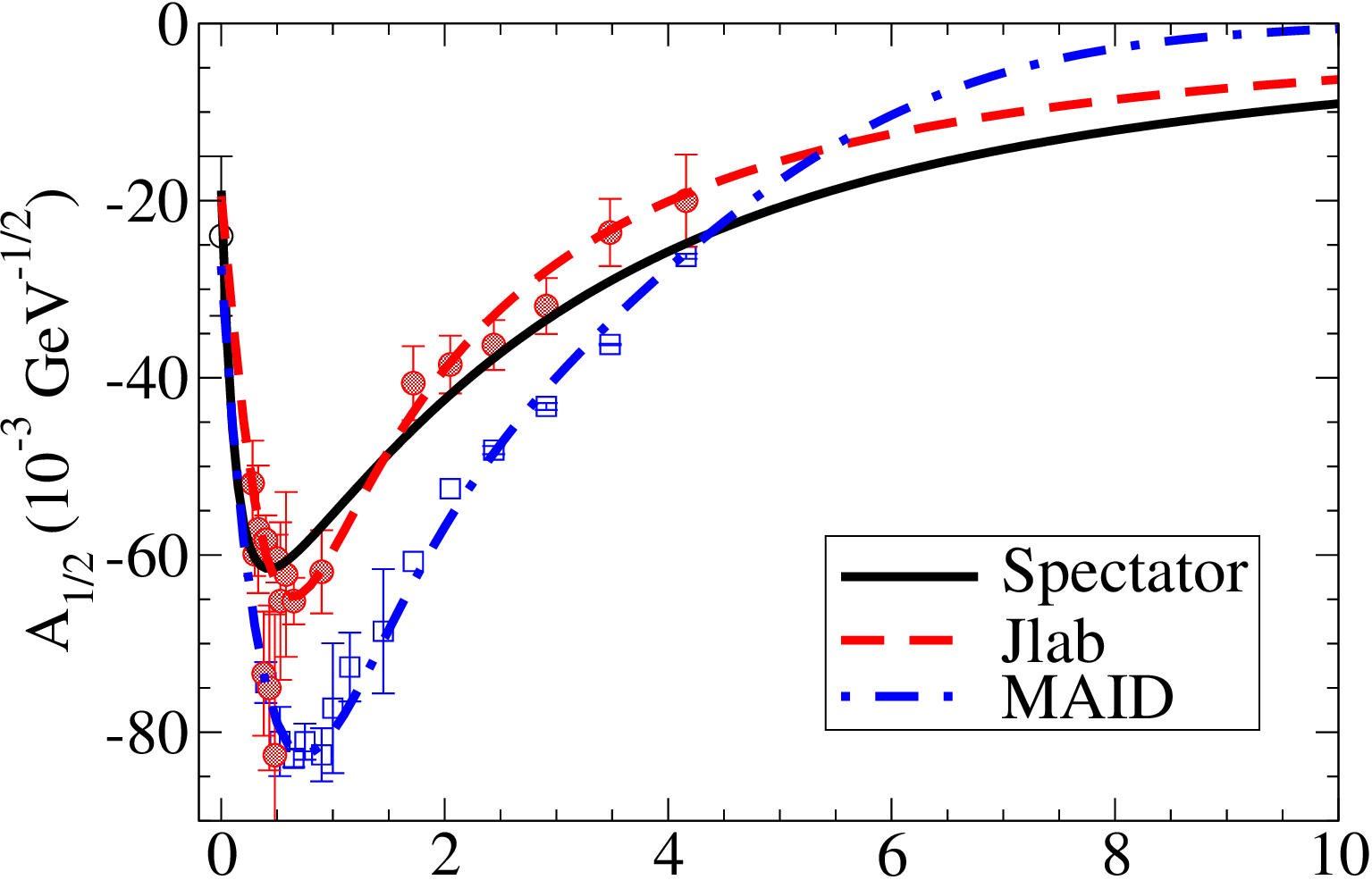}
}}
\centerline{
\vspace{.5cm} }
\centerline{
\mbox{
\includegraphics[width=2.8in]{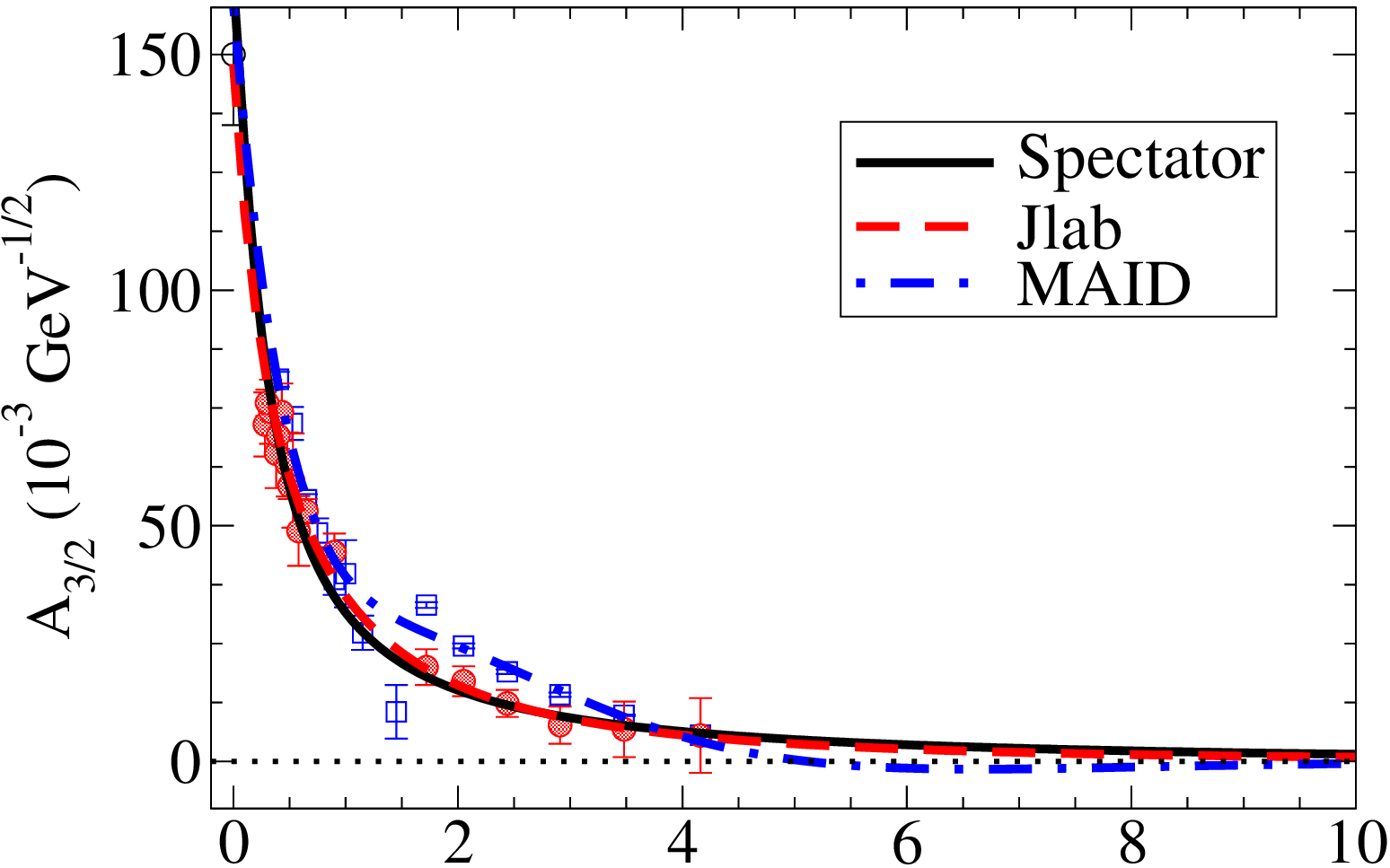}
}}
\centerline{
\vspace{.5cm} }
\centerline{
\mbox{
\includegraphics[width=2.8in]{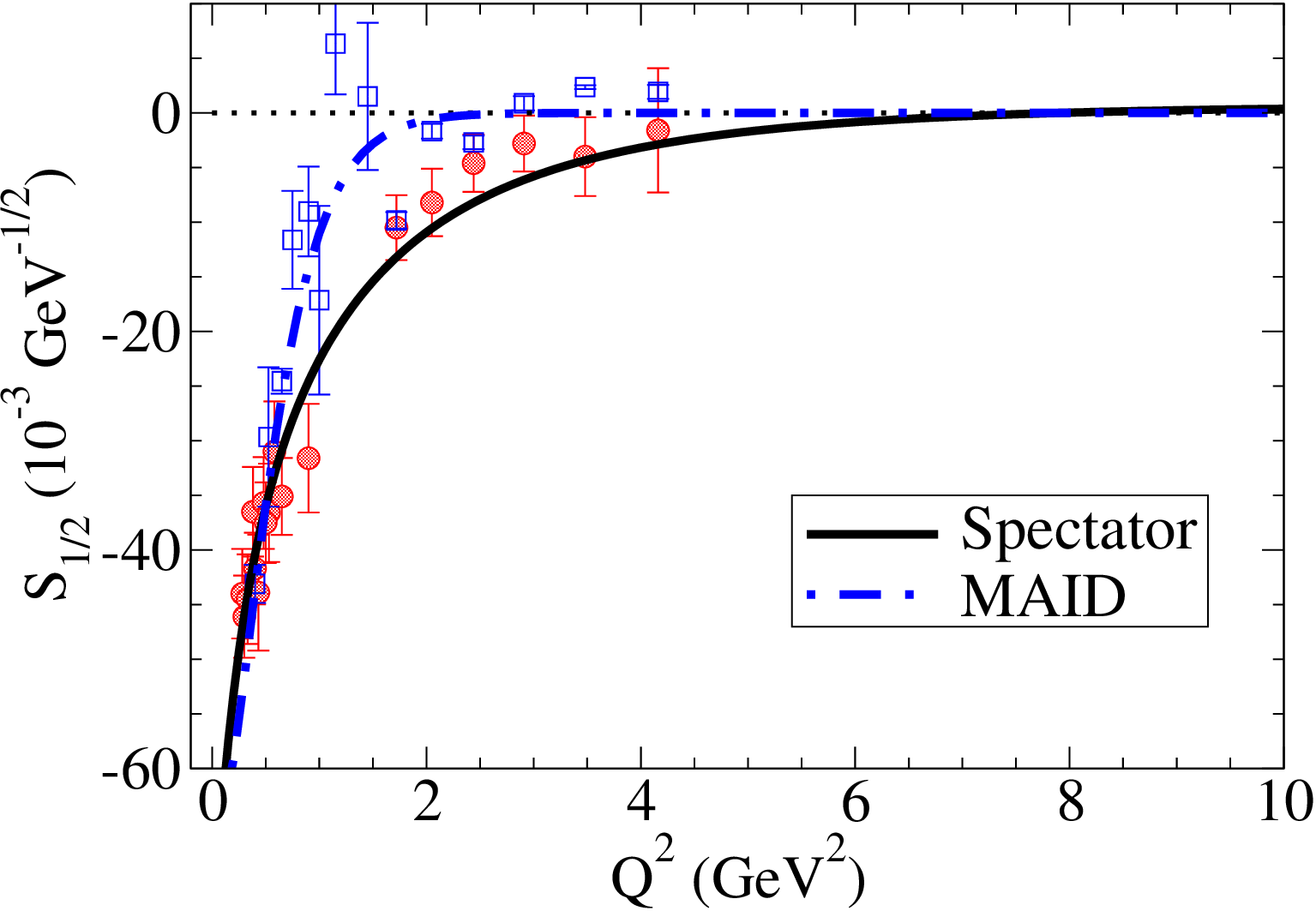}
}}
\caption{\footnotesize{
Helicity amplitudes up to 10 GeV$^2$.
Comparing the spectator model with the 
CLAS data (circles) \cite{Aznauryan09,Mokeev12} and 
the MAID analysis (squares) \cite{Drechse07,Tiator09}. 
The results of the Jlab \cite{Aznauryan12a} 
and MAID \cite{Tiator09} parametrizations 
are also shown.
}}
\label{figParam}
\end{figure}

\begin{figure}[t]
\vspace{.2cm}
\centerline{
\mbox{
\includegraphics[width=2.8in]{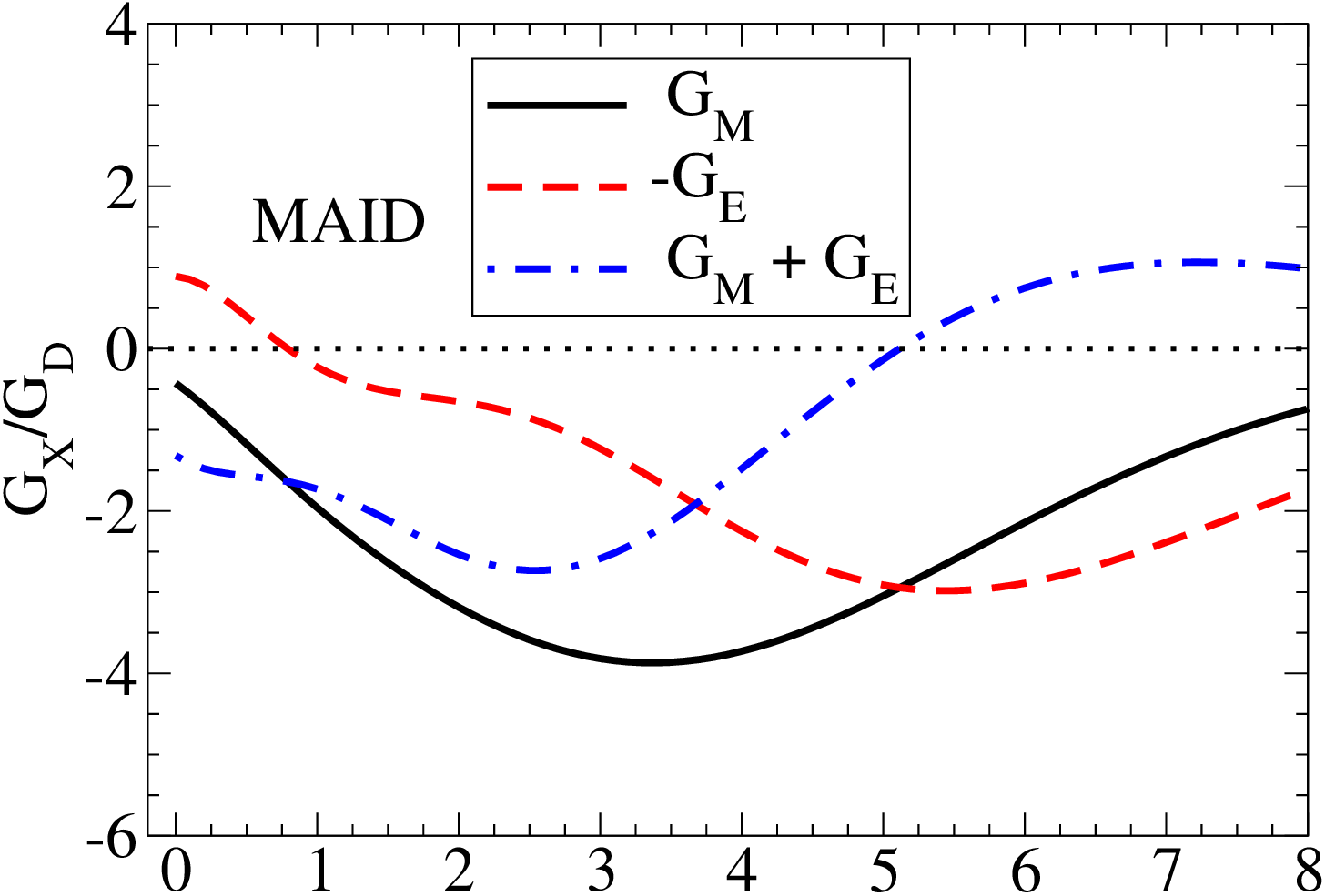}
}}
\centerline{
\vspace{.4cm} }
\centerline{
\mbox{
\includegraphics[width=2.8in]{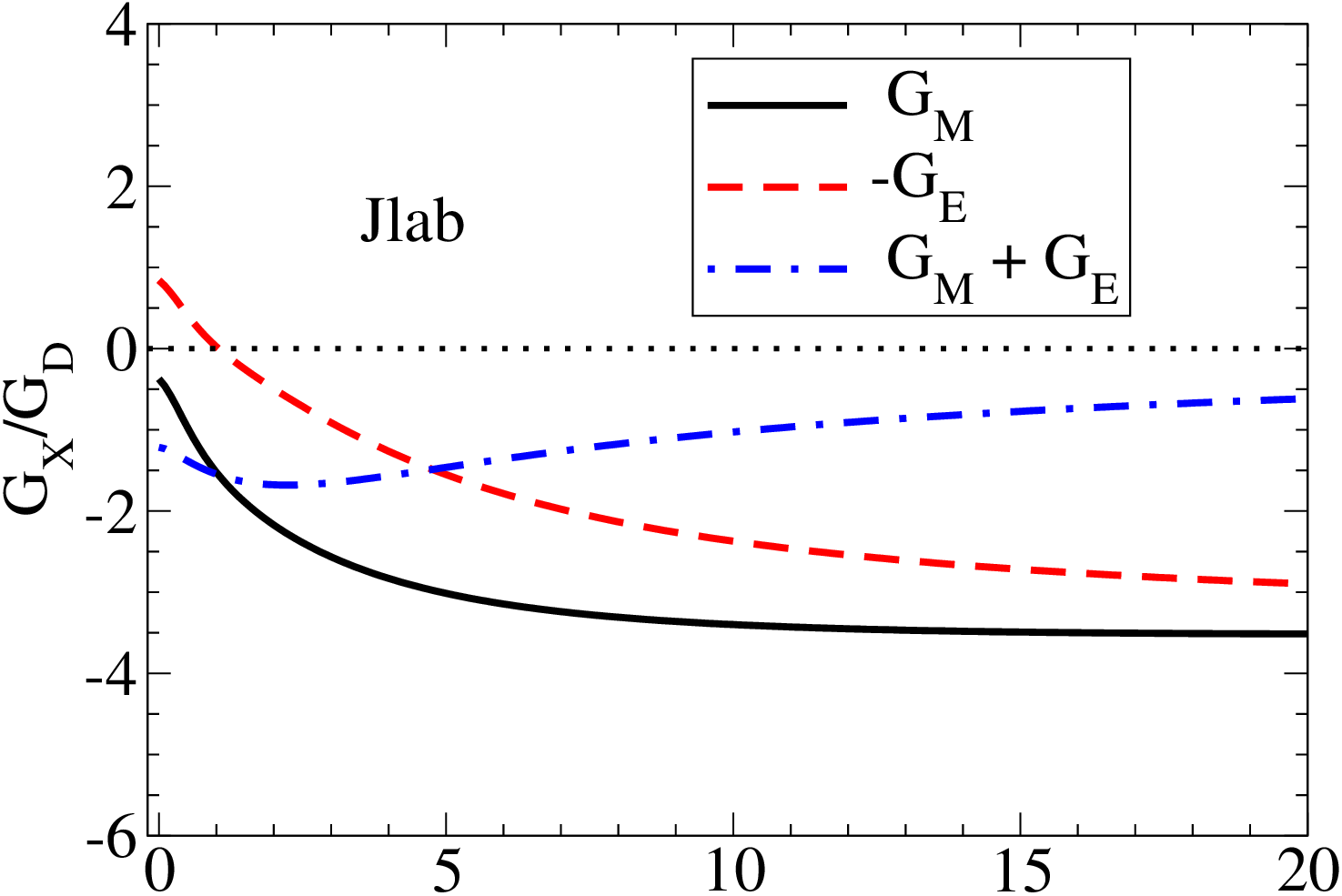}
}}
\centerline{
\vspace{.4cm} }
\centerline{
\mbox{
\includegraphics[width=2.8in]{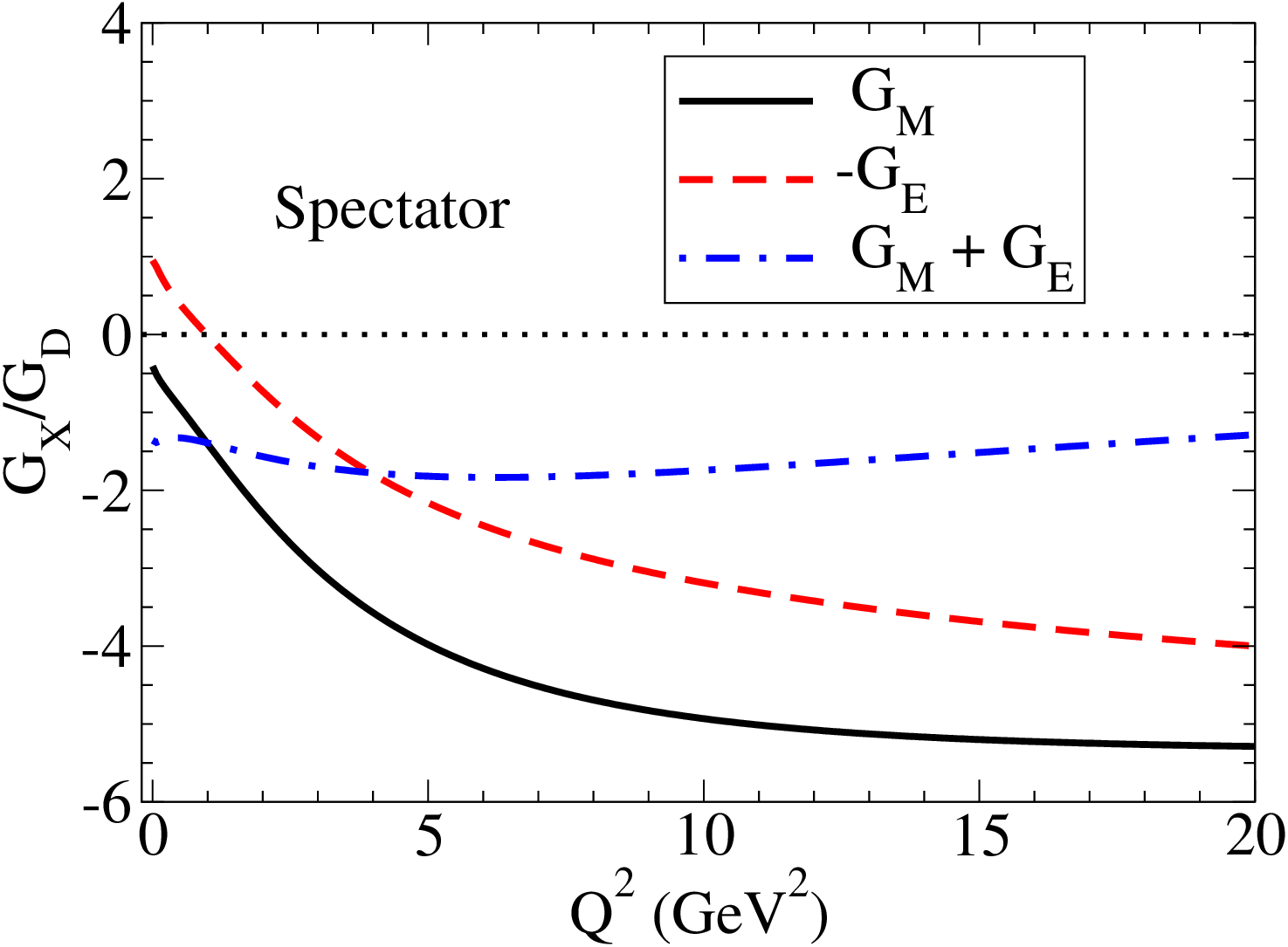}
}}
\caption{\footnotesize{
Form factors, $G_M$, $-G_E$ and $G_M+G_E$ 
normalized by dipole form factor $G_D$ 
given by Jlab~\cite{Aznauryan12a} and MAID~\cite{Tiator09} 
parametrizations,
and  by the Spectator model.
}}
\label{figFFparam}
\end{figure}

Besides the CLAS analysis~\cite{Aznauryan09,Mokeev12}
reported previously, there are also 
data from  the MAID 2007 analysis,
hereafter called MAID analysis.
The results from MAID include
the analysis from old data and 
recent Jlab data~\cite{Tiator09,Drechse07}.

In Ref.~\cite{Aznauryan12a}
a fit to the CLAS $A_{1/2},A_{3/2}$ data 
based on rational functions of $Q=\sqrt{Q^2}$ was  presented. 
We label it as Jlab parametrization.
The MAID parametrization is a fit to the MAID analysis 
presented in Ref.~\cite{Tiator09}, which instead is 
based on a combination of 
polynomials  and exponentials of $Q^2$.
Because the MAID parametrization is dominated 
by exponential falloffs for large $Q^2$,
it does not reproduce the expected 
power law behavior from pQCD. 
The range of application of the MAID parametrization
should then be restricted to the range 
of the available data.

The comparisons between the Jlab and MAID data, their parametrization fits 
and the results of the present model
(labeled as Spectator) are
shown in the panels of Fig.~\ref{figParam}
for the helicity amplitudes. 
For the $S_{1/2}$ amplitude there is no Jlab parametrization.
The two different  parametrizations 
describe the respective data, 
but differ significantly.
Note in particular for $A_{1/2}$
the difference of falloff 
between our model 
and the Jlab and MAID parametrizations 
when $Q^2> 5$ GeV$^2$.
This makes extremely interesting the data  for  higher $Q^2$ values,  
possible by the
forthcoming Jlab 12-GeV upgrade.

This is the reason why in this section 
we also look at the high $Q^2$ momentum transfer region,
that may eventually be explored with this upgrade. 
We extrapolated then our model to that region, and compared it with the Jlab 
and the MAID parametrizations,  to  predict when
the scaling between $G_M$ and $G_E$
in that high $Q^2$ region appears.
The different predictions for large $Q^2$,
in particular for $A_{1/2}$ and $A_{3/2}$, may be 
better analyzed by looking  at the form factors $G_M$ and $G_E$.
This is because as the ratio $|A_{3/2}|/|A_{1/2}|$ 
falls off very quickly when $Q^2$ increases,
it is expected that $-G_E$ approaches $G_M$ 
in that regime.
As both $G_M$ and $-G_E$ fall off very fast 
with $Q^2$, and assuming that $G_M$ and $G_E$ go with 
$1/Q^4$ for very large $Q^2$, it is advantageous 
in the study of their asymptotic behavior to normalize those form factors 
with the dipole form factor 
$G_D=\left( 1 + \frac{Q^2}{0.71}\right)^{-2}$,
with $Q^2$ given in GeV$^2$. 
This enables 
a better visualization of the falloff tail
and is usually done for the nucleon form factors.
The results for, $G_M$, $-G_E$ and 
$G_M+ G_E$, normalized by $G_D$, are 
presented in Fig.~\ref{figFFparam}.
It is interesting to look at the results for $G_M+ G_E$,
because pQCD predicts that it is  strongly 
suppressed \cite{Carlson,Carlson2}.

In Fig.~\ref{figFFparam} 
we restrict the results from the MAID parametrization
to the region $Q^2 < 8$ GeV$^2$,
due to the fast exponential falloff of the 
respective form factors.
For the Jlab parametrization of the data, and the Spectator model, 
we note the scaling of $G_M$ and $-G_E$ with $G_D$,
depicted by the almost flat lines obtained
for $Q^2> 15$ GeV$^2$,  especially for $G_M$.
In both cases we can observe also the 
falloff of $G_M + G_E$ with increasing $Q^2$.
Since $G_M + G_E$ is proportional 
to the amplitude $A_{3/2}$, the falloff of $G_M + G_E$
relative to $G_M$, is the sign of the suppression of $A_{3/2}$ 
relative to $A_{1/2}$.

Finally we note that in the large $Q^2$ region our model 
differs from the Jlab parametrization for $G_M$ and $-G_E$:
although our model and the data parametrization
fall with the same $1/Q^4$ power, 
our results are larger in absolute value.
That behavior can also be observed for $A_{1/2}$
in  Fig.~\ref{figParam}.

We conclude that future experiments, 
as the ones planned for the Jlab-12 GeV upgrade,
will be crucial to better constrain models  \cite{NSTAR,Jlab12}.

\section{Conclusions}
\label{secConclusions}

We have applied the covariant spectator quark model
to the $N^\ast(1520)$ system and to the 
$\gamma^\ast N \to N^\ast(1520)$ reaction.

Our formulation takes the wave function of the  $N^\ast(1520)$ 
as a combination of two components, $R1$  and $R3$, with core spin 1/2 and
3/2  respectively.
We conclude that the model with only quark core effects included
is particularly successful
in the description of the high $Q^2$ data ($Q^2 > 2$ GeV$^2$)
of the $A_{1/2}$ and $S_{1/2}$ helicity amplitudes.

In the small $Q^2$ region there is a  discrepancy between the data 
and our results for  helicity amplitudes and form factors. 
 This is not
surprising since for small $Q^2$  the photon is expected 
to couple to the baryon
as a whole and to the peripheral meson cloud,
given the small momentum resolution (long wavelengths) 
of the electromagnetic probe.

As we do not include in our quark core model
processes where the pion or heavier mesons collectively dress 
the three quarks, we have interpreted the deviations 
of our results from data in that region as meson cloud effects 
not present in our model, and we proceeded to obtain their parametrization.

The meson cloud parametrization used
in the present work is inspired in previous 
parametrizations of pion cloud effects.
However, since we obtain a 
good description of the overall data,
it can be regarded as an effective 
representation of all meson cloud effects
(including $\pi \pi N$ states).

In general we can say that 
our calculations are consistent 
with the data at relatively  high $Q^2$, 
the regime where the quark model is expected to work. Therefore,
we also used our model to predict the observables
in  the high $Q^2$ region, projected to the Jlab 12-GeV upgrade. 
As other quark models, we predict that for large $Q^2$:
$G_E \simeq -G_M$, equivalent to the 
condition $|A_{1/2}| \gg |A_{3/2}|$, 
although that asymptotic convergence is slow and 
$G_M+ G_E$ is still significant at $Q^2=20$ GeV$^2$.

Our constituent quark model is restricted to a quark-diquark 
picture which 
may not
include fully the orbital $P$-wave contributions 
to the resonance wave function. 
Nevertheless, this model
describes the main features of the
 the $\gamma^\ast N \to N^\ast(1520)$
transition form factors, and the parametrizations obtained in this work
may be very useful
for the study of this reaction, particularly 
in the timelike regime~\cite{NewPaper}.

\begin{acknowledgments}
This work was supported by the Brazilian Ministry of Science,
Technology and Innovation (MCTI-Brazil), and
Conselho Nacional de Desenvolvimento Cient{\'i}fico e Tecnol\'ogico
(CNPq), project 550026/2011-8.
This work was supported also 
by Portuguese national funds through
FCT -- Funda\c{c}\~ao para a Ci\^encia e a Tecnologia,
under Grant No.~PTDC/FIS/113940/2009,
``Hadron Structure with Relativistic Models'',
the project PEst-OE/FIS/UI0777/2011,
and partially by the European Union
under the HadronPhysics3 Grant No.~283286.
\end{acknowledgments}

\appendix

\section{$R1$-component}
\label{apStateP1}

We start with the nonrelativistic form of the state $\Psi_{R1}$.
Next we present the relativistic generalization.

\subsection{Nonrelativistic wave function}

For the case of the $R1$ component, 
for each total angular momentum projection,  $s=\pm \frac{1}{2},\pm \frac{3}{2}$
the orbital-spin states can be written as in Ref.~\cite{Capstick00}
\ba
\Psi_{R1} (s) = \frac{1}{2}
\left[
\phi_I^0 X_\rho + 
\phi_I^1 X_\lambda 
 \right]\tilde \psi_{R1},
\label{eqPsiNRP1}
\ea
where 
\ba
& & \hspace{-.5cm}
X_\rho(s) =
\sum_{m s'} \left< 1 \sfrac{1}{2};  m s' | \sfrac{3}{2} s \right>
 \left[ Y_{1m}(k_\rho) \left|s' \right>_\lambda + 
Y_{1m}(k_\lambda) \left|s' \right>_\rho
\right] \nonumber \\
& & \hspace{-.5cm}
X_\lambda(s) =
\sum_{m s'} \left< 1 \sfrac{1}{2};  m s' | \sfrac{3}{2} s \right>
\left[ Y_{1m}(k_\rho) \left|s' \right>_\rho - 
Y_{1m}(k_\lambda) \left|s' \right>_\lambda  \right], \nonumber 
\\
\label{eqP1s1}
\ea
and where the spin states 
$\left| s\right>_\rho$, $\left| s\right>_\lambda$: 
\ba
\left| s\right>_\rho =
\sum_{s_1} \left< 0 \sfrac{1}{2};  0 s_1 | \sfrac{1}{2} s \right> \chi_{s_1} 
\equiv \chi_s \\
\left| s\right>_\lambda^i =
\sum_{s_1} \left< 1 \sfrac{1}{2};  1 s_1 | \sfrac{1}{2} s \right>
\epsilon_{s-s_1}^i \chi_{s_1},
\ea 
for $s= \pm \sfrac{1}{2}$,
are antisymmetric and  symmetric, respectively,  
in the exchange of quarks 1 and 2. 
They are three-body coupled core spin states, $0 \oplus \frac{1}{2}$ 
and $1 \oplus \frac{1}{2}$ (labeled $\rho$- and $\lambda$-),
given in terms of the Pauli spinors $\chi_s$, 
and defined respectively, as the axial-scalar and axial-vector diquark terms. 
The vector $\epsilon_m^i$, (with Cartesian projections $i=1,2,3$)  
is a spin-1 state 
and corresponds to the vector diquark. 
These spin states are normalized according 
to
${_\rho}\!\left< s' | s \right>_\rho= \delta_{s' s}$,
${_\lambda}^i\!\!\left< s' | s \right>_\lambda^j= \frac{1}{3}\delta_{s' s} 
\delta_{ij}$, 
$\sum_i {_\lambda}^i\!\!\left< s' | s \right>_\lambda^i = \delta_{s' s}$ 
and 
${_\rho}\!\left< s' | s \right>_\lambda= 
{_\lambda}\!\!\left< s' | s \right>_\rho= 0$.

We restrict the  two  momentum projections $s$ to 
$s=+ \frac{1}{2},+ \frac{3}{2}$ since symmetries  
relate the remaining cases to those.
Therefore, with the notation
$\left| \pm \right>_\rho \equiv\left| \pm \sfrac{1}{2}\right>_\rho $,
$\left| \pm \right>_\lambda \equiv 
\left| \pm \sfrac{1}{2}\right>_\lambda$,
we write
\ba
X_\rho \left( + \sfrac{1}{2}\right) &=& +
\frac{1}{\sqrt{3}} 
\left[ 
Y_{1+1}(r)\left| -\right>_\lambda  + \sqrt{2} Y_{10}(r)
\left| +\right>_\lambda
\right]
\nonumber \\
& & + 
\frac{1}{\sqrt{3}} 
\left[ 
Y_{1+1}(k)\left| -\right>_\rho  + 
\sqrt{2} Y_{10}(k)
\left| +\right>_\rho
\right] \nonumber \\
X_\rho \left( + \sfrac{3}{2}\right) &=& 
Y_{1 +1}(r) \left| -\right>_\lambda +
 Y_{1 +1}(k) \left| -\right>_\rho, 
\ea
and
\ba
X_\lambda \left( + \sfrac{1}{2}\right) &=& +
\frac{1}{\sqrt{3}} 
\left[ 
Y_{1+1}(r)\left| -\right>_\rho  + \sqrt{2} Y_{10}(r)
\left| +\right>_\rho
\right]
\nonumber \\
& & - 
\frac{1}{\sqrt{3}} 
\left[ 
Y_{1+1}(k)\left| -\right>_\lambda  + 
\sqrt{2} Y_{10}(k)
\left| +\right>_\lambda
\right] \nonumber \\
X_\lambda \left( + \sfrac{3}{2}\right) &=& 
Y_{1 +1}(r) \left| +\right>_\rho -
 Y_{1 +1}(k) \left| +\right>_\lambda. 
\ea

\subsection{Relativistic generalization}

We collect now all the prescriptions, namely
 Eqs.~(\ref{eqIntR}) and (\ref{eqYrel1}), 
to obtain the relativistic generalization
the orbital-spin states $X_\rho(s)$ and $X_\lambda(s)$. 
In addition to those prescriptions 
the relativistic forms of 
$\left| s \right>_\rho $ and $\left| s \right>_\lambda$
are
\ba
& &
\left| s \right>_\rho \to u_R(P,s) \nonumber \\
& &
\left| s \right>_\lambda \to 
- \left( \varepsilon_P^\ast \right)_\alpha U_R^\alpha(P,s), 
\label{eqSpinS}
\ea
where $u_R$ is a Dirac spinor, and
$U_R^\alpha (P,s)$ is given by Eq.~(\ref{eqUR}).
The procedure was used already in previous 
applications \cite{Nucleon,Nucleon2,S11}.
One has, collecting all the transformations:

\ba
& &
\hspace{-.5cm}
X_\rho \left( + \sfrac{1}{2}\right) =
- \frac{1}{\sqrt{3}} 
\left[ 
\zeta_+^\nu  (\varepsilon_{\Lambda P}^\ast)_\alpha 
U_R^\alpha(-) +  
\sqrt{2} \, \zeta_0^\nu (\varepsilon_{\Lambda P}^\ast)_\alpha 
U_R^\alpha(+)
\right]
\nonumber \\
& & 
\hspace{-.5cm}
\qquad
-
N_{\tilde k}
\left[ \frac{1}{\sqrt{3}} (\varepsilon_- \cdot \tilde k) u_R(-) +
\sqrt{\frac{2}{3}} 
 (\varepsilon_0 \cdot \tilde k)
u_R(+)
\right] \nonumber \\
& & 
\hspace{-.5cm}
X_\rho \left( + \sfrac{3}{2}\right) = 
- \zeta_+^\nu (\varepsilon_{\Lambda P}^\ast)_\alpha 
U_R^\alpha(+)
- N_{\tilde k}
(\varepsilon_+ \cdot \tilde k) u_R(+), 
\ea
and
\ba
& &
\hspace{-.5cm}
X_\lambda \left( + \sfrac{1}{2}\right) = 
 \frac{1}{\sqrt{3}} 
\left[ 
\zeta_+^\nu \, u_R(-) + \sqrt{2}
\zeta_0^\nu \, u_R(+) 
\right]
\nonumber \\
& &
\hspace{-.5cm}
 - 
N_{\tilde k}
\left[ 
\frac{1}{\sqrt{3}} 
(\varepsilon_+ \cdot \tilde k) (\varepsilon_{\Lambda P}^\ast)_\alpha U_R^\alpha(-)
+
\sqrt{\frac{2}{3}} 
(\varepsilon_0 \cdot \tilde k) (\varepsilon_{\Lambda P}^\ast)_\alpha U_R^\alpha(+)
\right] \nonumber \\
& &
\hspace{-.5cm}
X_\lambda \left( + \sfrac{3}{2}\right)=
 \zeta_+^\nu u_R(+) -
N_{\tilde k}
(\varepsilon_+ \cdot \tilde k) 
(\varepsilon_P^\ast)_\alpha U_R^\alpha(+). 
\ea
Here, and in the following, for simplicity 
we adopt the notation 
$U_R^\alpha (\pm) \equiv U_R^\alpha \left(\pm \sfrac{1}{2} \right)$,
$u_R (\pm) \equiv u_R \left(\pm \sfrac{1}{2} \right)$,
and omit also the momentum $P$ from the labeling
of the spin states.

Finally, we may re-write these states in a short-hand notation by noting that
the Rarita-Schwinger vector spin 
is, in the rest frame~\cite{NDelta} 
\ba
u^\beta(s) &=& \sum_{s'} 
\left< 1 \sfrac{1}{2};  (s-s') \, s' | \sfrac{3}{2} s \right> 
\varepsilon_{s-s'}^\beta u_R(s')
\label{DD} \\
&=& 
\left\{
\begin{array}{ccc} \varepsilon_+^\beta u_R(+)& & s=+\frac{3}{2} \cr 
\sqrt{\frac{2}{3}} 
\varepsilon_0^\beta u_R(+)
+ \sqrt{\frac{1}{3}} 
\varepsilon_+^\beta u_R(-)
& & s=+\frac{1}{2} \cr 
\sqrt{\frac{2}{3}} 
\varepsilon_0^\beta u_R(-)
+ \sqrt{\frac{1}{3}} 
\varepsilon_-^\beta u_R(+)
& & s= -\frac{1}{2} \cr 
\varepsilon_-^\beta u_R(-)& & s=-\frac{3}{2} \cr
\end{array}
\right. ,
\nonumber 
\ea
using the state $u_\zeta^\nu(s')$ defined by Eq.~(\ref{eqZeta1}),
and
\ba
{\cal T}_R
= \frac{1}{\sqrt{3}}
(\varepsilon_P^\ast)_\alpha
\gamma_5 
\left( \gamma^\alpha - \frac{P^\alpha }{ M_R} \right).
\ea
We obtain then
\ba
X_\rho (s) &=&
- {\cal T}_R  u_\zeta^\nu (s) + 
N_{\tilde k}  
\tilde k^\beta u_\beta(s) \\
X_\lambda (s) &=&
u_\zeta^\nu (s) - 
N_{\tilde k}  {\cal T}_R
\tilde k^\beta u_\beta(s). 
\ea

Finally we can write the $R1$-state relativistic wave function as
\ba
\Psi_{R1}(s)&=&  N_{R1} \Big\{
(- {\cal T}_R \phi_I^0 + \phi_I^1) u_\zeta^\nu (s)  
 \nonumber \\
& &
-
N_{\tilde k}
\left(
\phi_I^0 + {\cal T}_R \phi_I^1\right) 
\tilde k^\beta
u_\beta(s)
\Big\} \psi_{R1}(P,k), 
\nonumber \\
\ea
where $\psi_{R1}$ generalizes the function $\tilde \psi_{R1}$.
The constant $N_{R1}$ was introduced by convenience 
and is related to the function $\psi_{R1}$.

If we want to suppress the diquarks 
with internal $P$-states (pointlike diquark limit),
we should remove the terms in $u_\zeta^\nu$.

By construction the wave function $\Psi_{R1}$ 
is a solution of the Dirac equation:
${\not \! P} \Psi_{R1} = M_R \Psi_{R1}$.

\subsection{Normalization}

The relativistic wave function, as the nonrelativistic one,
is normalized by the charge condition 
\ba
Q&=&\sum_{\nu \Gamma } \int_k \Psi_{R1}^\dagger (\bar P,k) (3j_1) 
\Psi_{R1}(\bar P,k), 
\nonumber \\
& =& \frac{1}{2}(1 + \tau_3),
\label{eqCharge}
\ea
defined at $Q^2=0$ for $\bar P=(M_R,0,0,0)$.
Note the inclusion of the index $\nu$ 
in order to take into account the contributions 
of the $P$-state diquarks. 
Using the previous definition, we obtain
\ba
Q= 6(j_1^A+ j_1^S)  
N_{R1}^2 \int_k |\psi_{R1}(\bar P,k)|^2,
\ea
with $j_1^A$ and $j_1^S$ defined by Eqs.~(\ref{eqJA})-(\ref{eqJS}) 
for $Q^2=0$.
As $3(j_1^A+ j_1^S) = (1+ \tau_3)$, 
and  imposing 
\ba
\int_k |\psi_{R1}(\bar P,k)|^2 =1,
\ea
we obtain the condition $N_{R1}^2= 1/4$, or 
\ba
N_{R1}= \frac{1}{2}.
\ea
We recover then the nonrelativistic normalization given by 
Eq.~(\ref{eqPsiNRP1}). 

Note however that if we suppress the diquarks
with internal $P$-states 
one obtain instead $N=1/\sqrt{2}$.
That was the option considered in 
the study of the $\gamma^\ast N \to N^\ast(1535)$ 
reaction~\cite{S11}.

\section{{$R3$-component}}
\label{apStateP3}

The $R3$ component of the  $N^\ast(1520)$ wave function,
which corresponds to core spin 3/2,
in the nonrelativistic framework, 
is defined as the coupled configuration $1 \oplus \frac{3}{2} \to \frac{3}{2}$.
One can write then~\cite{Capstick00} 
\ba
\Psi_{R3} (s)=
\frac{1}{\sqrt{2}}
\left[
\phi_I^0 X_\rho(s) + \phi_I^1 X_\lambda(s)
\right] \tilde \psi_{R3},
\label{eqPsiP3NR}
\ea 
where $\tilde \psi_{R3}= \tilde \psi_{R3}(r,k)$, and now
\ba
& &
X_\rho (s)=
\sum_{ms'} 
\left< 1 \sfrac{3}{2}; m s' | \sfrac{3}{2} s
\right> Y_{1m} (r)
\chi^S_{s'} 
\\
& &
X_\lambda (s)=
\sum_{ms'} 
\left< 1 \sfrac{3}{2}; m s' | \sfrac{3}{2} s
\right> Y_{1m} (k)
\chi^S_{s'}, 
\ea
where $\chi^S_{s'}$ is the totally symmetric
spin state.  

To obtain the relativistic extension of 
the wave function $R3$ component, we apply the replacements 
for the spherical harmonics (\ref{eqIntR}), (\ref{eqYrel1}) 
and the relativistic generalization of $\chi_{s'}^S$ 
given in terms of the Rarita-Schwinger vector spin $u_\beta$
\ba
\chi_{s}^S \to - (\varepsilon_{\Lambda P}^\ast)^\beta u_\beta(P,s).
\label{eqChiS}
\ea

With everything together, the relativistic 
wave function becomes the expression 
given by Eqs.~(\ref{eqPsiP3})-(\ref{eqPsiP3B}).
The last step was the replacement 
$\tilde \psi_{R3} \to \psi_{R3}(P,k)$.

The wave function (\ref{eqPsiP3}) is normalized
using the charge condition equivalent 
to (\ref{eqCharge}), with
\ba
\int_k |\psi_{R3}(\bar P,k)|^2=1,
\ea
and leads us to the factor $1/\sqrt{2}$, in 
Eq.~(\ref{eqPsiP3B}), identical 
to the nonrelativistic case \cite{Capstick00}
given by Eq.~(\ref{eqPsiP3NR}).

The state $\Psi_{R3}$  has 
the property ${\not \! P} \Psi_{R3} = - M_R \Psi_{R3}$.
The $-$ sign is a consequence 
of the factor $\gamma_5$ 
in Eq.~(\ref{eqPsiP3B}),
which contrasts with the state $R1$.

\section{Phenomenological radial functions}
\label{radial}

The phenomenological radial functions that are part of the baryon wave functions are
written
in terms of the scalar quantity $P \cdot k$. More specifically, they are written as functions of the dimensionless
variable $\chi$ which in the nonrelativistic limit
becomes proportional to ${\bf k}^2$, and is defined as

\ba
\chi_B = \frac{(M_B-m_D)^2-(P-k)^2}{M_B m_D}.
\label{eqChi}
\ea
In this formula $M_B$ is the baryon mass 
($M$ for the nucleon or $M_R$ for the excited state).

The nucleon radial wave function is 
\cite{Nucleon}
\ba
\psi_N(P,k)= \frac{N_0}{
m_D(\beta_1 + \chi_N)(\beta_2 + \chi_N)},
\label{eqPsiNrad}
\ea
where $\beta_1,\beta_2$ are two momentum range parameters
and $N_0$ is a normalization constant.
If $\beta_2 > \beta_1$, 
$\beta_2$ regulates the short range behavior and 
$\beta_1$ the long range  behavior in configuration space.

For the  $R1$-state component we use
\ba
\psi_{R1}(P,k)= 
 \frac{N_1}{ m_D (\beta_2 + \chi_R)}
\left[
\frac{1}{(\beta_1 + \chi_R)}-
\frac{\lambda_{R1}}{ (\beta_3 + \chi_R) } \right],
\nonumber \\
\label{eqPsiRP1}
\ea 
where $N_1$ is a normalization constant, 
$\beta_3$ is a new (short range) parameter,
and $\lambda_{R1}$ will be determined 
by the  orthogonality condition (\ref{eqOrthCondition})
between the nucleon and that component of the $N^*$ wave function.

For the state $R3$ we choose also a form with two terms, but with a
parametrization similar to the one used in another work 
for  the radial wave function of the $\Delta(1232)$ 
\cite{LatticeD} in the $S$-state. It reads
\ba
\psi_{R3}(P,k)= 
 \frac{N_3}{ m_D (\alpha_1 + \chi_R)^3}
\left( 1 - \frac{ \lambda_{R3} }{ \alpha_1 + \chi_R} \right),
\label{eqPsiP3R}
\ea 
where $N_3$ is a normalization constant,
$\alpha_1$ is the momentum range, taken to be the same as the one
of the $\Delta(1232)$ case,  
and $\lambda_{R3}$ a coefficient to be determined 
also by  the orthogonality condition (\ref{eqOrthCondition}).

At the moment then we introduce in our model 
only one more range parameter ($\beta_3$) 
to add to the already calibrated range parameters 
of the nucleon radial wave function.

\section{Transition current --- $R1$ component}
\label{apFF-P1}

Using the expressions given for 
the nucleon and $\Psi_{R1}$ wave functions,
we can calculate the transition current
in relativistic impulse approximation
\cite{Nucleon,Nucleon2,Omega}
\ba
J^\mu_{NR}   &\equiv &
3 \sum_{\Gamma} \int_k
\overline \Psi_{R1}(P_+,k) j_q^\mu \Psi_N(P_-,k).
\label{eqJ3}
\ea
Recall that 
\ba
j_q^\mu= j_1 \hat \gamma^\mu + j_2 {\cal O}^\mu,
\ea
where ${\cal O}^\mu = \frac{i \sigma^{\mu \nu} q_\nu}{2M}$.

To work the spin algebra
we project the quark current $j_q^\mu$ into 
the isospin states defining the coefficients
\ba
j_i^A= \left( \phi_I^0 \right)^\dagger j_i \phi_I^0 &=& j_i 
\label{eqJA}
\\
j_i^S= \left( \phi_I^1 \right)^\dagger j_i \phi_I^1 &=& \sfrac{1}{3}
\tau_j j_i \tau_j \nonumber \\
&=& \sfrac{1}{6}f_{i+} -\sfrac{1}{6} f_{i-} \tau_3.
\label{eqJS}
\ea

In both cases we include
the effect of the diquark polarization in the transition. This is done by 
summing  the initial and final polarization 
vectors, and one obtains
\cite{NDelta,FixedAxis}:
\ba
\Delta^{\alpha \beta} &\equiv& 
\sum_{\Lambda} (\varepsilon_{\Lambda P_+})^\alpha
 (\varepsilon_{\Lambda P_-}^\ast)^\beta 
\nonumber \\
&=& 
- \left( g^{\alpha \beta} 
- \frac{P_-^\alpha P_+^\beta}{x} \right) \nonumber \\
& & 
- a \left( P_- -\frac{x}{M_R^2} P_+ \right)^\alpha 
\left( P_+ -\frac{x}{M^2} P_- \right)^\beta, 
\label{eqDelta}
\ea
where $x= P_+ \cdot P_-$, and 
\ba
a= \frac{M M_R}{x \left[M M_R + x\right]}.
\ea

As the states $\Psi_{R1}$ and $\Psi_N$ are 
solutions of the Dirac equation and 
the the asymptotic states are on-mass-shell 
we can simplify the operators $\hat \gamma^\mu$ 
and ${\cal O}^\mu$ to 
\ba
& &
\hat \gamma^\mu \to 
\gamma^\mu - \frac{M_R-M}{q^2} q^\mu \nonumber \\
& & 
{\cal O}^\mu \to \frac{M_R + M}{2M} \gamma^\mu - \frac{P^\mu}{M},
\label{eqGordon}
\ea
recalling that $P=\frac{1}{2}(P_+ + P_-)$.

The direct calculation gives
\ba
\sum_{\Lambda} \overline \Psi_{R1} (3j_q^\mu) \Psi_N &=&
{\cal A}
 \left[
\bar u_\beta \tilde k^\beta
\left\{
j_1^A \hat \gamma^\mu + j_2^A {\cal O}^\mu \right\}u  
\right. \nonumber \\
& &
\left.
- \frac{1}{3}
\bar u_\beta \tilde k^\beta
\left\{
j_1^S
\gamma^\alpha \hat \gamma^\mu \gamma^\sigma \Delta_{\alpha \sigma}
\right\}u \right. \nonumber \\
& &
\left.
+ \frac{1}{3}
\bar u_\beta \tilde k^\beta
\left\{
j_2^S
\gamma^\alpha \tilde {\cal O}^\mu \gamma^\sigma \Delta_{\alpha \sigma}
\right\}u 
\right], \nonumber \\
& &
\label{eqSum1}
\ea
where $\Delta_{\alpha \sigma}$ 
is defined by Eq.~(\ref{eqDelta}),
${\cal A}= - \frac{3 N N_{\tilde k}}{\sqrt{2}} \psi_{R1} \psi_N$,
$\tilde {\cal O}^\mu = \gamma_5 {\cal O}^\mu \gamma_5$,
and $j_i^A$ and $j_i^S$ defined by Eqs.~(\ref{eqJA}) and (\ref{eqJS})
include the effect of the isospin.
In the derivation (\ref{eqSum1}) we 
take advantage also of the relations 
$P_+^\alpha \Delta_{\alpha \sigma}= \Delta_{\alpha \sigma}P_-^\sigma=0$.

We reduce therefore the calculation 
of the transition current $J_{NR}^\mu$ to the calculation 
of $\gamma^\alpha \hat \gamma^\mu \gamma^\sigma \Delta_{\alpha \sigma}$
and 
$\gamma^\alpha \tilde {\cal O}^\mu \gamma^\sigma \Delta_{\alpha \sigma}$,
where
\ba
\tilde {\cal O}^\mu  \to 
- \frac{M_R + M}{2M} \gamma^\mu - \frac{P^\mu}{M}.
\ea
We start calculating
\ba
\gamma^\alpha \gamma^\sigma \Delta_{\alpha \sigma} &=&-3 \nonumber \\
\gamma^\alpha \gamma^\mu \gamma^\sigma \Delta_{\alpha \sigma}& =&
- \gamma^\mu + 4 A (M_R + M) P^\mu \nonumber \\
& & 
- 2 A(M_R-M) q^\mu,
\label{eqContraction}
\ea
where 
$A= \frac{2}{(M_R+M)^2+Q^2}$.
The previous relations are 
valid when projected in the states $\bar u_\beta(P_+)$ and $u(P_-)$.

Using Eqs.~(\ref{eqContraction}) 
we derive 
\ba
\gamma^\alpha \hat \gamma^\mu \gamma^\sigma \Delta_{\alpha \sigma}&=&
-\hat \gamma^\mu + 4A(M_R+M) \left(P^\mu - \frac{P \cdot q}{q^2} q^\mu\right)
\nonumber \\
&=&- \gamma^\mu + 4A (M_R+M) P^\mu + B \frac{M_R-M}{q^2} q^\mu, \nonumber \\
& & 
\label{eqHgamma}
\\
\gamma^\alpha \tilde {\cal O}^\mu \gamma^\sigma \Delta_{\alpha \sigma}&=&
 \frac{M_R + M}{2M} \gamma^\mu + 
\left[ 
3- 2 A(M_R + M)^2   
\right] \frac{P^\mu}{M} \nonumber \\
& &
+ A \frac{M_R + M}{M} (M_R-M) q^\mu,
\label{eqOmu}
\ea
where $B= -\frac{3(M_R+M)^2- Q^2}{(M_R+M)^2+Q^2}$.
The relations (\ref{eqHgamma}) and (\ref{eqOmu})
are valid between
asymptotic states.

With the relations (\ref{eqGordon}), 
(\ref{eqHgamma}) and (\ref{eqOmu})
we can represent (\ref{eqSum1}) as
\ba
\sum_{\Lambda} \overline \Psi_{R1} (3j_q^\mu) \Psi_N =
{\cal A} \, 
 \bar u_\beta \tilde k^\beta
\left[
g_1 \gamma^\mu + g_2 P^\mu + g_3 q^\mu
\right] u, 
\nonumber \\
\ea
where
\ba
g_1&=&\left( j_1^A + \frac{1}{3} j_1^S\right)+ 
\frac{M_R+M}{2M}\left( j_2^A + \frac{1}{3} j_2^S\right) \\
g_2&=& -\frac{1}{M} 
\left[
j_2^A 
+ \frac{1}{3}  \frac{1- 3 \tau}{1 + \tau}   j_2^S
+ 
\frac{4}{3} \frac{2 M}{M_R +M} \frac{1}{1 + \tau} j_1^S 
\right] \nonumber \\
& & \\
g_3&=& 
\frac{M_R-M}{Q^2} \times \nonumber \\
& &
\left[
j_1^A + \frac{1}{3}\frac{\tau-3}{1+ \tau} j_1^S 
+ \frac{4}{3} \frac{M_R +M}{2M} \frac{\tau}{1+ \tau} j_2^S
\right] , \nonumber \\
\ea
with $\tau= \frac{Q^2}{(M_R+M)^2}$.

We can now calculate the current $J_{NR}^\mu$ 
given by Eq.~(\ref{eqJ3}), performing the integration in $k$:
\ba
J_{NR}^\mu = - 
\frac{3 N}{\sqrt{2}} 
{\cal I}^\beta \,
\bar u_\beta \left\{ 
g_1 \gamma^\mu + g_2 P^\mu + g_3 q^\mu
\right\} u ,  
\label{eqJ4}
\ea
where
\ba
{\cal I}^\beta = \int_k \tilde k^\beta N_{\tilde k} \psi_{R1}(P_+,k) 
\psi_N(P_-,k).
\ea

The previous integral is covariant,
therefore the result is frame independent.
We can write ${\cal I}^\beta$ in a covariant form as
\ba
{\cal I}^\beta &=&
\frac{\tilde q^\beta}{|{\bf q}|}
I_z^{R1} 
\label{eqIbeta}
\ea
where $\tilde q^\beta= q^\beta - \frac{P  \cdot q}{M_R^2}P^\beta$, 
$|{\bf q}|= \sqrt{-\tilde q^2}$ is an invariant, 
and
\ba
I_z^{R1}= - \int_k  N_{\tilde k} 
(\varepsilon_0 \cdot \tilde k) \psi_{R1}(P_+,k) \psi_N(P_-,k),
\ea
is an invariant scalar function.
In the resonance $R$ rest frame one has the simple form
\ba
I_z^{R1}= \int_k \frac{k_z}{|{\bf k}|} 
\psi_{R1} (P_+,k)\psi_N (P_-,k),
\ea
where
\mbox{$P_+= (M_R,0,0,0)$} and 
\mbox{$P_-=(E_N,0,0,-|{\bf q}|)$,} with
$E_N= \frac{M_R^2+ M^2+ Q^2}{2M_R}$ and
$|{\bf q}|$ is given by Eq.~(\ref{eqq2}).

Combining Eq.~(\ref{eqJ4}) with (\ref{eqIbeta})
we can write
\ba
J_{NR}^\mu =
\bar u_\beta (P_+)\left\{
G_1 q^\beta \gamma^\mu + G_2 q^\beta P^\mu 
+ G_3 q^\beta q^\mu 
\right\} u (P_-), \nonumber \\
\ea
where 
\ba
G_i = - \frac{3 N}{\sqrt{2}} \frac{I_z^{R1}}{|{\bf q}|} g_i,
\ea
for $i=1,2,3$.
As there is no term in $g^{\beta \mu}$, we conclude 
that 
\ba
G_4=0.
\ea

As the contribution of the $R1$-state in  
the $N^\ast(1520)$ wave function from Eq.~(\ref{eqPsiNX})
is proportional to
$\cos \theta_D$, the helicity amplitudes and the form factors are affected 
by the same weight.

\section{Transition current --- $R3$ component}
\label{apFF-P3}

Instead of considering the general expression 
for the current (\ref{eqJ2}), as for the $R1$-state, 
we will use in this case the 
definition of the helicity amplitudes 
(\ref{eqA12})-(\ref{eqS12}),
at the resonance rest frame.
Later we can use Eqs.~(\ref{eqA32b})-(\ref{eqS12b})
to extract the form factors.
The main reason for this procedure 
is that the components of the  
$R3$-state are not now 
written in terms of the Rarita-Schwinger 
states but there are more coefficients involved
[see Eqs.~(\ref{eqPsiP3})-(\ref{eqVeps})].

To calculate the helicity amplitudes 
we start with the current (\ref{eqJ2}):
\ba
J_{NR}^\mu (s',s)= 
\sum_\Lambda \overline\Psi_{R3}(P_+,k;s') 
(3 j_q^\mu) \Psi_N(P_-,k;s),
\ea
where the spin projections are explicitly included.
The sum symbol includes only the diquark 
polarization index ($\Lambda$) because 
only the isovector components of the 
wave functions contribute.
Next  we consider the projection with the photon polarization vector 
$\epsilon_+(q)$ and $\epsilon_0(q)$, 
not to be confused with the diquark 
polarization vectors $\varepsilon_{\Lambda P}$.
The calculations can be simplified using the 
Gordon decomposition for the quark current,
taking advantage of the relation${\not \! P_-} \Psi_N = M \psi_N$
and that ${\not \! P_+} \Psi_{R3} = - M_R \Psi_{R3}$ 
to obtain the simplification
\ba
j_q^\mu \to \left( j_1 - \frac{M_R-M}{2M} j_2 \right) \gamma^\mu
- j_2 \frac{P^\mu}{M} - \frac{M_R + M}{Q^2} j_1 q^\mu, 
\nonumber  \\
\label{eqJqTotal}
\ea 
where we recall that $P=\sfrac{1}{2}(P_+ + P_-)$.
We note that the calculations can be further reduced 
since $q \cdot \epsilon_{0,+}=0$ and 
$P \cdot \epsilon_+ =0$.
Therefore we will use $j_-$ to represent 
the effective term
\ba
j_-= j_1 - \frac{M_R-M}{2M} j_2.
\ea
Summing  in the (isovector) isospin states we obtain 
\ba
f_v &=&  
(\phi_I^1)^\dagger j_- (\phi_I^1) \nonumber \\
&=& 
j_1^S - \frac{M_R-M}{2M} j_2^S,
\ea
using the notation from Eq.~(\ref{eqJS}).

As for the state $R1$-state one can separate the 
dependence in the radial wave function into the covariant 
function
\ba
I_z^{R3} =
- \int_k N_{\tilde k} (\varepsilon_{0 P_+} \cdot \tilde k)
\psi_{R3}(P_+,k) \psi_N(P_-,k). 
\ea

The final calculation requires
the use of the Dirac spinors for the final state, $P_+= (M_R,0,0,0$)
\ba
u_R(s') = 
\left[
\begin{array}{c} 1\cr 0 \cr
\end{array}
 \right] 
\chi_{s'},
\ea
and for the initial state, $P_-= (E_N,0,0, -|{\bf q}|)$:
\ba
u(s) = N_q
\left[
\begin{array}{c} 1 \cr  -|\tilde q| \cr 
\end{array}
\right] 
\chi_{s},
\ea
where $N_q= \sqrt{\frac{E_N+ M}{2M}}$ and 
$|\tilde q|= \frac{|{\bf q}|}{M+ E_N}$.
The expressions for the final states 
are necessary because at the resonance 
rest frame we can represent 
the Rarita-Schwinger states in terms 
of the Dirac spinors using Eq.~(\ref{DD}).

For future reference we write $N_q$ in 
a covariant form
\ba
N_q= \sqrt{\frac{(M_R+M)^2+ Q^2}{4 M M_R}}.
\ea

\subsection{Amplitude $A_{3/2}$}

Considering the definition (\ref{eqA32})
and the procedures described previously, 
one can write,
using the 
wave functions $\Psi_{R3}\left( P_+,k; + \sfrac{3}{2} \right)$
and $\Psi_{N}\left( P_-,k; + \sfrac{1}{2} \right)$ 
\ba
A_{3/2}=
- \frac{3}{2} \sqrt{\frac{2\pi \alpha}{K}}
  C f_v I_z^{R3} \, T_{3/2},
\ea
where $C= \left< 1 \frac{3}{2}; 0 +\frac{3}{2} | 
\sfrac{3}{2} + \sfrac{3}{2} \right>= \sqrt{\frac{3}{5}}$, and 
\ba
T_{3/2} = 
- \frac{1}{\sqrt{3}}
\left[
\bar u_\beta \left( + \sfrac{3}{2} \right)
\gamma^\mu (\epsilon_+)_\mu \gamma_\alpha u \left( + \sfrac{1}{2} \right)
\right] \Delta^{\beta \alpha}.
\ea
Using the property $q^\beta \bar u_\beta\left( + \sfrac{3}{2} \right)=0$,
we can reduce the previous expression to
\ba
T_{3/2} = - \frac{2}{\sqrt{3}} N_q.
\ea

The final expression for the amplitude 
is then
\ba
A_{3/2} = \frac{3}{\sqrt{5}}
\sqrt{\frac{2 \pi \alpha}{K}}
 N_q
f_v I_z^{R3}. 
\ea

The corresponding expression for $G_4$,
given by Eq.~(\ref{eqA32b}),  is 
\ba
G_4= \frac{3}{\sqrt{5}}  f_v 
I_z^{R3}.
\ea

\subsection{Amplitude $A_{1/2}$}

We start with the expression (\ref{eqA12})
and use the
wave functions $\Psi_{R3}\left( P_+,k; + \sfrac{1}{2} \right)$
and $\Psi_{N}\left( P_-,k; - \sfrac{1}{2} \right)$.
Based in the previous discussion we reduce 
the calculation to
\ba
A_{1/2}=
- \frac{3}{2}
\sqrt{\frac{2\pi \alpha}{K}}
  C f_v \, I_z^{R3}\, T_{1/2},
\ea
where $C= \left< 1 \frac{1}{2}; 0 +\frac{1}{2} | 
\sfrac{3}{2} + \sfrac{1}{2} \right>$, and 
\ba
T_{1/2} = 
- \frac{1}{\sqrt{3}}
\left[
\bar u_\beta \left( + \sfrac{1}{2} \right)
\gamma^\mu (\epsilon_+)_\mu \gamma_\alpha u \left( + \sfrac{1}{2} \right)
\right] \Delta^{\beta \alpha}.
\ea

Using the expression
\ba
\bar u_\beta \left( + \sfrac{1}{2} \right) =
\sqrt{\frac{3}{2}} 
\bar u_R \left( + \sfrac{1}{2} \right)
(\varepsilon_0^\ast)_\beta 
+ \sqrt{\frac{3}{2}} 
\bar u_R \left( - \sfrac{1}{2} \right)
(\varepsilon_+^\ast)_\beta, \nonumber \\
\ea
we can continue with the calculation,
using the results $\varepsilon_0^\ast \cdot \epsilon_+ =0$,
$\varepsilon_+^\ast \cdot \epsilon_+ =-1$, 
$\varepsilon_+^\ast \cdot q=0$ and $\varepsilon_0^\ast \cdot q= -|{\bf q}|$.
One obtains then
\ba
T_{1/2} &=&
 \frac{2}{3} 
\left[
\bar u_R \left( - \sfrac{1}{2} \right) 
u \left( + \sfrac{1}{2} \right) \right] =0.
\ea
The final result is a consequence of the 
orthogonality of the states 
$u_R \left( - \sfrac{1}{2} \right)$ and
$u \left( + \sfrac{1}{2} \right)=0$.

In conclusion 
\ba
A_{1/2}=0.
\ea

Because $A_{3/2}=0$, we can write $G_E= 3G_M $
and \mbox{$A_{3/2} = - \frac{\sqrt{3}}{F} G_M$}
[see Eqs.~(\ref{eqA12d})-(\ref{eqA32d})].

\subsection{Amplitude $S_{1/2}$}

We consider now the amplitude $S_{1/2}$, as defined by Eq.~(\ref{eqS12}),
using the
wave functions $\Psi_{R3}\left( P_+,k; + \sfrac{1}{2} \right)$
and $\Psi_{N}\left( P_-,k; + \sfrac{1}{2} \right)$.
We start noting that when the current conservation 
is assured, as in the present case, we can replace 
$(\epsilon_0 \cdot J_{NR}) \frac{|{\bf q}|}{Q}$ by $J_{NR}^0$, 
defined by Eq.~(\ref{eqJ11}) with 
$J_{NR}^\mu$ replaced by $J^0_{NR}$.

We start with the calculation of 
the matrix element at the $R$ rest frame
\ba
J_{NR}^0= 
\sum_\Lambda \int_k
\overline\Psi_{R3}\left(P_+,k; + \sfrac{1}{2}\right) 
(3 j^\prime_-) \Psi_N \left(P_-,k; + \sfrac{1}{2}\right), \nonumber \\
\label{eqJ0x}
\ea
where 
\ba
j^\prime_-=
\left( 1- \frac{M_R +M}{Q^2} q^0 \right) j_1 \gamma^0 - 
\left( \frac{P^0}{M} + \frac{M_R-M}{2M} \right) j_2, 
\nonumber \\
\ea
as derived from Eq.~(\ref{eqJqTotal}).

We can simplify the calculation using 
the relation $\overline \Psi_{R3}(P_+,k, s') \gamma^0 =
\overline \Psi_{R3}(P_+,k, s')
\sfrac{
{\not  P_+}}{M_R} = - \overline \Psi_{R3}(P_+,k, s')$,
valid at the rest frame.
The components $P^0$ and $q^0$ refer to the $R$ rest frame
\ba
P^0= \frac{3M_R^2+ M^2+ Q^2}{4M_R} \\
q^0= \frac{M_R^2- M^2 - Q^2}{2M_R}.
\ea

In that case Eq.~(\ref{eqJ0x}) is still valid 
with
\ba
j^\prime_- &\to& - j^\prime_- \nonumber \\
  &=& - \left( 1- \frac{M_R +M}{Q^2} q^0 \right) j_1  - 
\left( \frac{P^0}{M} + \frac{M_R-M}{2M} \right) j_2. 
\nonumber \\
\ea

Taking in consideration the isospin effect, one has
\ba
f_v^\prime &=& (\phi_I^1)^\dagger j^\prime_- (\phi_I^1) \nonumber \\
&=& \frac{(M_R+M)^2 + Q^2}{2M_R} \left[
  \frac{M_R -M}{Q^2} j_1^S + \frac{j_2^S}{2M}
\right]. \nonumber  
\\
\ea
In these conditions we can write
\ba
S_{1/2} = - \frac{3}{2}
 \sqrt{ \frac{2\pi \alpha}{K}} 
 C f_v^\prime I_z^{R3} \, T_0,
\ea
where  
$C= \left< 1 \frac{1}{2}; 0 +\frac{1}{2} | 
\sfrac{3}{2} + \sfrac{1}{2} \right> = \sfrac{1}{\sqrt{15}}$, and 
\ba
T_0=  \frac{1}{\sqrt{3}} 
\left[  \bar u_\beta \left( + \sfrac{1}{2}\right)
\gamma_\alpha   u \left( + \sfrac{1}{2}\right)
\right] \Delta^{\beta \alpha}.
\ea
The explicit calculation gives 
\ba
T_0=
 \frac{4 \sqrt{2}}{3}  \frac{M_R |{\bf q}|}{(M_R+M)^2+ Q^2} N_q. 
\ea

At the end we obtain 
\ba
S_{1/2} = 
 - \sqrt{\frac{2}{15}}
\sqrt{\frac{2\pi \alpha}{K}} 
N_q |{\bf q}| 
\bar f_v^\prime I_z^{R3},
\ea
where
\ba
\bar f_v^\prime &=&
\frac{2 M_R}{(M_R+M)^2+ Q^2} f_v^\prime \nonumber \\
&=& \frac{M_R-M}{Q^2} j_1^S + \frac{j_2^S}{2M}.
\ea

As for $G_C$, one has
\ba
G_C= 
- 2   \sqrt{\frac{2}{15}}
\left( 
\frac{M M_R}{Q^2} j_1^S + \frac{M_R}{2(M_R-M)} j_2^S 
\right)  I_z^{R3}.
\nonumber \\
\ea

Because the contribution of the $R3$-state in  
the $N^\ast(1520)$ wave function from Eq.~(\ref{eqPsiNX})
is proportional to
$- \sin \theta_D$, the helicity amplitudes and the form factors are affected 
by the same weight.

\end{document}